\documentclass[aps,prd,nofootinbib,eqsecnum,10pt,superscriptaddress]{revtex4-2}

\usepackage{graphicx}
\usepackage{bm}
\usepackage{amsfonts}
\usepackage{amsmath,amsthm,amssymb}
\usepackage{xcolor}
\usepackage{dcolumn}
\usepackage{booktabs}
\usepackage{makecell}
\usepackage{multirow}
\usepackage{tabularx}
\usepackage{caption}
\usepackage{ragged2e}
\usepackage{nameref}
\usepackage{hyperref}

\allowdisplaybreaks[1]

\begin{document}

\newcommand{\be}{\begin{equation}}
\newcommand{\ee}{\end{equation}}
\newcommand{\ba}{\begin{eqnarray}}
\newcommand{\ea}{\end{eqnarray}}
\newcommand{\D}{\partial}
\newcommand{\Mpl}{M_{\rm pl}}
\newcommand{\tF}{\tilde{F}}
\newcommand{\tb}{\tilde{\beta}}

\title{
Dyonic hairy black holes in $U(1)$ gauge-invariant scalar-vector-tensor theories: Cubic and quartic SVT sectors
}

\author{Masaki Kitagawa}
\affiliation{Department of Physics, Faculty of Science,
Tokyo University of Science,
1-3 Kagurazaka, Shinjuku-ku, Tokyo 162-8601, Japan}

\author{Naoki Tsukamoto}
\affiliation{Department of Physics, Faculty of Science,
Tokyo University of Science,
1-3 Kagurazaka, Shinjuku-ku, Tokyo 162-8601, Japan}

\author{Ryotaro Kase}
\affiliation{Department of Physics, Faculty of Science,
Tokyo University of Science,
1-3 Kagurazaka, Shinjuku-ku, Tokyo 162-8601, Japan}
\affiliation{Research Center for Space System Innovation,
Research Institute for Science and Technology,
Tokyo University of Science,
2641 Yamazaki, Noda-shi, Chiba 278-8510, Japan}

\date{\today}

\begin{abstract}
We construct and classify asymptotically flat, static, and spherically symmetric hairy black hole solutions in $U(1)$ gauge-invariant scalar-vector-tensor (SVT) theories carrying both electric and magnetic charges. 
Extending previous analyses restricted to ${\cal L}_{\rm SVT}^{2}$, we incorporate the cubic and quartic SVT sectors, ${\cal L}_{\rm SVT}^{3}$ and ${\cal L}_{\rm SVT}^{4}$, respectively. 
At the covariant level, the quartic SVT sector can generate higher-order derivatives, and we derive a condition that removes them before specializing to dyonic backgrounds, where they are generically present.
We then classify the scalar hair according to the symmetry of the theory. 
In shift-symmetric theories, horizon regularity together with Noether-current conservation determines the scalar charge in terms of the remaining solution parameters, corresponding to secondary hair. 
When the couplings depend explicitly on the scalar field $\phi$, independent scalar integration constants appear in the asymptotic solutions, allowing branches with primary hair.
We also find that the magnetic charge activates the $\tilde f_3$ interaction in the cubic SVT sector, which does not contribute
in purely electric static and spherically symmetric configurations, thereby producing hairy solutions supported by the magnetic charge.
The scalar field also exhibits interaction-dependent asymptotic falloff rates. 
Combining these expansions with numerical integration, we connect the near-horizon and asymptotic regimes for all branches in the cubic sector and for one of the two quartic branches. 
For the remaining quartic branch, our result is restricted to the local near-horizon expansion.
\end{abstract}

\maketitle

\section{Introduction}
\label{sec1}

General relativity (GR) has passed high-precision tests in a weak-field regime, and its validity in the Solar System is well established~\cite{Will:2014kxa}.
Continuous advances in observational techniques are now increasing both the amount and the quality of data that probe strong-gravity environments, motivating systematic tests of GR in a strong-field regime.
The LIGO-Virgo Collaboration has directly detected gravitational waves from the merger of two black holes (BHs)~\cite{LIGOScientific:2016aoc}.
The Event Horizon Telescope (EHT) has achieved horizon-scale imaging of BH shadows: the shadow image of M87$^\ast$ and the shadow image of Sagittarius A$^\ast$ at the center of the Milky Way, obtained using very long baseline interferometry~\cite{EventHorizonTelescope:2019dse,EventHorizonTelescope:2019ths,EventHorizonTelescope:2022wkp}.
Together, these observational breakthroughs enable tests of GR and alternative theories of gravity in strong-field regimes.
BHs provide a natural laboratory for such tests.

In the Einstein-Maxwell system, stationary asymptotically flat BHs are characterized by only three parameters: mass, charge, and spin.
This is the standard no-hair paradigm~\cite{Israel:1967wq,Carter:1971zc,Ruffini:1971bza,Hawking:1971vc}.
No-hair results are also well established in scalar-tensor theories:
Bekenstein showed that a canonical scalar field does not yield independent BH hair even when coupled to gravity \cite{Bekenstein:1995un}, while Hui and Nicolis formulated no-hair theorems for shift-symmetric Galileon-type scalars coupled to gravity \cite{Hui:2012qt}.
Consequently, constructing hairy BH solutions within GR is challenging~\cite{Chew:2022enh,Chew:2023olq,Ghosh:2023kge,Chew:2024rin,Herdeiro:2015waa}.
However, numerous hairy solutions have been found both within GR and in modified gravity theories by relaxing the assumptions underlying these theorems. 
For instance, mechanisms such as spontaneous scalarization in Einstein-scalar-Gauss-Bonnet theories have been shown to induce scalar hair via curvature couplings~\cite{Doneva:2017bvd, Silva:2017uqg}. 
Of particular relevance to this paper is the coupling between a scalar field and the vector field.
Early examples include BHs coupled to a dilaton appearing in low-energy string effective theories~\cite{Gibbons:1982ih,Gibbons:1987ps,Garfinkle:1990qj,Reuter:1991cb,Monni:1995vu}
or to an axionlike scalar field~\cite{Campbell:1991rz,Lee:1991jw,Shapere:1991ta}, and more recent studies have explored axionlike scalar fields~\cite{Fernandes:2019kmh,Filippini:2019cqk,Boskovic:2018lkj}.
Nonlinear extensions of electrodynamics provide another well-defined setting where scalar–vector couplings yield nontrivial strong-field phenomenology~\cite{Stefanov:2007qw,Stefanov:2007bn,Stefanov:2007eq,Sheykhi:2014gia,Sheykhi:2014ipa,Sheykhi:2015ira,Dehghani:2019cuf}.
Within Einstein-Maxwell-scalar models, scalarization of charged BHs has been systematically studied for a broad class of couplings and potentials~\cite{Herdeiro:2018wub,Myung:2018vug,Myung:2019oua,Hod:2019ulh,Fernandes:2019kmh,Priyadarshinee:2023cmi,Promsiri:2023yda,Belkhadria:2023ooc}, and their stability has been extensively analyzed~\cite{Myung:2018vug,Myung:2019oua,Fernandes:2019kmh,Kase:2023kvq}.
Since scalar hair in dilatonic BH solutions leads to characteristic modifications of their physical properties \cite{Fernandes:2020gay,Tsukamoto:2024asy},
constructing such solutions provides an explicit route to compare GR with well-defined extensions in the strong-field regime.

In this paper, we explore hairy BH solutions in modified gravity theories where scalar and vector degrees of freedom are coupled to the gravitational sector.
To formulate systematic scalar-vector interactions free from Ostrogradsky-type instabilities, we impose that the resulting field equations contain no derivatives higher than second order~\cite{Woodard:2015zca}.
Applying this construction to a single scalar degree of freedom nonminimally coupled to gravity yields the Horndeski class, i.e., the most general scalar-tensor theory with second-order equations of motion~\cite{Horndeski:1974wa,Kolevatov:2016ppi,Kobayashi:2012wm}. 
Applying the same principle to a massive vector degree of freedom leads to the generalized Proca class~\cite{Heisenberg:2014rta,Heisenberg:2017xda,Heisenberg:2017hwb,Kase:2018owh,Kase:2018voo}.
Scalar-vector-tensor (SVT) theories provide a unified framework that contains both Horndeski- and generalized-Proca-type structures and, in addition, allows genuinely new couplings between the scalar and vector sectors~\cite{Heisenberg:2018acv}.
Imposing a \(U(1)\) gauge symmetry on the vector field further restricts this framework to the \(U(1)\) gauge-invariant SVT subclass, which includes Einstein-Maxwell-scalar theory as a special case \cite{Heisenberg:2018acv}.

Within this framework, a central ingredient of our analysis is the Dirac monopole sector~\cite{Dirac:1931kp}.
Magnetically charged BHs have been studied extensively in earlier works~\cite{Gibbons:1990um,Ortiz:1991eu,Lee:1991vy,Lee:1991qs,Breitenlohner:1991aa,Bronnikov:2000vy,Ayon-Beato:2000mjt,Goulart:2016cuv,Priyadarshinee:2021rch,Shepherd:2015dse,Meng:2018wza,Fernandes:2019kmh,Astefanesei:2019pfq,Shapere:1991ta}, and theoretical studies of how monopoles affect BHs have continued in recent years~\cite{Al-Badawi:2025urb,Liang:2025hzr,Chen:2025aom,Ahmed:2025did}.
In terms of formation and persistence, magnetically charged BHs serve as stable strong-gravity environments characterized by intense magnetic fields around a horizon.
Indeed, their magnetic charge is difficult to neutralize by the accretion of ordinary charged matter~\cite{Maldacena:2020skw}.
These points motivate a systematic study of the conditions under which the magnetic charge $P$ plays a crucial role in revealing interaction effects in the strong gravity regime, distinguishing its role from that of a trivial background charge.
The existence of the magnetic charge gives rise to nontrivial interactions that vanish or are degenerate in purely electric backgrounds, 
and also alters their perturbation structure. 
For instance, in general nonlinear electrodynamics with a Lagrangian $L(F, \tilde{F})$ depending on the two invariants, this structural change induces the mixing of odd- and even-parity sectors and affects the stability conditions against linear perturbations~\cite{Nomura:2020tpc}.
Similarly, in the quadratic SVT sector $\mathcal{L}^2_{\rm{SVT}}$ of $U(1)$ gauge-invariant SVT theories, a coupled treatment of odd- and even-parity sectors is required whenever the magnetic charge $P$ is nonzero or the Lagrangian depends on the parity-violating invariant $\tilde{F}$~\cite{Taniguchi:2025bmc}.
While Taniguchi et al.\ initiated the study of dyonic hairy BHs in this framework, their analysis was restricted to the quadratic SVT sector $\mathcal{L}^2_{\rm{SVT}}$~\cite{Taniguchi:2024ear}. 
In this paper, we extend the dyonic analysis to the full $U(1)$ gauge-invariant SVT action by incorporating the cubic and quartic SVT sectors, $\mathcal{L}^3_{\rm{SVT}}$ and $\mathcal{L}^4_{\text{SVT}}$, to systematically clarify how each interaction term induces BH hair in the presence of the magnetic charge. 
Here and throughout this paper, the terms ``cubic'' and ``quartic'' are used only as labels for the ${\cal L}_{\rm SVT}^{3}$ and ${\cal L}_{\rm SVT}^{4}$ sectors in the conventional classification of
$U(1)$ gauge-invariant SVT theories. They do not refer to the number of fields appearing in an individual
interaction vertex; in particular, terms belonging to these sectors can generate interaction vertices involving more than three or four fields.

Our main result demonstrates that $U(1)$ gauge-invariant SVT theories admit multiple families of static, spherically symmetric hairy BH solutions in the presence of a magnetic charge. 
Specifically, all hairy branches arising from the SVT coupling $f_3$ in the cubic sector can be regarded as dyonic extensions of hairy black hole solutions known in the purely electric case~\cite{Heisenberg:2018vti,Heisenberg:2018mgr,Zhang:2024cbw}.
The other SVT coupling $\tilde{f}_3$ in the cubic sector, which vanishes identically in the purely electric configuration, is also activated by the magnetic charge to induce nontrivial hair. 
Concerning the SVT couplings $f_4$ and $\tilde{f}_4$ in the quartic sector, we find that higher-order derivatives are generated already at the covariant level, before any spacetime symmetry is imposed. 
For the purely electric static and spherically symmetric configuration studied in Ref.~\cite{Heisenberg:2018vti}, these contributions cancel identically upon substituting the ansatz, so that no additional condition is required. 
This cancellation is specific to that configuration, and the corresponding terms survive on the dyonic background.
We therefore derive a condition\footnote{
Related results have recently been reported in Refs.~\cite{Gorji:2025lqr,Mironov:2025dzz}.
Their relation to the condition derived here is discussed in Sec.~\ref{sec:covariant_higher_derivatives}.
}
that eliminates these terms at the covariant level. 
Under this condition, $\phi$-dependent quartic couplings can yield branches that either retain scalar hair in a smooth $P\to0$ limit or fail to approach the Reissner--Nordstr\"om (RN) near-horizon expansion.

\begin{table*}[h]
  \centering
  \small
  \caption{Summary of dyonic hairy BH solutions classified by the interaction sectors and coupling types.}
  \label{tab:solution_summary}
  \renewcommand{\arraystretch}{1.2}
  \setlength{\tabcolsep}{0pt} 

  \begin{tabular*}{\linewidth}{@{\extracolsep{\fill}} c l l c c l l }
    \toprule
    Sector & Interaction & Coupling & Label & Section & Near-horizon limit ($P\to0$) & Remarks \\
    \midrule

    \multirow{10}{*}{$\mathcal{L}_3$} 
      & \multirow{4}{*}{$f_3$}
        & \multirow{2}{*}{Shift-symmetric}
          & $f_3$-Ia & ~\ref{f3Iasec} & \textbf{Hairy} & Extension of Ref.~\cite{Heisenberg:2018vti} \\
      & & & $f_3$-Ib & ~\ref{f3Ibsec} & \textbf{Hairy} & Extension of Ref.~\cite{Heisenberg:2018vti} \\
      \cmidrule{3-7}
      & & \multirow{2}{*}{$\phi$-dependent}
          & $f_3$-IIa & ~\ref{f3IIasec} & \textbf{Hairy} & Extension of Ref.~\cite{Heisenberg:2018vti} \\
      & & & $f_3$-IIb & ~\ref{f3IIbsec} & \textbf{Hairy} & Extension of Ref.~\cite{Heisenberg:2018vti} \\
      \cmidrule{2-7}

      & \multirow{4}{*}{$g_3$}
        & \multirow{2}{*}{Shift-symmetric}
          & $g_3$-Ia & ~\ref{g3Iasec} & RN (No scalar hair) & \textbf{New hair} \\
      & & & $g_3$-Ib & ~\ref{g3Ibsec} & RN (No scalar hair) & \textbf{New hair} \\
      \cmidrule{3-7}
      & & \multirow{2}{*}{$\phi$-dependent}
          & $g_3$-IIa & ~\ref{g3IIasec} & RN (No scalar hair) & \textbf{New hair} \\
      & & & $g_3$-IIb & ~\ref{g3IIbsec} & RN (No scalar hair) & \textbf{New hair} \\
    \midrule

    \multirow{5}{*}{$\mathcal{L}_4$} 
      & \multirow{2}{*}{$g_4$}
        & Shift-symmetric & \multicolumn{1}{c}{---} & \multicolumn{1}{c}{---} & \multicolumn{1}{c}{---} & \multicolumn{1}{c}{---} \\
      & & $\phi$-dependent & $g_4$-IIa & ~\ref{g4IIasec} & \textbf{Hairy} & \textbf{New hair} 
      \\
      \cmidrule{2-7}

      & \multirow{2}{*}{$\tilde{g}_4$}
        & Shift-symmetric & \multicolumn{1}{c}{---} & \multicolumn{1}{c}{---} & \multicolumn{1}{c}{---} & \multicolumn{1}{c}{---} \\
      & & $\phi$-dependent & $\tilde{g}_4$-IIb & ~\ref{tg4IIbsec} & No smooth RN limit & Near-horizon solution only\footnote{A connection to spatial infinity has not been established} \\

    \bottomrule
  \end{tabular*}
\end{table*}

This paper is organized as follows.
In Sec.~\ref{sec2}, we introduce the ${\cal L}_{\rm SVT}^{3}$ and ${\cal L}_{\rm SVT}^{4}$ sectors of the $U(1)$ gauge-invariant SVT theory and derive the covariant field equations together with the corresponding background equations for static and spherically symmetric dyonic configurations. 
We also introduce a parametrization of $\tilde{f}_3$ suitable for the asymptotically flat limit and derive the condition that eliminates higher-order derivatives at both the covariant and symmetry-reduced levels.
Section~\ref{sec3} analyzes the scalar-hair formation mechanism in shift-symmetric theories and identifies the $f_3$ and $g_3$ sectors as those supporting scalar hair.
In Secs.~\ref{shift_f3_sec}--\ref{phi_f4_sec}, we systematically construct asymptotic and near-horizon solutions in the relevant shift-symmetric and $\phi$-dependent cubic and quartic sectors and classify the resulting near-horizon families in Table~\ref{tab:solution_summary}.
The table also summarizes their behavior as $P\to0$ and indicates whether each family extends a known purely electric solution or is newly analyzed in the present dyonic setting.
Finally, Sec.~\ref{conclusion_sec} summarizes our main results.

Throughout the paper, we use geometrized units, setting the speed of light and the gravitational constant to unity, $c=G=1$.

\section{Action and field equation with condition avoiding higher-order derivatives}
\label{sec2}

Our starting point is the $U(1)$ gauge-invariant scalar-vector-tensor (SVT) theory described by the action~\cite{Heisenberg:2018acv}
\be
\mathcal{S} = \int d^4x\sqrt{-g}
\bigg(
  \frac{M_{\text{pl}}^2}{2}R
  +
  \sum_{i=2}^{4} \mathcal{L}_{\text{SVT}}^{i}
\bigg)\,,
\label{eq:specific_action}
\ee
where $g$ is the determinant of the metric tensor $g_{\mu\nu}$ and $R$ is the Ricci scalar. 
The first term represents the Einstein–Hilbert action with the reduced Planck mass $M_{\rm pl} = 1/ \sqrt{8\pi G}$. 
The Lagrangian densities $\mathcal{L}^2_{\rm SVT}$, $\mathcal{L}^3_{\rm SVT}$, and $\mathcal{L}^4_{\rm SVT}$ describe the interactions between the scalar and vector fields.
For later convenience, we introduce the following scalar quantities:
\be
X=-\frac{1}{2}\nabla_\mu \phi \nabla^\mu \phi\,,
\qquad
F=-\frac{1}{4}F_{\mu \nu} F^{\mu \nu}\,,
\qquad
\tilde{F}=-\frac{1}{4}F_{\mu \nu} \tilde{F}^{\mu \nu}\,,
\qquad
Y=\nabla _\mu \phi \nabla _\nu \phi F^{\mu \alpha}F^{\nu}{}_\alpha\,,
\label{eq:scalars_XF}
\ee
where $X$ is the kinetic term of the scalar field $\phi$, 
the field strength tensor and its dual are defined by
\be
F_{\mu\nu} = \nabla_{\mu} A_{\nu} - \nabla_{\nu} A_{\mu}\,,
\qquad
\tilde{F}^{\mu\nu} = \mathcal{E}^{\mu\nu\alpha\beta} F_{\alpha\beta}/2\,.
\label{eq:F_and_dual}
\ee
Here, $\mathcal{E}^{\mu\nu\alpha\beta}$ is the antisymmetric Levi-Civita tensor,
normalized as $\mathcal{E}_{\mu\nu\alpha\beta}\mathcal{E}^{\mu\nu\alpha\beta} = -4!$.
In terms of these building blocks, 
the SVT interaction Lagrangian densities are specified as
\begin{align}
    \mathcal{L}_{\text{SVT}}^2 &= f_2(\phi\,,X\,,F\,,\tilde{F}\,,Y)\,,
    \label{eq:L2_SVT}\\[2mm]
    \mathcal{L}_{\text{SVT}}^3 &= \mathcal{M}_3^{\mu \nu}\nabla_\mu\nabla_\nu\phi\,,
    \label{eq:L3_SVT}\\[2mm]
    \mathcal{L}_{\text{SVT}}^4 &= \mathcal{M}_4^{\mu \nu \alpha \beta}
      \nabla_\mu\nabla_\alpha \phi \nabla_\nu\nabla_\beta \phi
      + f_4(\phi\,,X)L^{\mu \nu \alpha \beta}F_{\mu \nu}F_{\alpha\beta}\,.
    \label{eq:L4_SVT}
\end{align}
The rank-2 tensor $\mathcal{M}_3^{\mu \nu}$ is given by
\be
 \mathcal{M}_3^{\mu \nu}
 = \left[
      f_3(\phi\,,X)g_{\rho \sigma}
      + \tilde{f}_3(\phi\,,X)\nabla_\rho \phi  \nabla _\sigma \phi
   \right]
   \tilde{F}^{\mu \rho}\tilde{F}^{\nu \sigma}\,,
\label{eq:M3_def}
\ee
where $f_{3}$ and $\tilde{f}_{3}$ are arbitrary functions of $\phi$ and $X$.
The rank-4 tensor $\mathcal{M}_4^{\mu \nu \alpha \beta}$ is defined by 
\be
 \mathcal{M}_4^{\mu \nu \alpha \beta}
 = \left[
      \frac{1}{2} f_{4,X}(\phi\,,X)
      + \tilde{f}_4(\phi)
   \right]
   \tilde{F}^{\mu \nu}\tilde{F}^{\alpha \beta}\,,
\label{eq:M4_def}
\ee
where $f_{4}=f_{4}(\phi\,,X)$ depends on $\phi$ and $X$, $f_{4,X} = \partial f_{4}/\partial X$ denotes differentiation with respect to $X$, while $\tilde{f}_{4}=\tilde{f}_{4}(\phi)$ is a function of $\phi$ only.
Throughout this paper, a comma in the subscript indicates a partial derivative with respect to the variable that follows.
For example, for $i\in\{2,3,4\}$, we use $f_{i,\phi}=\partial f_i(\phi\,,X)/\partial \phi$ and $f_{i,\phi\phi}=\partial^{2} f_i(\phi\,,X)/\partial \phi^{2}$.
The tensor $L^{\mu\nu\alpha\beta}$ entering Eq.~\eqref{eq:L4_SVT} is the double-dual of the Riemann tensor $R_{\rho\sigma\gamma\delta}$\,, defined by
\be
 L^{\mu \nu \alpha \beta}
 = \frac{1}{4}\mathcal{E}^{\mu \nu \rho \sigma}
   \mathcal{E}^{\alpha \beta \gamma \delta}
   R_{\rho \sigma \gamma \delta}\,.
\label{eq:double_dual}
\ee

We focus on static and spherically symmetric spacetime configurations with a line element given by 
\be
 ds^2
 = -f(r)dt^2
   + h^{-1}(r)dr^2
   + r^2  (d\theta^2 + \sin^2 \theta d\varphi^2)\,,
\label{eq:metric_ansatz}
\ee
where $t$ and $r$ are the time and radial coordinates. 
The event horizon is located at the radius $r=r_{h}$, satisfying
\be
 f(r_h) = 0\,,
 \qquad
 h(r_h) = 0\,.
\label{eq:horizon_def}
\ee
In this paper, we focus on the static exterior region outside the event horizon, $r > r_h$, where
\be
 f(r) > 0\,,
 \qquad
 h(r) > 0\,.
\label{eq:outside_region}
\ee
According to the spacetime symmetry, we adopt the ansatz for the scalar field and the $U(1)$ gauge potential as~\cite{Taniguchi:2024ear, Taniguchi:2025bmc}
\be
 \phi = \phi(r)\,,
 \qquad
 A_\mu = \left(V(r)\,,0\,,0\,,-P\cos\theta\right)\,,
\label{eq:field_ansatz}
\ee
where $\phi(r)$ and $V(r)$ are functions of the radial coordinate, and $P$ is a constant corresponding to the magnetic charge.
Substituting the metric and field ansatz, \eqref{eq:metric_ansatz} and \eqref{eq:field_ansatz}, into the scalar quantities defined in Eq.~\eqref{eq:scalars_XF}, we obtain
\be
 X = -\frac{h{\phi'}^{2}}{2}\,,
 \qquad
 F = \frac{h{V'}^{2}}{2f} - \frac{P^{2}}{2r^{4}}\,,
 \qquad
 \tilde{F} = -\frac{PV'}{r^{2}}\sqrt{\frac{h}{f}}\,,
 \qquad
 Y = -\frac{h^{2}{\phi'}^{2}{V'}^{2}}{f}\,.
\label{eq:scalars_reduced}
\ee
Hereafter, a prime denotes differentiation with respect to $r$, i.e., $\zeta'=d\zeta/dr$ for any function $\zeta(r)$. 
For background configurations without the magnetic charge ($P=0$), the invariant $\tilde{F}$ vanishes identically, and the derivative interaction reduces to $Y=4XF$.
In the presence of the magnetic charge, however, $\tilde{F}$ is nonzero, and the relation $Y = 4XF$ no longer holds.
Substituting these expressions into Eq.~\eqref{eq:specific_action}, the reduced action is given by
\begin{align}
\mathcal{S}=&4\pi \int dt  dr \sqrt{\frac{f}{h}} \Bigg[M_{\text{pl}}^2 (1-h-r h')+ r^2 f_2+\frac{2}{f} r h^{2} \phi' { V'}^2 f_3 -{\frac{h}{f}} {V'}^2 \{ 4(h-1) f_4 -h^2 {\phi'}{}^2(f_{4,X}+2 \tilde{f}_4) \}    \notag\\ 
 &+ \bigg\{\left[-2 (2 h {\phi''}+{h'} {\phi'}) f^{2} -2 \phi' f'fh \right]f_3-2{\phi'}^{3} f{f'} h^{2} {\tilde{f}_3}
+\left[(8 f''h+4 f' h')f-4 h {f'}{}^2\right] {f_4}\notag\\ 
&-(f_{4,X}+2\tilde{f}_{4})(2h\phi''+h'\phi')fhf'\phi' \bigg\}\frac{P^2}{4f^2 r^2} \Bigg]\,.
\label{reduced_action}
\end{align}
At this stage, we use this reduced action only to examine the asymptotic scaling of the interaction terms and to introduce a parametrization suitable for the limit $X\to0$.
The background equations used in the subsequent analysis are instead derived by first varying the full covariant action and then imposing the spacetime symmetry and field ansatz.

We search for BH solutions that are asymptotically flat at spatial infinity. 
This requires the metric functions and the fields to satisfy the asymptotic conditions $f\to1$, $h\to1$, $\phi\to$ constant, $V\to$ constant, for $r\to\infty$. In other words,
\be
f'(r) \to 0,\qquad h'(r) \to 0,\qquad \phi'(r) \to 0,\qquad V'(r) \to 0
\qquad (r \to \infty)\,.
\label{coninf}
\ee
Consequently, the kinetic term behaves as 
\be
X=-\frac{h\phi'{}^2}{2}\ \to\ 0
\qquad (r\to\infty)\,.
\ee
We restrict our analysis to interaction functions whose contributions to the reduced action remain regular in the limit $X\to0$. 
Within the class of models considered here, we therefore take $f_2$, $f_3$, and $f_4$ to be regular functions of $X$ around $X=0$, without negative powers of $X$. 
However, the situation is different for the interaction $\tilde{f}_3$. 
As seen in Eq.~\eqref{reduced_action}, this function appears in the combination ${\phi'}^3 \tilde{f}_3$. 
Since $|\phi'|^3\propto |X|^{3/2}$ asymptotically, this kinetic prefactor permits the integer Laurent behavior $\tilde{f}_3\propto X^{-1}$ while keeping the corresponding contribution regular as $X\to0$.
To parametrize this case systematically, we introduce a regular function $g_3(\phi,X)$ through
\be
\tilde{f}_3(\phi\,,X)=\frac{g_3(\phi\,,X)}{X}\,.\label{tldef3_to_g3}
\ee
With this definition, we can treat the set of functions $\{f_2, f_3, g_3, f_4, \tilde{f}_4\}$ as being regular at spatial infinity, provided they contain only non-negative powers of $X$.

\subsection{Higher-order derivatives in the covariant field equations}
\label{sec:covariant_higher_derivatives}

To examine the derivative structure of the theory without imposing any spacetime symmetry, we derive the covariant field equations from the action~\eqref{eq:specific_action}. 
We vary the action with respect to the metric $g^{\mu\nu}$, the scalar field $\phi$, and the vector field $A^{\mu}$.
The complete field equations are presented in the Supplemental Material. 
Here, we display only the terms containing derivatives higher than second order.

The metric field equation receives higher-derivative contributions from the $\mathcal{L}_{\rm SVT}^{4}$ sector.
In particular, it contains third covariant derivatives of the scalar field.
These terms are given by
\begin{align}
\left.
\mathcal{E}^{(g)}_{\mu\nu}
\right|_{\rm HD}
={}&
\frac{1}{2}
\left(
2\tilde{f}_{4}-3f_{4,X}
\right)
\nabla^{\lambda}\phi
\Bigl[
F_{\nu}{}^{\rho}F_{\rho}{}^{\sigma}
\nabla_{\sigma}\nabla_{\lambda}\nabla_{\mu}\phi
+F_{\mu}{}^{\rho}
\Bigl(
F_{\rho}{}^{\sigma}
\nabla_{\sigma}\nabla_{\lambda}\nabla_{\nu}\phi
-
F_{\nu}{}^{\sigma}
\nabla_{\sigma}\nabla_{\rho}\nabla_{\lambda}\phi
+
F_{\nu\rho}
\nabla_{\sigma}\nabla^{\sigma}\nabla_{\lambda}\phi
\Bigr) \notag \\
&+
F_{\rho}{}^{\kappa}F^{\rho\sigma}g_{\mu\nu}
\nabla_{\kappa}\nabla_{\sigma}\nabla_{\lambda}\phi
+\frac{1}{2}
F_{\rho\sigma}F^{\rho\sigma}
\Bigl(
\nabla_{\lambda}\nabla_{\nu}\nabla_{\mu}\phi
-
g_{\mu\nu}
\nabla_{\kappa}\nabla^{\kappa}\nabla_{\lambda}\phi
\Bigr)
\Bigr]\,,
\label{eq:covariant_HD_metric}
\end{align}
where $\mathcal{E}^{(g)}_{\mu\nu}=0$ denotes the metric
field equation.
The subscript ``HD'' denotes the contributions containing derivatives
higher than second order.

The scalar field equation also contains higher-derivative terms.
In this case, they appear through covariant derivatives of the curvature
tensors:
\begin{align}
\left.
\mathcal{E}^{(\phi)}
\right|_{\rm HD}
={}&
\frac{1}{4}
\left(
2\tilde{f}_{4}-3f_{4,X}
\right)
F^{\nu\lambda}
\Bigl(
4F_{\nu}{}^{\rho}\nabla_{\mu}R_{\lambda\rho}
-
F_{\nu\lambda}\nabla_{\mu}R
-F^{\rho\sigma}
\nabla_{\mu}R_{\nu\lambda\rho\sigma}
\Bigr)
\nabla^{\mu}\phi \,,
\label{eq:covariant_HD_scalar}
\end{align}
where $\mathcal{E}^{(\phi)}=0$ denotes the scalar field equation.
The derivatives of the curvature tensors in
Eq.~\eqref{eq:covariant_HD_scalar} correspond to third derivatives of
the metric.

Equations~\eqref{eq:covariant_HD_metric} and
\eqref{eq:covariant_HD_scalar} show that higher-derivative terms are
already present before imposing any spacetime symmetry.
Both contributions are proportional to the same combination,
$2\tilde{f}_{4}-3f_{4,X}$.
To eliminate these contributions identically at the covariant level, without relying on cancellations specific to a particular configuration, we impose
\begin{equation}
\tilde{f}_4=\frac{3}{2}f_{4,X}\,.
\label{eq:covariant_HD_condition}
\end{equation}
This condition removes the higher-derivative contributions from both
the metric and scalar field equations.
We note that, for the purely electric configuration considered in \cite{Heisenberg:2018vti}, these higher-derivative contributions vanish identically upon imposing the ansatz, even without invoking the condition above. This cancellation is specific to the purely electric configuration. 
Related aspects of higher-order derivatives in $U(1)$ gauge-invariant SVT theories have recently been discussed in Refs.~\cite{Gorji:2025lqr,Mironov:2025dzz}.
Gorji et al.~\cite{Gorji:2025lqr} pointed out that the contribution proportional to $\tilde f_4$ in the original construction generates higher-order derivatives in the metric field equations. 
Within the overlap between the restricted class considered in Ref.~\cite{Gorji:2025lqr} and the quartic sector considered here, for which $f_{4,X}=0$, our condition reduces to $\tilde f_4=0$. 
More generally, the complete quartic sector considered here allows a nonvanishing $\tilde f_4$ contribution to be canceled by the terms proportional to $f_{4,X}$ when $\tilde f_4=3f_{4,X}/2$. 
Mironov et al.~\cite{Mironov:2025dzz} constructed a different second-order SVT theory through a dimensional reduction of five-dimensional Horndeski theory.
After translating conventions, the coefficient relation appearing in the corresponding quartic structure is consistent with the condition derived here.
This agreement provides a complementary consistency check, whereas our condition is obtained directly from the covariant equations of motion of the four-dimensional SVT action considered in this work.

\subsection{Condition avoiding higher-order derivatives in the presence of the magnetic charge}
Having obtained the condition from the covariant field equations, we now use the reduced action as an independent diagnostic of the higher-order radial-derivative structure in static and spherically symmetric configurations carrying a magnetic charge.
As evident from Eq.~\eqref{reduced_action}, the presence of the magnetic charge $P$ introduces second-order derivatives, $f''$ and $\phi''$, into the action. Extracting these terms, the part of the Lagrangian responsible for generating higher-order derivatives in the field equations can be written schematically as
\be
\mathcal{L}_{P^2}
= P^2\left[
A_{f_3}(\phi,X,f, h)\phi''
+ B_{f_4}(\phi,X,f,h)f''
+ C_{f_4\tilde{f}_4}(\phi,\phi',X,f,h,f')\phi''
\right]\,,
\label{LP2_second_deriv_struct}
\ee
where the functions $A_{f_3}$\,, $B_{f_4}$\,, and $C_{f_4\tilde{f}_4}$ can be read off directly from Eq.~\eqref{reduced_action}.
For instance, the coefficient of the $\phi''$ term in the $f_3$ sector is explicitly given by
\be
A_{f_3}(\phi\,,X\,,f\,,h)
=-\frac{4\pi}{r^2}\sqrt{fh}\,f_3(\phi,X)\,.
\ee
To identify the origin of the higher-derivative terms, we isolate the relevant contribution in the Lagrangian as
\be
\mathcal{L}_{P^2 f_3}=P^2\,A_{f_3}(\phi,X,f,h)\phi''\,.
\ee
For this isolated Lagrangian, the Euler-Lagrange equation for $\phi$ leads to
\be
\frac{d^2}{dr^2}\!\left(\frac{\partial \mathcal{L}_{P^2f_3}}{\partial \phi''}\right)
- \frac{d}{dr}\!\left(\frac{\partial \mathcal{L}_{P^2f_3}}{\partial \phi'}\right)
+ \frac{\partial \mathcal{L}_{P^2f_3}}{\partial \phi}
= P^2\frac{d^2 A_{f_3}}{dr^2}
- P^2\frac{d}{dr}\bigl(A_{f_3,\phi'}\,\phi''\bigr)
+P^2 A_{f_3,\phi}\,\phi''\,.
\label{ELeq_for_Af3}
\ee
The first term on the right-hand side, $P^2 d^2 A_{f_3}(\phi,X,f,h)/dr^2$, involves the second radial derivative of $X$. 
Since $d^2X/dr^2$ contains $\phi'''$, it generates the third-order derivative of the form $P^2A_{f_3,\phi'}\phi'''$ along with lower-order terms. Crucially, however, the second term on the right-hand side of Eq.~\eqref{ELeq_for_Af3} yields a contribution $-P^2 A_{f_3, \phi'} \phi'''$, which exactly cancels the leading term. Consequently, third-order derivatives are entirely eliminated from the scalar field equation. Furthermore, since $A_{f_3}$ is independent of $f'$, $h'$, and $V'$, variations with respect to the metric and vector fields remain strictly second-order. Thus, the $f_3$ interaction sector does not generate higher-order derivatives in the field equations, requiring no additional conditions on the theory.

In contrast to the $f_3$ sector, the terms involving $f_4$ and $\tilde{f}_4$ exhibit a more complex coupling structure.
Extracting the coefficients of the $f''$- and $\phi''$-dependent terms in the $f_4$ sector from Eq.~\eqref{reduced_action}, we find
\be
B_{f_4}(\phi\,,X\,,f\,,h)\,f''
=
\frac{8\pi}{r^{2}}\sqrt{\frac{h}{f}}\,f_4(\phi,X)\,f''\,,
\qquad
C_{f_4\tilde{f}_4}(\phi\,,\phi'\,,X\,,f\,,h\,,f')\,\phi''
=
-\frac{2\pi h f'\phi'}{ r^{2}}  \sqrt{\frac{h}{f}}
\bigl(f_{4\,,X}+2\tilde{f}_4\bigr)\,\phi'' \,.
\label{BC_coefficients}
\ee
Crucially, $B_{f_4}$ couples to $f''$ but depends on $\phi'$ via $X$, whereas $C_{f_4\tilde{f}_4}$ couples to $\phi''$ but depends explicitly on $f'$. 
Due to this mixed dependence, the higher-order derivative terms generated by variations do not automatically cancel.
To isolate the higher-order contributions from the $f_4$ and $\tilde{f}_4$ sectors, it is sufficient to analyze the reduced Lagrangian density containing second radial derivatives of the form
\be
\mathcal{L}_{P^2 f_4 \tilde{f}_4}
=
P^2 B_{f_4}(\phi,X,f,h)\,f''
+
P^2 C_{f_4\tilde{f}_4}(\phi,\phi',X,f,h,f')\,\phi''\,.
\ee
The corresponding Euler--Lagrange operators, $\mathcal{E}_q[\mathcal{L}] = (d^2/dr^2) (\partial_{q''} \mathcal{L}) - (d/dr) (\partial_{q'} \mathcal{L}) + \partial_q \mathcal{L}$ for $q=f,\phi$, acting on this Lagrangian yield
\begin{align}
\mathcal{E}_f\!\left[\mathcal{L}_{P^2f_4\tilde{f}_4}\right]
=&
P^2\frac{d^{2}B_{f_4}}{dr^{2}}
-P^2\frac{d}{dr}\!\left(\frac{\partial C_{f_4\tilde{f}_4}}{\partial f'}\,\phi''\right)
+\cdots, \label{ELeq_of_f_forBC_term}\\
\mathcal{E}_\phi\!\left[\mathcal{L}_{P^2f_4\tilde{f}_4}\right]
=&
P^2\frac{d^{2}C_{f_4\tilde{f}_4}}{dr^{2}}
-P^2\frac{d}{dr}\!\left(
\frac{\partial B_{f_4}}{\partial \phi'}\,f''
+
\frac{\partial C_{f_4\tilde{f}_4}}{\partial \phi'}\,\phi''
\right)
+\cdots,\label{ELeq_of_phi_for_BC_term}
\end{align}
where the ellipses denote terms involving at most second derivatives.

First, we find a cancellation of higher-order derivatives in the scalar field equation. 
Focusing on the $\phi'''$ contributions in Eq.~\eqref{ELeq_of_phi_for_BC_term}, we have
\be
   \frac{d^2 C_{f_4\tilde f_4}}{dr^2}
\supset
\frac{\partial C_{f_4\tilde f_4}}{\partial\phi'}\phi''' \,,
\qquad
-\frac{d}{dr}
\left(
\frac{\partial C_{f_4\tilde f_4}}{\partial\phi'}\phi''
\right)
\supset
-\frac{\partial C_{f_4\tilde f_4}}{\partial\phi'}\phi''' \,.
\label{phi3_cancel}
\ee

Here, the notation $A \supset B$ indicates that $B$ represents the third-order derivative terms extracted from $A$. 
The above equation shows that the terms proportional to $\phi'''$ in the $\phi$ equation cancel exactly, analogously to the $f_3$ sector.

However, the remaining third derivatives arise from two sources:
(i) the chain-rule terms in $d^{2}B_{f_4}/dr^{2}$ and $d^{2}C_{f_4\tilde{f}_4}/dr^{2}$ involving the other field,
and (ii) the outer derivative $d/dr$ acting on the mixed-derivative couplings.
Specifically, the equation for $f$ receives $\phi'''$ contributions from
\be
\frac{d^{2}B_{f_4}}{dr^{2}}
\supset
B_{f_4,X}\,X''
\supset
B_{f_4,X}\,\frac{\partial X}{\partial\phi'}\,\phi''',
\qquad
-\frac{d}{dr}\!\left(\frac{\partial (C_{f_4\tilde{f}_4}\phi'')}{\partial f'}\right)
=
-\frac{d}{dr}\!\left(\frac{\partial C_{f_4\tilde{f}_4}}{\partial f'}\,\phi''\right)
\supset
-\frac{\partial C_{f_4\tilde{f}_4}}{\partial f'}\,\phi''' ,
\ee
so that
\be
\mathcal{E}_f\!\left[\mathcal{L}_{P^2f_4 \tilde{f}_4}\right]\supset
-P^2\left(
\frac{\partial C_{f_4\tilde{f}_4}}{\partial f'}
-B_{f_4,X}\,\frac{\partial X}{\partial\phi'}
\right)\phi'''\,.
\label{Efhigh}
\ee
Similarly, the $\phi$ equation receives $f'''$ contributions from
\be
\frac{d^{2}C_{f_4\tilde{f}_4}}{dr^{2}}
\supset
\frac{\partial C_{f_4\tilde{f}_4}}{\partial f'}\,f''',
\qquad
-\frac{d}{dr}\!\left(\frac{\partial (B_{f_4}f'')}{\partial \phi'}\right)
=
-\frac{d}{dr}\!\left(\frac{\partial B_{f_4}}{\partial \phi'}\,f''\right)
\supset
-B_{f_4,X}\,\frac{\partial X}{\partial\phi'}\,f'''\,,
\ee
which leads to
\be
\mathcal{E}_\phi\!\left[\mathcal{L}_{P^2f_4 \tilde{f}_4}\right]\supset
P^2\left(
\frac{\partial C_{f_4\tilde{f}_4}}{\partial f'}
-B_{f_4,X}\,\frac{\partial X}{\partial\phi'}
\right)f''' .
\label{Ephigh}
\ee
From Eqs.~\eqref{Efhigh} and \eqref{Ephigh}, we find that the potentially dangerous higher-order derivatives in both equations share the same prefactor, ${\partial C_{f_4\tilde{f}_4}}/{\partial f'}-B_{f_4,X}({\partial X}/{\partial\phi'})$.
Substituting the explicit expressions from Eq.~\eqref{BC_coefficients} into this prefactor, we find that the third-order derivative contribution is proportional to the combination $3 f_{4,X} - 2 \tilde{f}_4$.
To ensure that the field equations remain strictly second-order, we require this combination to vanish, i.e., 
\be
\tilde{f}_4=\frac{3}{2}f_{4,X}\,.
\label{degeneracy_condition}
\ee
Since the $f_3$ contributions do not generate any third-order derivatives, Eq.~\eqref{degeneracy_condition} constitutes the sole condition required to eliminate all higher-order derivatives in the symmetry-reduced system with the magnetic charge. 
Notably, this condition is identical to the condition \eqref{eq:covariant_HD_condition} obtained directly from the
fully covariant field equations before imposing any spacetime
symmetry.
\subsection{Field equations}

In the general $U(1)$ gauge-invariant SVT theory, the quadratic coupling function $f_2$ depends on the full set of scalar invariants, $f_2 = f_2(\phi, X, F, \tilde{F}, Y)$. 
The fully covariant equations presented in the Supplemental Material retain this general dependence.
While the quadratic sector has been extensively studied, for instance in Refs.~\cite{Gibbons:1987ps,Shapere:1991ta,Herdeiro:2018wub,Taniguchi:2024ear}, the roles of the higher-order couplings $f_3$, $\tilde f_3$, $f_4$, and $\tilde f_4$ remain less explored.
For compactness, the background equations displayed below are specialized to the subclass $f_2=f_2(\phi,X,F)$. 
For the concrete solution analysis from Sec.~\ref{sec3} onward, we further adopt the canonical choice
\be
f_2=X+F \,,
\label{f2choice}
\ee
following the standard formulation in Ref.~\cite{Heisenberg:2018vti}, and treat the higher-order couplings as the primary source of modifications.

Starting from the covariant equations presented in the Supplemental Material, we impose Eq.~\eqref{eq:covariant_HD_condition}, substitute the static and spherically symmetric dyonic ansatz, and adopt the reparametrization~\eqref{tldef3_to_g3}. 
This procedure yields second-order background equations for $f$, $h$, $\phi$, and $V$. 
In general, variation after symmetry reduction need not reproduce all components of the full covariant equations of motion. 
As an independent consistency check, we have therefore also varied the reduced action~\eqref{reduced_action} while retaining the full dependence $f_2=f_2(\phi,X,F,\tilde F,Y)$. 
For this comparison, we impose the same condition and reparametrization and account for the normalization factors associated with the symmetry reduction. 
The resulting independent equations are algebraically equivalent to the corresponding symmetry-reduced components of the full covariant equations. 
The background equations can be written as
\begin{align}
{\Mpl}^{2} f r {h'} =&{\Mpl}^{2} f (1-h )+r^{2} (-h {V'}^{2} {f_{2,F}} +f {f_2} )-2 {f_3} h^{2} r {V'}^{2} {\phi'} -4 h {V'}{}^{2} \{{f_4} (1-h )+{f_{4,X}} h^{2} {\phi'}^{2}\}  \notag  \\
&+\bigg[-4 r {\phi'} h f {f_3}+2 f h r^{2} {\phi'}^{2} {f_{3,\phi}}-f {\phi'}^{2} r^{2} h (2 h {\phi''}+{h'} {\phi'}) {f_{3,X}}-2 f r (2 h r {\phi''}+{\phi'} {h'} r-4 {\phi'} h) {g_3}  \notag \\
&-4fhr^{2} {\phi'}^{2} {g_{3,\phi}}+2 f {\phi'}^{2} r^{2} h (2 h {\phi''}+{h'} {\phi'}) {g_{3,X}}-8f({h'} r-6 h) {f_4}-4 f r (-2 h r {\phi''}-{\phi'} {h'} r+8 {\phi'}h) {f_{4,\phi}}  \notag \\
&+8 f hr^{2} {\phi'}{}^{2} {f_{4,\phi\phi}}+4fh\phi'r(2 h {\phi''}+{h'} {\phi'})(2f_{4,X}-\phi'rf_{4,\phi X})\bigg] \frac{P^{2}}{2r^{4}}\,,\label{eq_f_all_function}\\
{\Mpl}^{2} r h {f'}=&{\Mpl}^{2} f (1-h)+r^{2} (f {f_2}+fh{\phi'}^{2}{f_{2,X}}-h{V'}{}^{2}{f_{2,F}})-2r h^{2}{\phi'} {V'}^{2} (3{f_3}-h{\phi'}^{2}{f_{3,X}}) \notag \\
&-4{V'}^{2} h (-3 h+1) {f_4} 
-4 {V'}^{2} h (6 h^{2}-h) {\phi'}^{2} {f_{4,X}}
-\bigg[-{2 {\phi'} h f {f_3}}+{f h {\phi'}^{2}r {f_{3,\phi}}}+\frac{{\phi'}^{3} {f'} h^{2} r{f_{3,X}}}{2}  \notag \\
&+{hf'\phi'r(g_3-h{\phi'}^2 g_{3,X})}
+{2 h{f'} \{2 {f_4}-{f_{4,\phi}}r{\phi'}+{\phi'}^{2} ({f_{4,\phi X}} r {\phi'}-2 {f_{4,X}}) h\} } \Bigg] \frac{P^{2}}{r^3}\,,\label{eq_h_all_function}\\
J_{\phi}'=&\mathcal{P}_\phi\,
\label{eq_J_phi_P_phi_all_function}\\
J_V'=&0\,,\label{eq_J_V_all_function}
\end{align}
where the effective currents and source terms are defined as
\begin{align}
    J_\phi=&-\sqrt{\frac{h}{f}}\left[ fr^{2} {\phi'} {f_{2,X}}-2 h {V'}^{2} r {f_3}+2 h^{2} {V'}^{2} {\phi'}^{2} {f_{3,X}} r-4{\phi'}{V'}^{2} h (3 h-1) {f_{4,X}}\right]  \notag \\
+&\frac{P^{2}}{2 r^{3}} \sqrt{\frac{h}{f}}\left[-4f{f_3}+2fr{\phi'}{f_{3,\phi}}+h{f'}r{\phi'}^{2}{f_{3,X}}+2{f'}r({g_3}-h{\phi'}^{2}{g_{3,X}})-8h{f'}{\phi'}{f_{4,X}}+4hr{f'}{\phi'}^{2}{f_{4,\phi X}}\right]\,,
 \label{def_J_phi_all_function}\\
    P_\phi=&\frac{1}{\sqrt{f h}}\left[r^{2}f {f_{2,\phi}}+2 {V'}^{2} r {\phi'} h^{2} {f_{3,\phi}}+4 h {V'}^{2} \{ h^{2} {f_{4,\phi X}} {\phi'}^{2}
+(1-h)f_{4,\phi}\}\right]
+\frac{1}{2f^{\frac{3}{2}}\sqrt{h}r^{2}}\Big[-(2 f h {\phi''}+f {h'} {\phi'}\notag \\
+&{f'} {\phi'} h) f{f_{3,\phi}} 
+2f{f'}{\phi'} h {{g}_{3,\phi}}+2(2 f h {f''}+f {f'} {h'}-h {f'}^{2}) {f_{4,\phi}}-2fh{f'}{\phi'}(2 h {\phi''}+{h'} {\phi'}) {f_{4,\phi X}}\Big] P^{2}\,,\label{def_P_phi_all_function}\\
    J_V=&\sqrt{\frac{h}{f}} \left[r^{2} {V'} {f_{2,F}}+4 h r {\phi'} {V'} {f_3}-8 (h-1) {V'} {f_4}+8 {f_{4,X}} h^{2} {\phi'}^{2} {V'}\right]\,.
    \label{defJ_V}
\end{align}
Eq.~\eqref{eq_J_V_all_function} implies that $J_V$ is constant reflecting the conservation of the current associated with the $U(1)$ gauge symmetry. 
As expected, the function $\tilde{f}_4$ is completely absent from these field equations as a consequence of the condition~\eqref{degeneracy_condition}. 
In contrast, the equations explicitly retain the interaction 
$\tilde{f}_3$ via $g_3=X\tilde{f}_3$. 
While this term decouples in purely electrically charged configurations~\cite{Heisenberg:2018vti}, the presence of the magnetic charge $P$ activates the $\tilde{f}_3$ sector, as evident from the reduced action~\eqref{reduced_action}.

Since the reduced action does not retain $g_{\theta\theta}$ and $A_\varphi$ as independent variables, we verify that the corresponding covariant equations do not provide additional independent background equations. We denote the metric, scalar-field, and vector-field equations by $\mathcal{E}^{(g)}_{\mu\nu}=0$, $\mathcal{E}^{(\phi)}=0$, and $\mathcal{E}^{(A)}_{\mu}=0$, respectively. Here, $\mathcal{E}^{(g)}_{\mu\nu}$ and $\mathcal{E}^{(A)}_{\mu}$ denote the lower-index equations obtained by varying the action with respect to $g^{\mu\nu}$ and $A^\mu$, respectively.

We first consider the angular component of the metric equation. For the background ansatz~\eqref{eq:metric_ansatz} and \eqref{eq:field_ansatz}, the covariant equations satisfy the off-shell identity
\begin{align}
0\equiv{}&
\left(h\mathcal{E}^{(g)}_{rr}\right)'+\frac{f'}{2f}
\left(h\mathcal{E}^{(g)}_{rr}+\frac{\mathcal{E}^{(g)}_{tt}}{f}\right)
+\frac{2}{r}\left(h\mathcal{E}^{(g)}_{rr}-\frac{\mathcal{E}^{(g)}_{\theta\theta}}{r^2}\right)
-\frac{\phi'}{2}\mathcal{E}^{(\phi)}+\frac{V'}{2f}\mathcal{E}^{(A)}_{t}\,.
\label{eq:EOM_identity_theta}
\end{align}
Although the angular metric sector is fixed before constructing the reduced action, the corresponding angular equation can be recovered by temporarily generalizing the angular part of the metric as
\begin{equation}
r^2
\left(
d\theta^2+\sin^2\theta\,d\varphi^2
\right)
\ \longrightarrow\
r^2 e^{2\zeta(r)}
\left(
d\theta^2+\sin^2\theta\,d\varphi^2
\right)\,.
\label{eq:angular_metric_zeta}
\end{equation}
We vary the resulting reduced action with respect to $\zeta(r)$ and subsequently set $\zeta=0$. We have explicitly verified that the resulting angular equation is equivalent to the $\theta\theta$ component of the covariant metric equation evaluated on the background ansatz. We have also confirmed that this angular equation, together with the remaining equations obtained from the reduced action, satisfies the identity~\eqref{eq:EOM_identity_theta}. 
Therefore, once the remaining background equations are imposed, the identity~\eqref{eq:EOM_identity_theta} guarantees that the $\theta\theta$ component of the metric equation is automatically satisfied. This shows that $\mathcal{E}^{(g)}_{\theta\theta}=0$ is dependent on the other background equations and need not be imposed independently.

We next consider the azimuthal component of the vector-field equation. Direct substitution of the background ansatz~\eqref{eq:metric_ansatz} and \eqref{eq:field_ansatz} into the covariant vector-field equation, without using any of the other background equations, gives
\be
\mathcal{E}^{(A)}_\varphi\equiv0\,.
\label{eq:EOM_identity_Aphi}
\ee
We therefore conclude that neither the angular metric equation nor the azimuthal vector-field equation introduces an independent background equation beyond those obtained from the reduced action.

\section{Conditions for scalar hair formation in shift-symmetric theories}
\label{sec3}

In this section, we investigate the subclass of theories possessing shift symmetry, invariant under the transformation $\phi \to \phi + c$, where $c$ is a constant.
In this case, the interaction functions depend solely on the kinetic term $X$ as
\be
f_3 = f_3(X)\,,\qquad
g_3 = g_3(X)\,,\qquad
f_4 = f_4(X)\,.
\ee
The shift symmetry enforces the vanishing of the source term, $P_\phi=0$, so that the scalar field equation~\eqref{eq_J_phi_P_phi_all_function} reduces to the conservation of the Noether current, i.e., $J_\phi=$ constant.
Substituting the canonical choice $f_2 = X+F$ given in Eq.~\eqref{f2choice} into Eq.~\eqref{def_J_phi_all_function}, the conserved current simplifies to
\begin{align}
J_\phi =&
-\sqrt{\frac{h}{f}}
\bigg[
f r^{2} \phi'
-2 h V'{}^{2} f_3\, r
+ 2 h^{2} V'{}^{2} r f_{3,X} \phi'{}^{2}
- 4 h V'{}^{2} \phi' (3 h - 1) f_{4,X}
\notag \\
&
+\frac{P^{2}}{2 r^{3}}\left\{4f{f_3}
-h{f'}r{\phi'}{}^{2}{f_{3,X}}
-2{f'}r({g_3}-h{\phi'}{}^{2}{g_{3,X}})
+8h{f'}{\phi'}{f_{4,X}}\right\}
\bigg] \notag \\
=&\,\,\text{constant}\,.
\label{J_phi_shift_all_function}
\end{align}
%

\subsection{Current conservation and no-hair argument in the \texorpdfstring{$f_4$}{f4} sector}

We begin by briefly reviewing the no-hair argument for shift-symmetric scalar fields in the context of Galileon gravity formulated by Hui and Nicolis~\cite{Hui:2012qt}.
To distinguish the conserved current in their general argument from the specific current derived in our model, we denote the former by the calligraphic symbol $\mathcal{J}_\phi$. 
Due to the shift symmetry, the scalar field equation integrates to the conservation law $\mathcal{J}_\phi=$ constant. 
Requiring regularity at the horizon with finite $\phi'$ and $V'$, and assuming that only positive-power metric components appear there, implies that the conserved current must vanish at the horizon, $\mathcal{J}_\phi(r_h)=0$. 
Since $\mathcal{J}_\phi$ is constant, this boundary condition immediately leads to $\mathcal{J}_\phi=0$ for all $r$.
If the current factorizes as
\be
\mathcal{J}_\phi=\mathcal{C}(r)\,\phi'(r)\,,
\label{Jphi_factorized_shift}
\ee
with $\mathcal{C}(r)$ being a non-vanishing regular function, then the condition $\mathcal{J}_\phi=0$ reduces to $\phi'(r)=0$ everywhere.
Consequently, the only solution compatible with the asymptotic condition $\phi'(r)\to 0$ at spatial infinity is the trivial configuration
\be
\phi'(r)=0 \quad \text{for all } r\,,
\ee
which implies the absence of scalar hair.

In light of this argument, let us examine the structure of the current~\eqref{J_phi_shift_all_function} in our model to see if the same factorization~\eqref{Jphi_factorized_shift} holds.
We assume that the background geometry describes a static, spherically symmetric BH satisfying asymptotic flatness~\eqref{coninf} at spatial infinity, and regularity at the horizon~\eqref{eq:horizon_def}, characterized by the finiteness of $\phi'$, $V'$, and consequently $J_\phi$. 
Specifically, to evaluate the behavior of each term in Eq.~\eqref{J_phi_shift_all_function} around the horizon, we consider deviations from the dyonic Reissner-Nordstr\"om (RN) solution given by 
\begin{equation}
f_{\rm RN}(r)=h_{\rm RN}(r)
=\Bigl(1-\frac{r_h}{r}\Bigr)\Bigl(1-\frac{\mu r_h}{r}\Bigr)
= 1-\frac{2M}{r}+\frac{Q^2+P^2}{2 M_{\rm Pl}^2 r^2}\,,
\end{equation}
where $\mu$ is a constant in a range $0<\mu<1$.
Expanding the metric functions around the horizon $r_h$, we have
\begin{equation}
f(r)=\sum_{i=1}^\infty f_i (r-r_h)^i \,, \qquad
h(r)=\sum_{i=1}^\infty h_i (r-r_h)^i \,. 
\label{fhexpand}
\end{equation}
The dyonic RN limit corresponds to the leading coefficients $f_1=h_1=(1-\mu)/r_h$. 
Consequently, the ratio of the metric functions approaches unity at the horizon, i.e., $h/f\to 1$ as $r\to r_h$.

We first focus on the $f_4$ sector by setting $f_3=g_3=0$.
Evaluating Eq.~\eqref{J_phi_shift_all_function} under this choice, we obtain the explicit expression 
\be
J_{\phi}=-\sqrt{\frac{h}{f}}\left[fr^2-4h{V'}^2(3h-1)f_{4,X}+\frac{4P^2}{r^3}(hf'f_{4,X})\right]\phi'. 
\label{reduced_act_f4}
\ee
Noting that the metric functions vanish at the horizon ($f=h=0$), this current $J_{\phi}$ vanishes at the horizon, as long as $\phi'$ and $V'$ are finite. 
Consequently, the conservation law $J_{\phi}=$ constant leads to $J_{\phi}=0$ everywhere. 
Moreover, the above expression is clearly factorized into the form given in Eq.~\eqref{Jphi_factorized_shift} as $J_\phi=\mathcal{C}_4(r)\,\phi'(r)$ where $\mathcal{C}_4(r)$ is a regular function of $r$ determined by the background and the functional form of $f_{4,X}$.
With the asymptotic boundary condition $\phi'(r)\to0$ ($r\to\infty$), we now conclude that $\phi'(r)=0$ everywhere.
Therefore, the $f_4$ sector by itself does not give rise to scalar hair. 

\subsection{Scalar hair formation in the \texorpdfstring{$f_3$}{f3} and \texorpdfstring{$g_3$}{g3} sectors}
\label{shift_hair_con}

%
\subsubsection{\texorpdfstring{$f_3$}{f3} sector}
\label{shift_hair_con_f3}

We next focus on the $f_3$ sector by setting $g_3=f_4=0$. 
In this case, the current~\eqref{J_phi_shift_all_function} reduces to 
\be
J_{\phi}=-\sqrt{\frac{h}{f}}\left[
f r^{2} \phi'
-2 h V'{}^{2} f_3\, r
+ 2 h^{2} V'{}^{2} r f_{3,X} \phi'{}^{2}
+\frac{P^{2}}{2 r^{3}}(4f{f_3}-h{f'}r{\phi'}{}^{2}{f_{3,X}})
\right]. 
\label{reduced_act_f3}
\ee
Since all terms in the above expression are multiplied by positive powers of $f$ and $h$, we obtain $J_\phi(r_h)=0$ which leads to $J_\phi=0$ everywhere.
Away from the horizon, however, a crucial structural difference appears compared to the $f_4$ sector. 
If $f_3(X)$ contains a constant term, the $f_3$ contribution, specifically the term $-2 h V'{}^{2} f_3\, r$ inside the square brackets, enters the equation without an overall factor of $\phi'$.
Consequently, even though the total current must be zero ($J_\phi=0$), this does not necessitate $\phi'(r)=0$.
Instead, it yields a nontrivial algebraic relation between $\phi'$ and the background geometry.
Through this mechanism, the $f_3$ interaction can support nontrivial scalar profiles, i.e., scalar hair, even though the conserved current is constrained to vanish.

To extract the asymptotic constraint from $J_\phi=0$, we expand the current in powers of $1/r$. 
Using the asymptotic expansion of the scalar field at spatial infinity,
\be
\phi(r)=\phi_\infty+\sum_{i=1}^{\infty}\frac{{\phi}_{[i]}}{r^i}\,, 
\label{phiexpinf}
\ee
we find that the leading-order term of $J_\phi$ under this expansion is proportional to the coefficient ${\phi}_{[1]}$\,.
Therefore, the condition $J_\phi=0$ enforces ${\phi}_{[1]}=0$.
The fact that the scalar charge is uniquely determined by the horizon boundary condition is a characteristic feature of secondary hair, where the scalar configuration is sourced not by an independent free parameter, but effectively by the gravitational and electromagnetic environment.

\subsubsection{\texorpdfstring{$g_3$}{g3} sector}
\label{shift_hair_con_g3}

The situation is distinct in the $g_3$ sector defined by setting $f_3=f_4=0$.  In this case, Eq.~\eqref{J_phi_shift_all_function} reduces to
\be
J_\phi = -\sqrt{\frac{h}{f}}\left[
f r^2 \phi' -\frac{P^2}{r^2} f'(g_3 - h{\phi'}^2 g_{3,X})
\right]\,.
\label{reduced_act_g3}
\ee
Crucially, unlike the previous cases, the term proportional to $g_3$ does not carry an overall factor of $f$ or $h$. 
Since the derivative of the metric function $f'(r_h)$ is finite for non-extremal BHs and $\sqrt{h/f} \to 1$ at the horizon, the term originating from $g_3(X)$ remains non-vanishing. 
Specifically, noting that the kinetic term vanishes at the horizon ($X \to 0$), the current approaches a finite, nonzero constant as $J_\phi(r_h) \propto P^2 g_3(0)$ at the horizon. 
Due to the current conservation, this nonzero value must be maintained globally, i.e., $J_\phi(r) = J_\phi(r_h)$. 
Consequently, the current cannot be written in the factorized form~\eqref{Jphi_factorized_shift}, and the presence of a nonzero conserved charge provides a mechanism to evade the no-hair theorem while satisfying the asymptotic condition $\phi'(r)\to 0$. 
Evaluating the conserved current at the horizon by using Eq.~\eqref{fhexpand} with $f_1=h_1=(1-\mu)/r_h$, the corresponding charge is given by $J_{\phi} ={(1-\mu)\,g_3(0)\,P^{2}}/{r_h^{3}}$.
On the other hand, expanding $J_\phi$ at spatial infinity in powers of $1/r$, we find that the leading order contribution coincides with the coefficient $\phi_{[1]}$ in the expression~\eqref{phiexpinf}.
Equating the horizon current with the asymptotic one by virtue of the conservation law, we obtain the relation
\begin{equation}
{\phi}_{[1]}=\frac{(1-\mu)g_3(0)P^{2}}{r_h^{3}}\,.
\label{relation_g3_shift_charge}
\end{equation}
Similar to the $f_3$ sector, the scalar charge is determined by the boundary condition at the horizon, classifying the scalar hair into the secondary type.

\paragraph*{Summary of classification}
In summary, in the shift-symmetric theories, the hair formation mechanism depends on the interaction type: 
\begin{itemize}
\item 
The $f_4$ sector possesses the factorized structure~\eqref{Jphi_factorized_shift} with a vanishing current, leading to no scalar hair.
\item 
The $f_3$ and $g_3$ sectors both feature a scalar current that cannot be factorized by $\phi'$.
A common characteristic feature of these sectors is that the scalar charge is not a free parameter but is uniquely determined by the boundary condition at the horizon, i.e., the hair is of secondary type. 
\item 
A key distinction between the latter two sectors is the following. 
In the $f_3$ sector, the horizon condition constrains the conserved charge to vanish ($J_\phi=0$), yet hair exists due to the algebraic structure of the current equation. 
In the $g_3$ case, the conserved charge is generically nonzero ($J_\phi \neq 0$) and is fixed to a specific value by the horizon boundary condition. 
\end{itemize}
In the following two sections~\ref{shift_f3_sec} and~\ref{shift_g3_sec}, we analyze the latter two sectors separately, first for $f_3(X)$ and then for $g_3(X)$, in detail. 

\section{Shift-symmetric solutions in the \texorpdfstring{$f_3$}{f3} sector}
\label{shift_f3_sec}

In this section, we engage in the detailed construction of hairy BH solutions within the $f_3$ sector, defined by the interaction choice 
\be
f_3=f_3(X)\,,\qquad g_3=0\,,\qquad f_4=0\,. 
\label{f3-shift-function}
\ee
As discussed in the general classification in Sec.~\ref{sec3}, this sector is characterized by the vanishing of the conserved scalar current, $J_\phi=0$.
Nevertheless, the interaction term modifies the metric and vector field equations in a way that supports a nontrivial scalar profile, manifesting as secondary hair.
We derive the asymptotic and near-horizon solutions, classify them as the Ia and Ib solutions, and use numerical integration to examine smooth connections between these regimes.

\subsection{Asymptotic solution}
\label{shift_f3_inf}

We seek asymptotically flat configurations satisfying $f(r)\to 1$, $h(r)\to 1$, $V(r)\to V_\infty$, $\phi(r)\to \phi_\infty$ as $r\to\infty$, where $V_\infty$ and $\phi_\infty$ are constants.
To obtain the solutions in this regime, we expand the background functions in inverse powers of $r$ as
\be
f = 1 + \sum_{i=1}^{\infty} \frac{f_{[i]}}{r^i}\,, \qquad 
h = 1 + \sum_{i=1}^{\infty} \frac{h_{[i]}}{r^i}\,, \qquad
V = V_\infty + \sum_{i=1}^{\infty} \frac{ V_{[i]}}{r^i}\,, \qquad
\phi = \phi_\infty + \sum_{i=1}^{\infty} \frac{ \phi_{[i]}}{r^i}\,.
\label{expand_f_h_V_phi}
\ee
Substituting these series into the field equations \eqref{eq_f_all_function}-\eqref{eq_J_V_all_function} and solving them order by order in $1/r$, we determine the coefficients $f_{[i]}$, $h_{[i]}$, $V_{[i]}$, and $\phi_{[i]}$ in terms of the integration constants and the value of the coupling function at spatial infinity.
Choosing the interaction functions as Eq.~\eqref{f3-shift-function}, 
we obtain the following iterative solutions:
\begin{align}
 f=&1-\frac{2 M}{r}+\frac{P^{2}+Q^{2}}{2 {\Mpl}^{2} r^{2}}+\frac{3 f_3^{2} (P^2-Q^2)^{2}}{14 {\Mpl}^{2} r^{8}}-\frac{M f_3^{2} (P^2-Q^2)^{2}}{2 {\Mpl}^{2} r^{9}}+\mathcal{O}\bigg(\frac{1}{r^{10}}\bigg)\,,\label{solf3-shift-inf-f} \\
h=&1-\frac{2 M}{r}+\frac{P^{2}+Q^{2}}{2 {\Mpl}^{2} r^{2}}-\frac{2 f_3^{2} (P^2-Q^2)^{2}}{7 {\Mpl}^{2} r^{8}}+\frac{M f_3^{2} (P^2-Q^2)^{2}}{2 {\Mpl}^{2} r^{9}}+\mathcal{O}\bigg(\frac{1}{r^{10}}\bigg) \,,\\
V=&V_{\infty}+\frac{Q}{r}+\frac{8 Q f_3^{2} (P^2-Q^2)}{7 r^{7}}-\frac{2 M f_3^{2} (P^2-Q^2) Q}{r^{8}} \notag \\
+&\frac{ \{{3(P^2-Q^2)/4}+  29P^{2}+27 Q^{2}\}f_3^{2}(P^2-Q^2) Q}{63 {\Mpl}^{2} r^{9}}+\mathcal{O}\bigg(\frac{1}{r^{10}}\bigg)\,, \\
\phi=&\phi_{\infty}+\frac{f_3 (P^2-Q^2)}{2 r^{4}} +\mathcal{O}\bigg(\frac{1}{r^{5}}\bigg)\,,
\label{solf3-shift-inf-phi}
\end{align}
where the coupling function $f_3$ is evaluated at spatial infinity ($X\to 0$), and we have identified the integration constants as $ f_{[1]}= h_{[1]}=-2M$ and $ V_{[1]}=Q$.

The above asymptotic solutions highlight several important features.
First, taking the limit $P\to 0$ with $f_3=\text{constant}$ reproduces the results of Ref.~\cite{Heisenberg:2018vti}.
The effect of the $f_3$ interaction appears at $\mathcal{O}(1/r^8)$ in the metric functions and at $\mathcal{O}(1/r^7)$ in the vector field, while it enters the scalar field already at $\mathcal{O}(1/r^4)$.
Crucially, these decay rates are identical to those found in the purely electric case~\cite{Heisenberg:2018vti}, implying that the magnetic charge modifies only the amplitude of the corrections without altering their asymptotic radial dependence.
Second, the amplitude of these corrections is governed by the combination $P^2-Q^2$.
This dependence contrasts with the dyonic Reissner-Nordstr\"om solution, which depends only on the sum $P^2+Q^2$, indicating that the $f_3$ interaction breaks the electromagnetic duality rotation symmetry.
Analogous to the dyonic solutions in the Einstein-Maxwell-dilaton theory~\cite{Gibbons:1987ps,Horowitz:1992jp}, the interaction terms vanish identically when $P=Q$, reducing the solution exactly to the dyonic Reissner-Nordstr\"om spacetime with a constant scalar field.
Finally, Eq.~\eqref{solf3-shift-inf-phi} confirms that the scalar profile decays as $1/r^4$ (implying $\phi_{[1]}=0$), which is consistent with the absence of a conserved scalar charge and classifies this configuration as secondary hair~\cite{Herdeiro:2015waa}.

\subsection{Near-horizon solution for the \texorpdfstring{$f_3$}{f3}-\rm{Ia} case}
\label{f3Iasec}

We derive the near-horizon solution for the specific choice of coupling, 
\be
f_3=\beta_3.
\label{f3-sol1-shift-function-horizon}
\ee
Hereafter, we refer to the solution satisfying this choice as the $f_3$-$\mathrm{I}$a solution.
To obtain the analytic solutions to the field equations around the horizon $r=r_h$, we expand $f$, $h$, $V$, and $\phi$ in power series of $(r-r_h)$. 
Imposing the regularity of these functions and the vanishing of the metric functions at the horizon, $f(r_h)=h(r_h)=0$, we write
\be
f=\sum_{i=1}^\infty f^{(i)}(r-r_h)^i\,,\quad 
h=\sum_{i=1}^\infty h^{(i)}(r-r_h)^i\,,\quad 
V=V^{(0)}+\sum_{i=1}^\infty V^{(i)}(r-r_h)^i\,,\quad 
\phi=\phi^{(0)}+\sum_{i=1}^\infty \phi^{(i)}(r-r_h)^i\,.
\label{expand_horizon_fh_V_phi}
\ee
Without loss of generality, we set $V^{(0)}=0$ since the field equations depend on $V$ only through its derivative by virtue of the $U(1)$ gauge symmetry.
We substitute the ansatz~\eqref{expand_horizon_fh_V_phi} and the coupling choice~\eqref{f3-sol1-shift-function-horizon} into the field equations~\eqref{eq_f_all_function}-\eqref{eq_J_V_all_function}, and expand the resultant equations around $r=r_h$. 
Solving them order by order, we obtain the solution that generalizes the purely electric one found in Ref.~\cite{Heisenberg:2018vti} to the dyonic case. 
At linear order, we obtain
\be
f^{(1)}= h^{(1)}=\frac{1-\mu}{r_h}\,,\qquad 
V^{(1)}=\frac{\sqrt{2 \Mpl^2 r_h^2\mu- P^2}}{r_h^2}\,,\qquad 
\phi^{(1)}=\frac{4 ({\Mpl}^{2} {r_h}^{2} \mu -P^{2}) \beta_3}{r_h^{5}}\,,
\label{f3-sol1-shift-f1h1V1phi1}
\ee
where we have chosen the branch $V^{(1)} >0$. 
At second order, the coefficients are given by
\begin{align}
f^{(2)}=&\frac{12 (\mu-1 ) ({\Mpl}^{2} {r_h}^{2} \mu -P^{2})^{2} {\beta_3}^{2}}{{r_h}^{10} {\Mpl}^{2}}+\frac{2 \mu -1}{{r_h}^{2}}\,, \\
h^{(2)}=&-\frac{4 (\mu-1 ) ({\Mpl}^{2} {r_h}^{2} \mu -P^{2})^{2} {\beta_3}^{2}}{{r_h}^{10} {\Mpl}^{2}}+\frac{2 \mu -1}{{r_h}^{2}}\,, \\
V^{(2)}=&
\frac{\sqrt{2 {\Mpl}^{2} {r_h}^{2} \mu -P^{2}}}{{r_h}^{3}}
\left[\frac{4 \{{\Mpl}^{2} {r_h}^{2} (\mu -2)+P^{2}\} ({\Mpl}^{2} {r_h}^{2} \mu -P^{2}){\beta_3}^{2}}{{r_h}^{8} {\Mpl}^{2}}
-1\right]\,, \label{f3-sol1-shift-V2}\\
\phi^{(2)}=&\frac{2 ({\Mpl}^{2} {r_h}^{2} \mu -P^{2}) {\beta_3}}{{r_h}^{6}}\left[\frac{16 (2 {\Mpl}^{2} {r_h}^{2} \mu -P^{2}) (\mu-1 )  {\beta_3}^{2}}{{r_h}^{6}}-5\right]\,,
\label{f3-sol1-shift-secondorder} 
\end{align}
which correctly reduces to the known solution~\cite{Heisenberg:2018vti} in the limit $P\to 0$.

Let us check the relation between the near-horizon parameters appearing in the above expressions and the physical charges defined in the asymptotic solutions \eqref{solf3-shift-inf-f}-\eqref{solf3-shift-inf-phi}. 
First, the magnetic charge $P$ is an intrinsic parameter of the ansatz~\eqref{eq:field_ansatz} and thus appears identically in both the near-horizon and asymptotic regions. 
Regarding the remaining parameters, the near-horizon parameters $(r_h, \mu)$ should be related to the asymptotic charges $(M,Q)$ provided that the solution around the horizon regularly connects to that at spatial infinity. 
The conserved current $J_V$ allows us to find the explicit relation between them. 
Substituting the asymptotic solutions~\eqref{solf3-shift-inf-f}-\eqref{solf3-shift-inf-phi} and the near-horizon solutions~\eqref{f3-sol1-shift-f1h1V1phi1}-\eqref{f3-sol1-shift-secondorder} into Eq.~\eqref{defJ_V}, we obtain $J_V=-Q$ and $J_V=\sqrt{2\Mpl^{2}\mu r_h^{2}-P^{2}}$, respectively, as the leading-order contributions. 
Equating these expressions via the current conservation law~\eqref{eq_J_V_all_function}, we obtain the relation
\be
\sqrt{2\Mpl^{2}\mu r_h^{2}-P^{2}}=-Q\,, 
\label{J_Vcurrent_relation_f3-sol1-shift}
\ee
which generalizes a result found in Ref.~\cite{Heisenberg:2018vti} to the dyonic case. 
Regarding the scalar field, we note that the constant modes $\phi_{\infty}$ and $\phi^{(0)}$ have no physical significance under the shift symmetry $\phi\to\phi+c$. 
Moreover, as we discussed in Sec.~\ref{shift_hair_con}, the vanishing scalar current $J_{\phi}=0$ leads to $\phi_{[1]}=0$. 
Consequently, the scalar field solutions, at the spatial infinity~\eqref{solf3-shift-inf-phi} and around the horizon~\eqref{f3-sol1-shift-f1h1V1phi1} and~\eqref{f3-sol1-shift-secondorder}, do not possess any free parameter. 
This confirms that the scalar hair in this case is of the secondary type.

\begin{figure}[!htbp]
\centering
\begin{minipage}[t]{0.48\textwidth}
  \captionsetup{width=\linewidth}
  \includegraphics[width=0.85\linewidth]{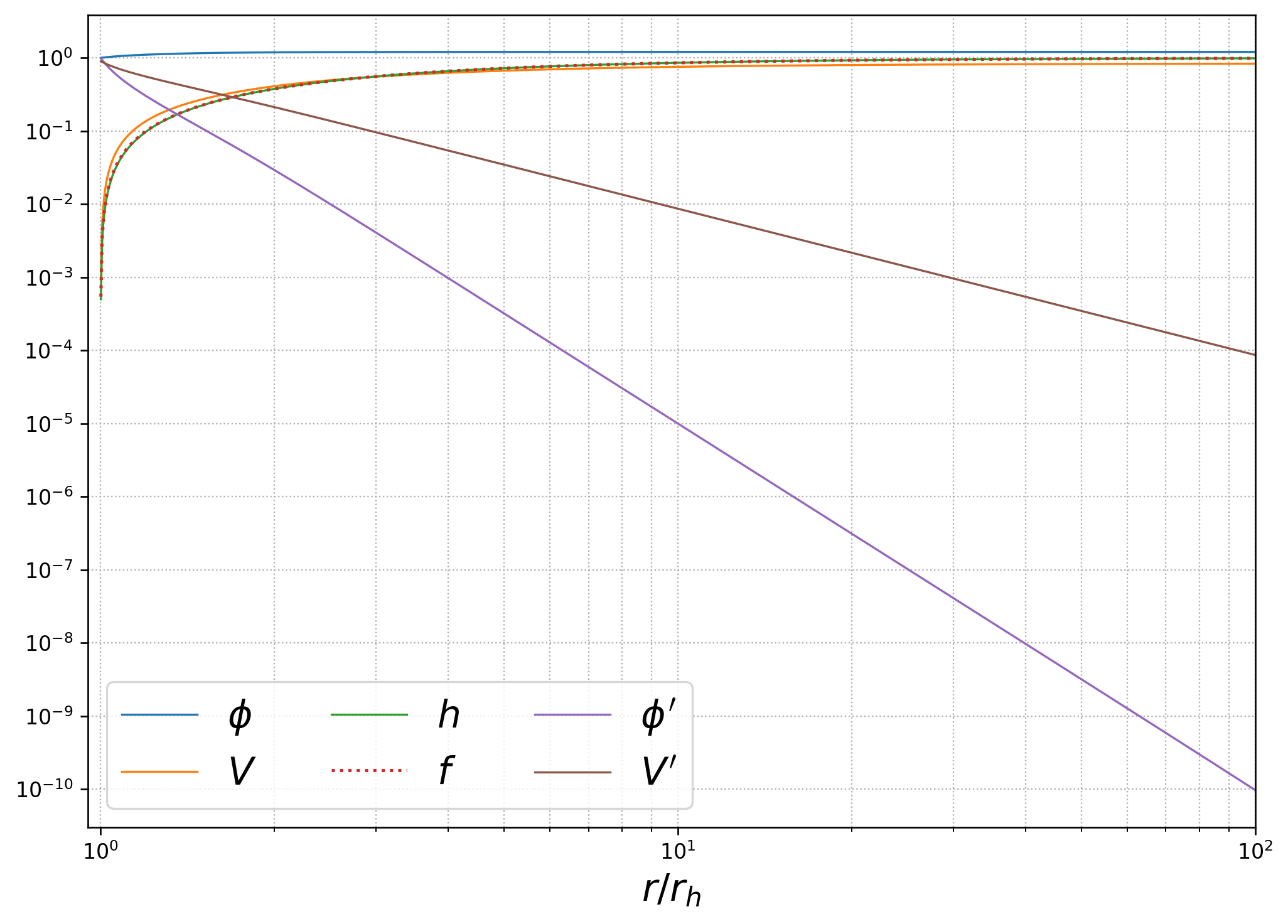}
  \caption{\RaggedRight
   Radial profiles of $f$, $h$, $V$, $V'$, $\phi$, and $\phi'$ for the \texorpdfstring{$f_3$}{f3}-Ia solution. 
   The fields $V$ and $\phi$ are normalized by $\Mpl$, while their derivatives are normalized by $\Mpl/r_h$.
   Boundary conditions are imposed at $r=1.001\,r_h$ based on the near-horizon expansion Eqs.~\eqref{f3-sol1-shift-f1h1V1phi1}-\eqref{f3-sol1-shift-secondorder}, with parameters $\mu=0.5$, $\phi^{(0)}=\Mpl$, $P=0.5\,\Mpl\,r_h$, and $c_3=1.0$.
}
\label{Fig-f3-1a-1}
\end{minipage}\hfill
\begin{minipage}[t]{0.48\textwidth}
  \captionsetup{width=\linewidth}
  \includegraphics[width=0.86\linewidth]{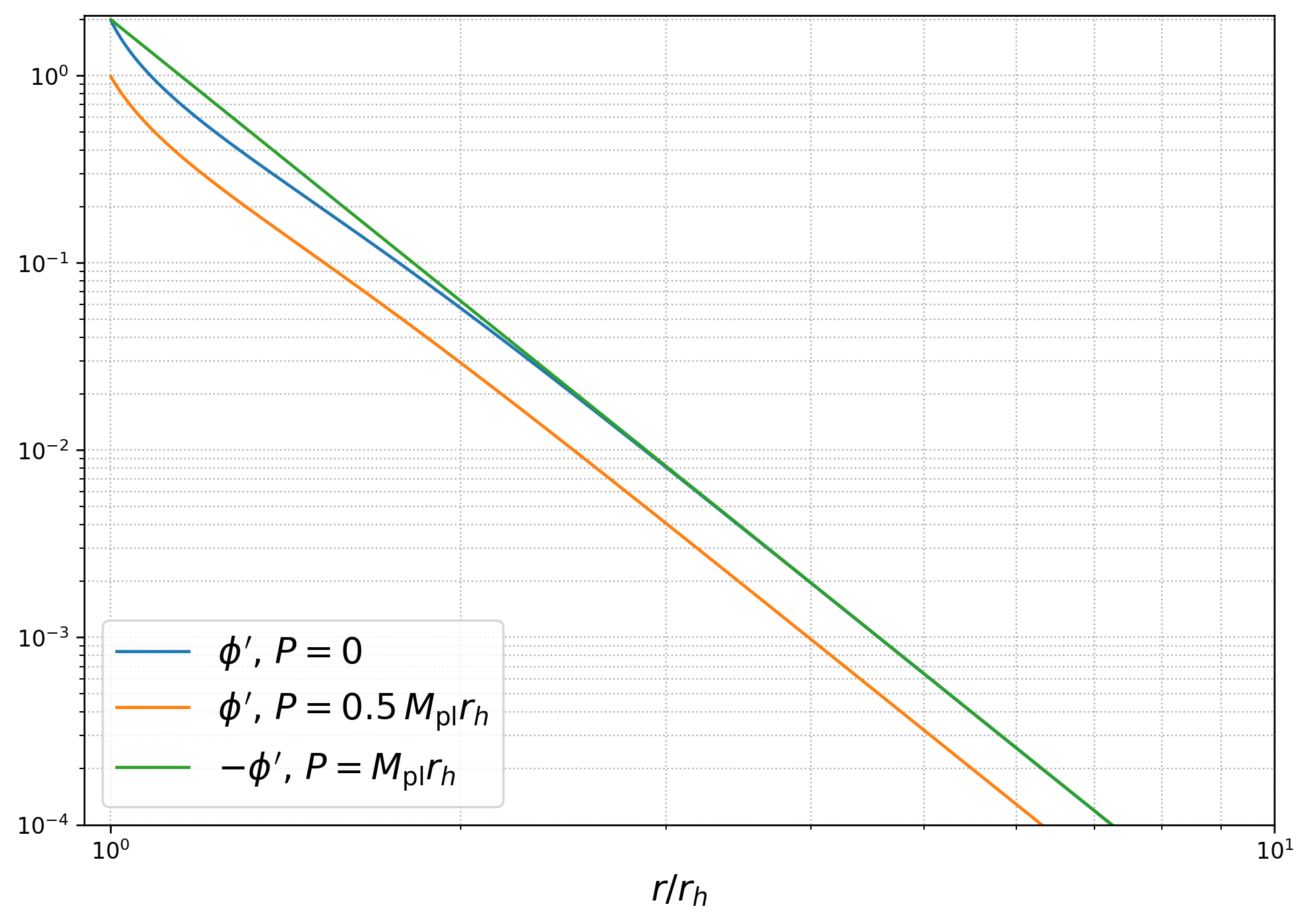}
  \caption{\RaggedRight
  Scalar field derivative profiles.
  We show the radial profile of $\phi'$ normalized by $\Mpl/r_h$ for three representative magnetic charges, $P=0$ (purely electric), $P=0.5\,\Mpl\,r_h$, and $P=\Mpl\,r_h$.
  Note that the sign of $\phi'$ is reversed (negative) only for the case $P=\Mpl\,r_h$. 
  Other parameters are fixed as $\mu=0.5$, $\phi^{(0)}=\Mpl$, and $c_3=1.0$.}
  \label{Fig-f3-1a-2}
\end{minipage}
\end{figure}

We examine the connection between the near-horizon and asymptotic solutions by numerical integration.
Figs.~\ref{Fig-f3-1a-1} and~\ref{Fig-f3-1a-2} show the numerical results for the solution $f_3$-$\mathrm{I}$a, where we specify the interaction function as
\be
f_3=\beta_3=\frac{r_h^2}{M_{\mathrm{pl}}}c_3\,,
\ee
with $c_3$ being a dimensionless parameter.
To enforce $f \to 1$ as $r \to \infty$ by rescaling the time coordinate, we employ a two-step integration procedure utilizing the time-rescaling freedom~\cite{DeFelice:2024bdq}.
First, we integrate Eqs.~\eqref{eq_f_all_function}-\eqref{eq_J_V_all_function} from $r=1.001\,r_h$ imposing the boundary conditions consistent with the near-horizon expansions~\eqref{f3-sol1-shift-f1h1V1phi1}-\eqref{f3-sol1-shift-secondorder}.
We extract the asymptotic value of the metric function, $f_\infty$, at large radii.
Then, we rescale the metric function as $f \to f/f_\infty$ and the vector field as $V \to V/\sqrt{f_\infty}$ via the freedom of rescaling the time coordinate, ensuring that the physical solution satisfies the asymptotic flatness condition.
Throughout this subsection, we set the parameters as $\mu=0.5$, $\phi^{(0)}=M_{\mathrm{pl}}$, and $c_3=1.0$ (unless stated otherwise), and normalize $\phi, V$ by $M_{\mathrm{pl}}$ and $\phi', V'$ by $M_{\mathrm{pl}}/r_h$.

The results, shown in Fig.~\ref{Fig-f3-1a-1}, exhibit the same qualitative behavior as the purely electric case~\cite{Heisenberg:2018vti}.
For the representative solutions shown, the field profiles remain finite and smooth over the numerical integration domain and approach the asymptotic expansions in Eqs.~\eqref{solf3-shift-inf-f}-\eqref{solf3-shift-inf-phi} at large radii.
Our parameter scan finds the same behavior for coupling constants in the range $|c_3|\lesssim\mathcal{O}(10)$.

In Fig.~\ref{Fig-f3-1a-2}, we focus on the dependence of the scalar hair on the magnetic charge $P$.
We plot $\phi'$ for $P=0$, $P=0.5\,M_{\mathrm{pl}}\,r_h$, and $P=M_{\mathrm{pl}}\,r_h$.
Crucially, the sign of $\phi'$ flips for the large magnetic charge case ($P=M_{\mathrm{pl}}\,r_h$).
This behavior is analytically predicted by the linear-order coefficient $\phi^{(1)}$ in Eq.~\eqref{f3-sol1-shift-f1h1V1phi1}.
For small magnetic charges satisfying $P<\sqrt{\mu}\,M_{\mathrm{pl}} r_h$, we have $\phi^{(1)}>0$ (assuming $\beta_3 > 0$), meaning $\phi'$ is positive around the horizon.
In contrast, for large magnetic charges with $P>\sqrt{\mu}\,M_{\mathrm{pl}} r_h$, the coefficient becomes negative ($\phi^{(1)}<0$), resulting in $\phi'<0$.
We also note that, from Eqs.~\eqref{f3-sol1-shift-f1h1V1phi1} and \eqref{f3-sol1-shift-V2}, the regularity of the vector field imposes an upper bound on the magnetic charge, $P\le\sqrt{2\mu}\,M_{\mathrm{pl}} r_h$.
At the saturation point $P=\sqrt{2\mu}\,M_{\mathrm{pl}} r_h$, we numerically confirmed that the boundary conditions $V^{(1)}=V^{(2)}=0$ enforce the vanishing of $V$ everywhere. However, the scalar field still possesses a nontrivial profile, exhibiting secondary hair supported solely by the magnetic charge.

\subsection{Near-horizon solution for the \texorpdfstring{$f_3$}{f3}-\rm{Ib} case}
\label{f3Ibsec}

We now consider a second branch of near-horizon solutions, which we refer to as the $f_3$-Ib solution. 
This solution arises for the general interaction function $f_3(X)$ given in Eq.~\eqref{f3-shift-function}. 
As discussed in Sec.~\ref{shift_hair_con}, the vanishing of the conserved current $J_\phi=0$ permits a nontrivial scalar profile ($\phi' \neq 0$) only if the interaction function $f_3$ contains a constant term that acts as a source in the scalar field equation.
Accordingly, we assume $f_3(0) \neq 0$.
In addition, to obtain a solution branch distinct from the $f_3$-Ia case, we require the function to exhibit explicit dependence on the kinetic term at the horizon, i.e., $f_{3,X}(0) \neq 0$.

Substituting the near-horizon ansatz~\eqref{expand_horizon_fh_V_phi} into the field equations under these conditions and solving them iteratively, we find the following solution coefficients. 
At linear order, we obtain
\be
f^{(1)}= h^{(1)}=\frac{1-\mu}{r_h}\,,\quad V^{(1)}=\frac{\sqrt{2 \Mpl^2 r_h^2\mu- P^2}}{r_h^2}\,,\quad 
\phi^{(1)}=\frac{{r_h}^{5}-\sqrt{{r_h}^{10}+8 P^{2} {f_3} {f_{3,X}} (1-\mu )(P^2-\mu\Mpl^2r_h^2)}}{(1-\mu ) {f_{3,X}} P^{2}}\,,
\label{f3_shift_sol2_f1h1V1}
\ee
where we have chosen the branch $V^{(1)} >0$.
In these expressions, $f_3$ and $f_{3,X}$ denote their values at the horizon ($X=0$). 
At second order, the coefficients are given by
\begin{align}
f^{(2)}=& 
\frac{ (24f_3-f_{3,X}f^{(1)}{\phi^{(1)}}{}^2{r_h}) f^{(1)}\phi^{(1)}P^2 }{4{\Mpl^2}{r_h^4}}
+\frac{6 {f_3}  {f^{(1)}} ({r_h}  {f^{(1)}}-1) \phi^{(1)}}{{r_h}^{2}}
-\frac{2 {r_h}  {f^{(1)}}-1}{{r_h}^{2}}
+\frac{3 {\phi^{(1)}}{}^{2} {r_h}  {f^{(1)}}}{4 {\Mpl}^{2}}\,, \\
h^{(2)}=&
-\frac{ (8f_3+f_{3,X}f^{(1)}{\phi^{(1)}}{}^2{r_h}) f^{(1)}\phi^{(1)}P^2 }{4\Mpl^2r_h^4} 
-\frac{2 {f_3}  {f^{(1)}} ({r_h}  {f^{(1)}}-1) \phi^{(1)}}{{r_h}^{2}}
-\frac{2 {r_h}  {f^{(1)}}-1}{{r_h}^{2}}
-\frac{{\phi^{(1)}}{}^{2} {r_h}  {f^{(1)}}}{4 {\Mpl}^{2}} \, ,\\
V^{(2)}=&\frac{2 V^{(1)} {f_3} \phi^{(1)} P^{2}}{{r_h}^{4} {\Mpl}^{2}}-\frac{2 V^{(1)} {f_3} \phi^{(1)}}{{r_h}^{2}}-\frac{V^{(1)}}{{r_h}}+\frac{V^{(1)} {r_h} {\phi^{(1)}}{}^{2}}{4 {\Mpl}^{2}}\,, \\
\phi^{(2)}=&\frac{1}{8 (P^{2} {f_{3,X}}  {f^{(1)}} \phi^{(1)}-{r_h}^{4})  {f^{(1)}} {r_h}^{2}}  \bigg[\bigg\{{\phi^{(1)}}{}^{4} {r_h}^{2}  {f^{(1)}}{}^{3} {f_{3,XX}}+(14 {r_h} f^{(2)}+6 {r_h} h^{(2)}-8  {f^{(1)}}) {f_3}\notag \\
-&\frac{11 {f_{3,X}} {r_h}^{2}  {f^{(1)}} f^{(2)} {\phi^{(1)}}{}^{2}}{2} 
-\frac{7 {f_{3,X}} {r_h}^{2}  {f^{(1)}} h^{(2)} \phi^{(1)}{}^{2}}{2}-4 {f_{3,X}} {r_h}  {f^{(1)}}{}^{2} {\phi^{(1)}}{}^{2}\bigg\} P^{2}\notag \\
+&(-6 {r_h}^{5} {V^{(1)}}{}^{2} f^{(2)}-14 {V^{(1)}}{}^{2} {r_h}^{5} h^{(2)}-32 V^{(1)} {r_h}^{5} V^{(2)}  {f^{(1)}}-24 {r_h}^{4} {V^{(1)}}{}^{2}  {f^{(1)}}) {f_3} \notag \\
+&7 \phi^{(1)} {r_h}^{6} f^{(2)}+3 \phi^{(1)} h^{(2)} {r_h}^{6}+4  {f^{(1)}} \phi^{(1)} (3 {f_{3,X}} {V^{(1)}}{}^{2}  {f^{(1)}} \phi^{(1)}+4) {r_h}^{5}\bigg]\,.\label{f3-sol2-second-order}
\end{align}
Direct substitution of $P=0$ into the expression for the linear coefficient $\phi^{(1)}$ leads to an indeterminate form. Expanding the square root before taking
this limit yields
\begin{equation}
\lim_{P\to 0}\phi^{(1)}
=
\frac{4\mu M_{\rm pl}^{2}f_3}{r_h^{3}}\,,\label{f3_sol2_phi1_P_to_0}
\end{equation}
which is finite and nonvanishing for $f_3\neq0$. 
The second-order coefficients also remain finite in this limit. 
In particular, $\phi^{(2)}$ retains a contribution proportional to $f_{3,X}$, while the metric coefficients $f^{(2)}$ and $h^{(2)}$ contain finite $f_3$-dependent corrections. 
Hence, at the level of the near-horizon expansion, the $P\to 0$ limit retains a nontrivial scalar profile together with finite modifications to the metric functions, rather than reducing to the RN configuration.
Likewise, we consider a functional limit in theory space in which the $X$ dependence of $f_3(X)$ is removed while the constant value $f_3(0)$ is held fixed.
Accordingly, all derivatives with respect to $X$, including $f_{3,X}$ and $f_{3,XX}$, vanish consistently in this limit. 
Although direct substitution into the expression for $\phi^{(1)}$ seems to produce an indeterminate form, expanding the coefficients along this functional limit shows that the near-horizon coefficients through second order reduce smoothly to those of the $f_3$-Ia solution given in Eqs.~\eqref{f3-sol1-shift-f1h1V1phi1}-\eqref{f3-sol1-shift-secondorder}. 
Thus, the explicit factor of $f_{3,X}$ in the denominator does not represent a divergence when the kinetic dependence of the interaction function is removed.
Finally, using the current conservation law~\eqref{eq_J_V_all_function}, we recover the same relation between the horizon and asymptotic parameters as in Eq.~\eqref{J_Vcurrent_relation_f3-sol1-shift}. 

We examine the connection between the near-horizon and asymptotic solutions by numerical integration.
For this purpose, we specify the interaction function as $f_3=(r_h^2/\Mpl)\!\left(\bar{c}_3+\bar{d}_3 r_h^2 X/\Mpl^2\right)$, where $\bar{c}_3$ and $\bar{d}_3$ are dimensionless parameters, and set the parameters as $\mu=0.7$, $\phi^{(0)}=M_{\mathrm{pl}}$, $P=0.7\,M_{\mathrm{pl}}\,r_h$, $\bar{c}_3=1.0$, and $\bar{d}_3=1.0$.
We used boundary conditions derived from the series expansions~\eqref{f3_shift_sol2_f1h1V1}-\eqref{f3-sol2-second-order} and applied the same two-step rescaling procedure described in Sec.~\ref{f3Iasec}. 
As a result, we obtained profiles qualitatively similar to those shown in Fig.~\ref{Fig-f3-1a-1}. 
For the parameter values examined, the field profiles remain finite and smooth over the numerical integration domain and approach the asymptotic behavior in Eqs.~\eqref{solf3-shift-inf-f}-\eqref{solf3-shift-inf-phi} at large radii.

\section{Shift-symmetric solutions in the \texorpdfstring{$g_3$}{g3} sector}
\label{shift_g3_sec}

We now turn our attention to the $g_3$ sector, specified by the couplings 
\be
f_3=0\,,\qquad g_3=g_3(X)\,,\qquad f_4=0\,.
\label{g3-shift-function}
\ee
In contrast to the $f_3$ sector, the scalar hair in this case is associated with a non-vanishing conserved current, the value of which is determined by the magnetic charge $P$ and the horizon parameter as shown in Eq.~\eqref{relation_g3_shift_charge}.
We first derive the asymptotic solution and then analyze two near-horizon branches distinguished by whether $g_3$ is constant or has explicit kinetic dependence in addition to a nonvanishing constant term.

\subsection{Asymptotic solution}
\label{shift_g3_inf}

%
While Ref.~\cite{Heisenberg:2018vti} found that this interaction never affects the geometry in the purely electric configuration, our analysis reveals that the nonzero magnetic charge $P$ activates the $g_3$ term, realizing a novel hairy BH solution.
This property follows directly from the structure of the field equations~\eqref{eq_f_all_function}-\eqref{eq_J_V_all_function}, where $g_3$ contributes only when coupled with $P$.

Substituting the functional choice \eqref{g3-shift-function} and the asymptotic expansion~\eqref{expand_f_h_V_phi} into the field equations~\eqref{eq_f_all_function}-\eqref{eq_J_V_all_function}, we obtain the following iterative solutions at spatial infinity:
\begin{align}
f=&1-\frac{2 M}{r}+\frac{P^{2}+Q^{2}}{2 {\Mpl}^{2} r^{2}}+\frac{M \phi_{[1]}^{2}}{6 {\Mpl}^{2} r^{3}}+\frac{\phi_{[1]}^{2} \{4 M^{2} {\Mpl}^{2}-(P^{2}+Q^{2})\}}{12 {\Mpl}^{4} r^{4}} 
+\mathcal{O}\bigg(\frac{1}{r^{5}}\bigg)\,,\label{g3-shift-inf-f} \\
h=&1-\frac{2 M}{r}+\frac{P^{2}+Q^{2}+\phi_{[1]}^{2}}{2 {\Mpl}^{2} r^{2}}+\frac{M \phi_{[1]}^{2}}{2 {\Mpl}^{2} r^{3}}+\frac{\phi_{[1]}^{2} \{8 M^{2} {\Mpl}^{2}-(P^{2}+Q^{2})\}}{12 {\Mpl}^{4} r^{4}} 
+\mathcal{O}\bigg(\frac{1}{r^{5}}\bigg)\,, \\
V=&V_{\infty}+\frac{Q}{r}-\frac{Q \phi_{[1]}^{2}}{12 {\Mpl}^{2} r^{3}}-\frac{M Q \phi_{[1]}^{2}}{6 {\Mpl}^{2} r^{4}} 
+\mathcal{O}\bigg(\frac{1}{r^{5}}\bigg)\,,\label{g3-shift-inf-V} \\
\phi=&\phi_{\infty}+\frac{\phi_{[1]}}{r}+\frac{\phi_{[1]} M}{r^{2}}+\frac{\phi_{[1]} \{16 M^{2} {\Mpl}^{2}-2( P^{2}+Q^{2})-\phi_{[1]}^{2}\}}{12 {\Mpl}^{2} r^{3}} \notag \\ 
&+\frac{M \phi_{[1]} \{12 M^{2} {\Mpl}^{2}-3 (P^{2}+Q^{2})-2 \phi_{[1]}^{2}\}}{6 {\Mpl}^{2} r^{4}}
+\mathcal{O}\bigg(\frac{1}{r^{5}}\bigg)\,,
\label{g3-shift-inf-phi}
\end{align}
where $\phi_{[1]}$ is given by
\be
{\phi}_{[1]}=\frac{(1-\mu)g_3(0)P^{2}}{r_h^{3}}\,, 
\notag
\ee
as derived in Eq.~\eqref{relation_g3_shift_charge}, and we have identified the integration constants as $f_{[1]}= h_{[1]}=-2M$ and $ V_{[1]}=Q$.
The leading scalar coefficient $\phi_{[1]}$ is non-vanishing and is uniquely determined by the magnetic charge $P$ and the near-horizon parameters $r_h$ and $\mu$ through the scalar current conservation, implying that this hair is of the secondary type.
The nonzero value of $\phi_{[1]}$ gives rise to a crucial physical distinction from the $f_3$ sector where the scalar field given in Eq.~\eqref{solf3-shift-inf-phi} decays rapidly as $1/r^4$ due to the vanishing scalar current ($\phi_{[1]}=0$). 
Indeed, in the $g_3$ sector, the interaction generates a scalar profile decaying as $1/r$, which is the standard fall-off rate for a massless field with a source, but here it arises without an independent scalar charge. 
This slow decay results in the effect of $g_3$ on the metric and vector fields appearing at lower orders ($1/r^2$ for $h$, $1/r^3$ for $f$ and $V$) compared to the $f_3$ case given in Eqs.~\eqref{solf3-shift-inf-f}-\eqref{solf3-shift-inf-phi}. 
We also note that the explicit appearance of $P^2 g_3$ in the coefficient $\phi_{[1]}$ confirms the breaking of electromagnetic duality.
We next derive the two near-horizon solution families, both of which can be matched to the asymptotic behavior in Eqs.~\eqref{g3-shift-inf-f}-\eqref{g3-shift-inf-phi}.

\subsection{Near-horizon solution for the \texorpdfstring{$g_3$}{g3}-\rm{Ia} case}
\label{g3Iasec}

We first consider the case where the interaction function is constant. 
We specify the function $g_3$ as
\be
g_3=\tilde{\beta}_3\,.
\label{g3-sol1-shift-horizon-function}
\ee
Hereafter, we refer to the solution satisfying this condition as the $g_3$-Ia solution.

Substituting the ansatz~\eqref{expand_horizon_fh_V_phi} and the choice~\eqref{g3-sol1-shift-horizon-function} into the field equations~\eqref{eq_f_all_function}-\eqref{eq_J_V_all_function}, we expand them around $r=r_h$ and solve iteratively in powers of $(r-r_h)$.
At linear order, we obtain
\be
f^{(1)}= h^{(1)}=\frac{1-\mu}{r_h}\,,\quad 
V^{(1)}
=\frac{1}{r_h^2}\sqrt{\frac{2 \Mpl^{2} r_h^2[\Mpl^2 r_h^8\mu+2P^4\tb_3^2(2-\mu)]}{\Mpl^2 r_h^8+4P^4\tb_3^2}-P^2} \,,\quad
\phi^{(1)}=\frac{ 2(3\mu -2) {\Mpl}^{2} \tilde{\beta}_3 P^{2} {r_h}^{3}}{({\Mpl}^{2} {r_h}^{8}+4 P^{4} \tilde{\beta}_3^{2}) (1-\mu)}\,,  \label{g3-sol1-shift-first-order}\\
\ee
where we have chosen the branch $V^{(1)} >0$.
At second order, the coefficients are determined as
\begin{align}
f^{(2)}=
&\frac{1}{ (\mu-1 ){r_h}^{2} ({\Mpl}^{2} {r_h}^{8}+4 P^{4} \tilde{\beta}_{3}^{2})^{2} (2 P^{4} \tilde{\beta}_{3}^{2}-{\Mpl}^{2} {r_h}^{8})}\bigg[
{\Mpl}^{6} (1-2\mu) (\mu-1 ) {r_h}^{24}
+(25\mu -{17}) {\Mpl}^{4} (\mu-1 ) P^{4} \tilde{\beta}_{3}^{2} {r_h}^{16} \notag \\
&+ {\Mpl}^{2} P^{8} \tilde{\beta}_{3}^{4} (98\mu^{2}-156 \mu +56) {r_h}^{8}
- (\mu-1 ) P^{12} \tilde{\beta}_{3}^{6} (64\mu -80)\bigg] \,,\\
h^{(2)}=
&\frac{1}{ (\mu-1 ){r_h}^{2} ({\Mpl}^{2} {r_h}^{8}+4 P^{4} \tilde{\beta}_{3}^{2})^{2} (2P^4\tilde{\beta}_3^2-{\Mpl}^{2} {r_h}^{8})} 
\bigg[{\Mpl}^{6} (1-2\mu ) (\mu-1 ) {r_h}^{24}-{\Mpl}^{4} (\mu-1 ) (75\mu -{43}) P^{4} \tilde{\beta}_{3}^{2} {r_h}^{16} \notag \\
&-{\Mpl}^{2} (198\mu^{2}-308 \mu +104) P^{8} \tilde{\beta}_{3}^{4} {r_h}^{8}
+ (\mu-1 ) P^{12} \tilde{\beta}_{3}^{6} (64\mu -112) \bigg] \,, \\
V^{(2)}=
&\frac{V^{(1)}}{(\mu-1 )^{2} (2 P^{4} \tilde{\beta}_{3}^{2}-{\Mpl}^{2} {r_h}^{8}) ({\Mpl}^{2} {r_h}^{8}+4 P^{4} \tilde{\beta}_{3}^{2})^{2} {r_h}} 
\bigg[{\Mpl}^{6} (\mu-1 )^{2} {r_h}^{24}
-{\Mpl}^{4} (\mu-1 ) P^{4} \tilde{\beta}_{3}^{2} (19\mu -9) {r_h}^{16} \notag \\
&
- {\Mpl}^{2}(74\mu^{2}-116 \mu +40) P^{8} \tilde{\beta}_{3}^{4} {r_h}^{8}
-16 P^{12} \tilde{\beta}_{3}^{6} (\mu-1 )\bigg]\,, \\
\phi^{(2)}=
&-\frac{{r_h}^{2} {\Mpl}^{2} P^{2} \tilde{\beta}_{3}}{(\mu-1 )^{3}({\Mpl}^{2} {r_h}^{8}+4 P^{4} \tilde{\beta}_{3}^{2})^{3} (2 P^{4} \tilde{\beta}_{3}^{2}-{\Mpl}^{2} {r_h}^{8})}  
\bigg[ {r_h}^{24} ({10} \mu^{2}-18 \mu +{7}) (\mu-1 ) {\Mpl}^{6}\notag \\
&-2{r_h}^{16} \tilde{\beta}_{3}^{2} (\mu - 1)(15\mu^2 + \mu - 4)
{\Mpl}^{4} P^{4}
- {r_h}^{8} (108 \mu^{3}-176 \mu^{2}+80 \mu -16) \tilde{\beta}_{3}^{4} {\Mpl}^{2} P^{8}\notag \\
&+(\mu -1) P^{12} \tilde{\beta}_{3}^{6} (256\mu^{2}-416 \mu +128)\bigg] \,.
\label{g3-sol1-shift-second-order}
\end{align}
Unlike the $f_3$ case, the interaction $g_3$ modifies the vector field coefficient $V^{(1)}$ even at linear order.
The solution exhibits nontrivial scalar hair carried by $\phi$.
In the limit $P\to 0$, Eq.~\eqref{g3-sol1-shift-first-order} reduces to $V^{(1)} = \sqrt{2\Mpl^2 r_h^2\mu}/r_h^2$ corresponding to the electric RN configuration, and $\phi^{(1)}$ vanishes. 
Thus, this branch smoothly reduces to the RN limit in the absence of magnetic charge, confirming that this hair is indeed generated by the magnetic charge. 
Using the current conservation law~\eqref{eq_J_V_all_function}, the physical charge $-Q$ at spatial infinity is related to the near-horizon parameters via
\be
r_h^{2}V^{(1)}=-Q\,.
\label{J_Vcurrent_relation_g3-sol1-shift}
\ee

\begin{figure}[!htbp]
\includegraphics[width=0.55\linewidth]{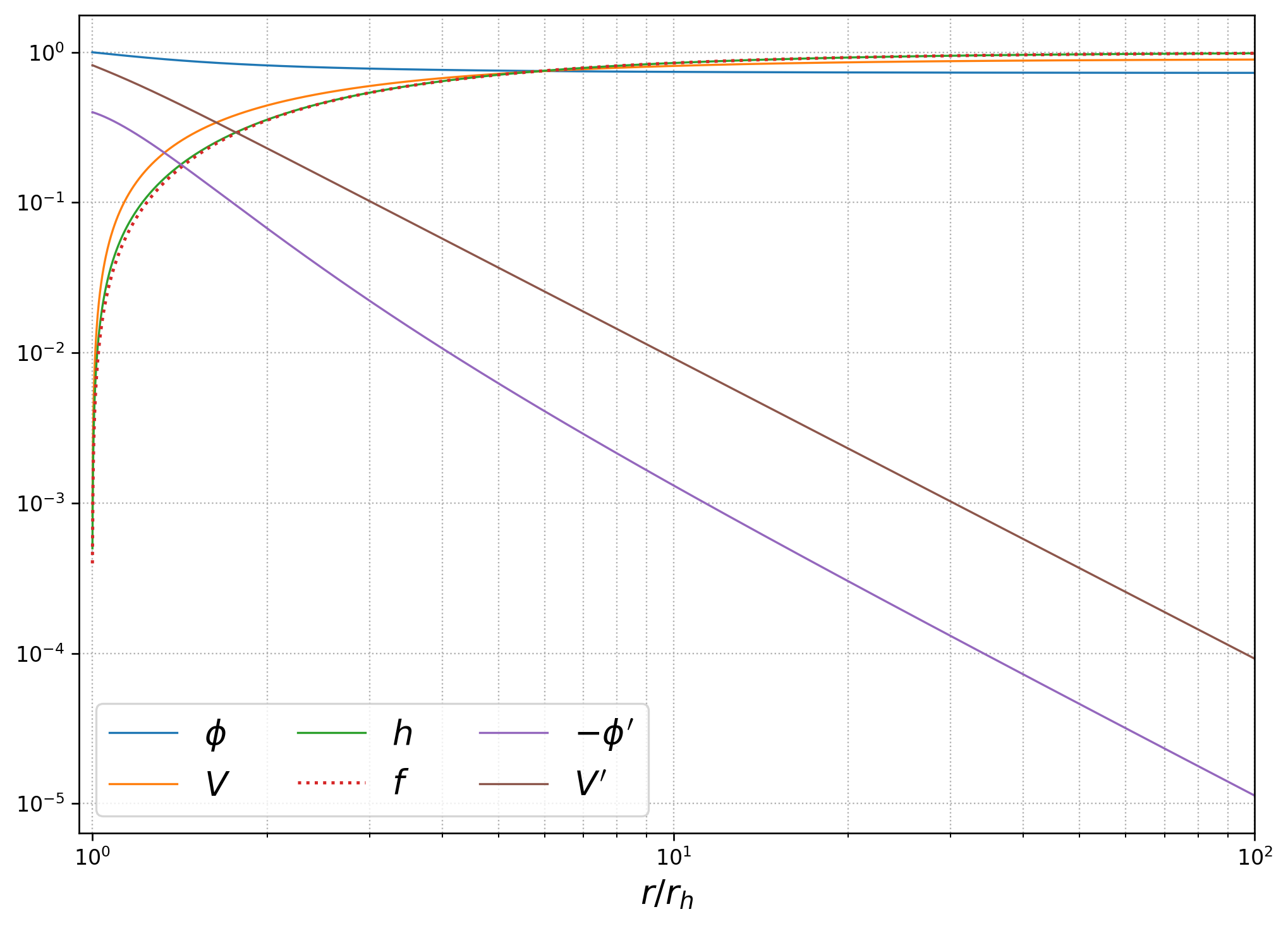}
\caption{\RaggedRight
Radial profiles of the metric functions ($f, h$), vector field ($V$), and scalar field ($\phi$), along with their derivatives ($V', \phi'$), in the exterior region for the $g_3$-$\mathrm{I}$a solution.
Normalization is the same as in Fig.~\ref{Fig-f3-1a-1}.
Boundary conditions are set at $r=1.001\,r_h$ consistent with Eqs.~\eqref{g3-sol1-shift-first-order}-\eqref{g3-sol1-shift-second-order}, with parameters $\mu=0.5$, $\phi^{(0)}=\Mpl$, $P=0.5\,\Mpl\,r_h$, and $\tilde{c}_3=1.0$.
Note the slower decay of $\phi'$ compared to the $f_3$-Ia case (Fig.~\ref{Fig-f3-1a-1}).
}
\label{Fig-g3-1a}
\end{figure}

Figure~\ref{Fig-g3-1a} displays the numerical results for the $g_3$-$\mathrm{I}$a solution obtained with the coupling choice $g_3=\tilde{\beta}_3={r_h^2}\tilde{c}_3/{M_{\mathrm{pl}}}$, where $\tilde{c}_3$ is a dimensionless parameter. 
To ensure asymptotic flatness, we employ the same two-step integration and time-rescaling procedure described in Sec.~\ref{f3Iasec}. 
We integrate the field equations outwards from $r=1.001\,r_h$, using boundary conditions derived from the near-horizon expansion in Eqs.~\eqref{g3-sol1-shift-first-order}-\eqref{g3-sol1-shift-second-order}. 
The profiles shown in Fig.~\ref{Fig-g3-1a} correspond to the parameters $\mu=0.5$, $\phi^{(0)}=M_{\mathrm{pl}}$, $P=0.5\,M_{\mathrm{pl}}\,r_h$, and $\tilde{c}_3=1.0$. 
For coupling constants in the range $|\tilde{c}_3|\lesssim\mathcal{O}(1)$, the numerical profiles remain finite and smooth over the integration domain and approach the asymptotic expansions in Eqs.~\eqref{g3-shift-inf-f}-\eqref{g3-shift-inf-phi} at large radii.

Compared with Fig.~\ref{Fig-f3-1a-1}, the numerical profile in Fig.~\ref{Fig-g3-1a} shows that $\phi'$ decays more slowly in the $g_3$-Ia solution than in the $f_3$-Ia solution.
This behavior agrees with the analytic falloffs $\phi\sim1/r$ in Eq.~\eqref{g3-shift-inf-phi} and $\phi\sim1/r^4$ in Eq.~\eqref{solf3-shift-inf-phi}.

\subsection{Near-horizon solution for the \texorpdfstring{$g_3$}{g3}-\rm{Ib} case}
\label{g3Ibsec}

We now consider the second branch of near-horizon solutions, referred to as the $g_3$-Ib solution. 
This solution arises for the general interaction function $g_3=g_3(X)$ in Eq.~\eqref{g3-shift-function}. 
Analogous to the $f_3$-Ib solution, the existence of a nontrivial scalar profile in this branch requires specific conditions on the coupling function. 
First, as discussed in Sec.~\ref{shift_hair_con}, the scalar hair must be sourced by a non-vanishing constant term, $g_3(0) \neq 0$, at the horizon. 
Second, to distinguish this branch from the Ia case, we require explicit dependence on the kinetic term at the horizon, $g_{3,X}(0) \neq 0$.

Substituting the ansatz~\eqref{expand_horizon_fh_V_phi} into the field equations under these conditions and solving them iteratively, we obtain the following series solution. 
At linear order, the coefficients are
\begin{align}
f^{(1)}
=& h^{(1)}
=\frac{1-\mu}{r_h}\,,\qquad
V^{(1)}
=\pm\frac{\sqrt{\left[2g_3(\mu-1)\phi^{(1)}-r_h\right]P^{2}
+2{\Mpl}^{2}\mu r_h^{3}}}{r_h^{5/2}}\,,
\label{g3-sol2-shift-f1h1V1}\\
\phi^{(1)}
=&
\frac{
{\Mpl}^{2}r_h^{8}+4P^{4}g_3^{2}
-\sqrt{\left({\Mpl}^{2}r_h^{8}+4P^{4}g_3^{2}\right)^{2}
+12{\Mpl}^{4}r_h^{6}P^{4}(3\mu-2)g_3g_{3,X}}
}{
3{\Mpl}^{2}P^{2}g_{3,X}r_h^{3}(\mu-1)
}\,.
\label{g3-sol2-shift-phi1}
\end{align}
Here, $g_3$ and $g_{3,X}$ denote their values at the horizon, where $X=0$. 
Eq.~\eqref{g3-sol2-shift-phi1} corresponds to the algebraic branch obtained by choosing the minus sign in front of the square root. 
We denote the scalar coefficient on this branch by $\phi^{(1)}_{-}$. 
The same algebraic equation also admits the plus-sign branch, whose scalar coefficient is denoted by $\phi^{(1)}_{+}$. As explained below, we restrict the subsequent analysis to the branch that is continuously connected to the purely electric configuration, and we therefore do not analyze the plus-sign branch in detail.
The corresponding second-order coefficients are lengthy and are provided in the Supplemental Material.
For the two choices of sign in $V^{(1)}$, the second-order metric and scalar-field coefficients are identical, whereas $V^{(2)}$ differs only by an overall sign.
Using the current conservation law~\eqref{eq_J_V_all_function}, we recover the same relation between the horizon and asymptotic parameters as in Eq.~\eqref{J_Vcurrent_relation_g3-sol1-shift},
\begin{align}
r_h^2V^{(1)}
=-Q\,.
\label{J_Vcurrent_relation_g3-sol2-shift}
\end{align}
The two signs of $V^{(1)}$ are related by the charge-conjugation transformation $V\to-V$ and $Q\to-Q$. 
They should not be confused with the two algebraic branches of the scalar coefficient, $\phi^{(1)}_{-}$ and $\phi^{(1)}_{+}$.

Before discussing the two roots, we note a structural property of the field equations~\eqref{eq_f_all_function}--\eqref{defJ_V} where the coupling $g_3$ enters the background equations only through the combination $P^{2}g_3$ and its derivatives. 
Consequently, the $g_3$ sector decouples completely at $P=0$, where the background equations reduce to those of Einstein--Maxwell theory with the canonical scalar field~\eqref{f2choice}, whose only static, spherically symmetric, asymptotically flat solution with a regular event horizon is the RN solution with a constant scalar field. 
In this sector we therefore restrict our analysis to the near-horizon branch that is continuously connected to this purely electric configuration. 

For the branch characterized by $\phi^{(1)}_{-}$, the numerator cancels at leading order as $P\to0$ as long as $g_3$ and $g_{3,X}$ are kept fixed. 
This cancellation occurs for either sign of $g_3$, because the two terms in the numerator of $\phi^{(1)}_{-}$ approach the same
positive value, $\Mpl^2r_h^8$, as $P\to0$. 
Explicitly, keeping the interaction function fixed, the two roots behave as
\begin{align}
\phi^{(1)}_{-}=\frac{2(3\mu-2)g_3}{(1-\mu)r_h^{5}}P^{2}+\mathcal{O}(P^{6})\,,
\qquad
\phi^{(1)}_{+}=-\frac{2r_h^{5}}{3(1-\mu)g_{3,X}}\frac{1}{P^{2}}+\mathcal{O}(P^{2})\,.
\end{align}
The sign of $g_3$ affects the leading coefficient of $\phi^{(1)}_{-}$, but not its scaling or its smooth connection to the RN solution. 
On this branch the vector-field coefficient approaches its RN value,
\begin{align}
V^{(1)}\to\pm\frac{\sqrt{2\mu}\,\Mpl}{r_h}\,,
\label{g3-sol2-shift-V1-limit}
\end{align}
and we have also confirmed that $f^{(1)}$, $h^{(1)}$, $f^{(2)}$, and $h^{(2)}$ approach their RN values. 
We therefore retain this branch as the one continuously connected to the purely electric RN solution.

For the other branch, characterized by $\phi^{(1)}_{+}$, the near-horizon expansion does not approach that of the RN solution. 
It should be emphasized, however, that this failure is not signaled by the divergence of $\phi^{(1)}_{+}$ itself. 
Substituting the corresponding near-horizon coefficients into the Ricci scalar evaluated at the horizon, we obtain
\begin{align}
R_{h,-}=\frac{8(3\mu-2)g_3^{2}}{\Mpl^{2}r_h^{10}}P^{4}+\mathcal{O}(P^{8})\,,
\qquad
R_{h,+}=-\frac{8g_3}{3\Mpl^{2}g_{3,X}}+\mathcal{O}(P^{4})\,,
\label{g3-sol2-shift-Ricci-limits}
\end{align}
so that $R_{h,-}$ approaches the RN value $R_h^{\rm RN}=0$, whereas $R_{h,+}$ remains finite and generically nonzero even though $\phi^{(1)}_{+}$ diverges. 
Likewise, the electric charge determined by the conserved current through Eq.~\eqref{J_Vcurrent_relation_g3-sol2-shift} remains finite on this branch. 
Since the factor $P^{2}$ multiplying $\phi^{(1)}$ in Eq.~\eqref{g3-sol2-shift-f1h1V1} compensates the $P^{-2}$ behavior of $\phi^{(1)}_{+}$, the $P\to0$ limit gives $V^{(1)}\to\pm\sqrt{({2\mu \Mpl^{2}}/{r_h^{2}})+{4g_3}/({3g_{3,X})}}$. 
This differs from the RN value~\eqref{g3-sol2-shift-V1-limit} by the finite, $P$-independent offset $4g_3/(3g_{3,X})$. 
We note that this offset also controls $R_{h,+}$ in Eq.~\eqref{g3-sol2-shift-Ricci-limits}, and that a real near-horizon configuration requires the argument of the square root in $V^{(1)}$ to be nonnegative. 
The divergence of the scalar-field coefficient therefore does not by itself indicate a curvature singularity or a divergent charge. 
What distinguishes this branch is that it is non-perturbative in $g_{3,X}$. 
Since it is the second root of an algebraic equation whose leading coefficient is proportional to $P^{2}g_{3,X}$, it consequently leaves the solution space in the limit $P\to0$, in which the $g_3$ sector decouples. 
Accordingly, we regard the exclusion of this branch as a restriction of the scope of the present analysis to solutions
continuously connected to the purely electric configuration, rather than as a statement that it is physically singular. 

This classification according to the sign chosen in front of the square root is independent of the sign of $g_3$. 
It is also distinct from taking an additional theory limit such as $g_3\to0$ or $g_{3,X}\to0$. 
Since the present analysis is based only on the local near-horizon expansion, we make no claim about the global existence of either branch, and the finiteness of $R_h$ alone does not establish the regularity of the full spacetime.

To examine the radial profiles outside the horizon, we numerically solve the field equations~\eqref{eq_f_all_function}-\eqref{eq_J_V_all_function}. 
We impose boundary conditions at $r=1.001r_h$ consistent with Eqs.~\eqref{g3-sol2-shift-f1h1V1} and \eqref{g3-sol2-shift-phi1}, together with the second-order corrections given in the Supplemental Material. 
We select the scalar branch characterized by $\phi^{(1)}_{-}$ and perform the numerical integration for both signs of $V^{(1)}$. 
The two choices yield identical metric and scalar-field profiles, while their vector-field profiles differ only by an overall sign, so the following discussion applies to both.
As a concrete model, we specify the interaction function as
\begin{align}
g_3=\frac{r_h^2}{\Mpl}\left(\hat{c}_3+\frac{r_h^2\hat{d}_3}{\Mpl^2}X\right),
\label{g3-sol2-concrete-model}
\end{align}
where $\hat{c}_3$ and $\hat{d}_3$ are dimensionless parameters. The existence of real near-horizon boundary data requires the argument of the square root in Eq.~\eqref{g3-sol2-shift-phi1} to be nonnegative. For $\mu=0.5$ and $P=0.5\Mpl r_h$, this condition reduces to
\begin{align}
\left(1+\frac{\hat{c}_3^2}{4}\right)^2-\frac{3}{8}\hat{c}_3\hat{d}_3\geq0\,.
\label{g3-sol2-reality-condition}
\end{align}
For $\hat{c}_3\hat{d}_3<0$, the left-hand side of Eq.~\eqref{g3-sol2-reality-condition} is manifestly positive, and hence this parameter region is not excluded by the near-horizon reality condition~\eqref{g3-sol2-reality-condition}. By contrast, for $\hat{c}_3\hat{d}_3>0$, a part of the parameter region is already excluded by Eq.~\eqref{g3-sol2-reality-condition}.
This condition is necessary, but not sufficient, for a smooth connection to the asymptotic solution.
For $\hat{c}_3\hat{d}_3<0$, our parameter scan finds numerical profiles that remain finite and smooth over the integration domain and approach the asymptotic expansions at large radii for $|\hat{c}_3|\lesssim\mathcal{O}(10^{-1})$ and $|\hat{d}_3|\lesssim\mathcal{O}(10^3)$. 
For $\hat{c}_3\hat{d}_3>0$, the same behavior is found in the more restricted range $|\hat{c}_3|\lesssim\mathcal{O}(10^{-1})$ and $|\hat{d}_3|\lesssim\mathcal{O}(10)$.
For mild kinetic dependence, $|\hat{d}_3|\lesssim\mathcal{O}(10)$, the radial profiles are qualitatively similar to those of the $g_3$-Ia solution. 
As $|\hat{d}_3|$ increases to $\mathcal{O}(10^2)$, the scalar gradient $\phi'$ develops an approximately constant plateau immediately outside the horizon.
This plateau behavior is confined to the near-horizon region and can be understood analytically. 
For both signs of $\hat{d}_3$ in the parameter range considered, the magnitude of the second-order coefficient shown in the Supplemental Material decreases approximately as $|\phi^{(2)}|\propto |\hat{d}_3|^{-1/2}$.
Hence, increasing $|\hat{d}_3|$ suppresses the leading radial variation of $\phi'$, resulting in the nearly constant $\phi'$ in the immediate vicinity of the horizon.

\section{\texorpdfstring{$\phi$}{phi}-dependent solutions in the \texorpdfstring{$f_3$}{f3} sector}
\label{phi_f3_sec}

In the following three Secs.~\ref{phi_f3_sec}--\ref{phi_f4_sec}, we relax the shift-symmetry assumption and explore hairy BH solutions in a more general setup where the interaction functions explicitly depend on the scalar field $\phi$. 
Such dependence introduces a source term $\mathcal{P}_\phi$ in the scalar field equation~\eqref{eq_J_phi_P_phi_all_function}, which is expected to generate primary hair driven by the interaction itself, distinct from the secondary hair discussed in Secs.~\ref{sec3}--\ref{shift_g3_sec}. 
Throughout the following sections, we adopt a context-dependent shorthand notation for the interaction functions and their derivatives.
In the analysis of asymptotic solutions at spatial infinity, symbols such as $f_i$, $f_{i,\phi}$, and $f_{i,X}$ denote values evaluated at the asymptotic background, e.g., $f_i(\phi_\infty, 0)$.
Conversely, in the analysis of near-horizon solutions, they denote values evaluated at the horizon, e.g., $f_i(\phi^{(0)}, 0)$. 

Hereafter, in this section, we focus on the $f_3$ sector and examine the properties of the resulting hairy solutions. 
In doing so, we specify the interaction functions as
\be
f_3=f_3(\phi,X)\,,\qquad g_3=0\,,\qquad f_4=0\,.
\label{f3-phi-function}
\ee
%

\subsection{Asymptotic solution}
\label{phi_f3_inf}

Let us derive the asymptotic solution at spatial infinity for the $f_3$ sector. 
Substituting the choice~\eqref{f3-phi-function} into the field equations~\eqref{eq_f_all_function}-\eqref{eq_J_V_all_function} and using the asymptotic expansions~\eqref{expand_f_h_V_phi}, we obtain the following iterative solutions at spatial infinity:
\begin{align}
f=
&1-\frac{2M}{r}+\frac{P^2+Q^2}{2\Mpl^2r^2}+\frac{M \phi_{[1]}^2}{6\Mpl^2r^3}+\frac{\phi_{[1]}^2\{4M^2\Mpl^2-(P^2+Q^2)\}}{12\Mpl^4r^4}+\frac{M \phi_{[1]}^{2} \{144 M^{2} {\Mpl}^{2}-52 (P^{2}+ Q^{2})-9 \phi_{[1]}^{2}\}}{240 {\Mpl}^{4}r^5}\label{f3-sol1-inf-f} \notag \\
&+\frac{ \phi_{[1]}^{2} \bigg[24\{16M^{4}- (P^2+ Q^{2}) {f_{3,\phi}}\}{\Mpl}^{4}-4M^{2} \{45(P^{2}+Q^{2})+{16 \phi_{[1]}^{2}}\} {\Mpl}^{2}+{ (P^{2}+Q^{2}) \{9(P^{2}+Q^{2})+8 \phi_{[1]}^{2}})\}\bigg]}{360 {\Mpl}^{6} r^{6}}\notag \\
&+\mathcal{O}\bigg(\frac{1}{r^{7}}\bigg)\,,\\
h=
&1-\frac{2M}{r}+\frac{P^2+Q^2+\phi_{[1]}^2}{2\Mpl^2r^2}+\frac{M \phi_{[1]}^2}{2\Mpl^2r^3}+\frac{\phi_{[1]}^2\{8M^2\Mpl^2-(P^2+Q^2)\}}{12\Mpl^4r^4}+\frac{M \phi_{[1]}^2\{48M^2\Mpl^2-12(P^2+Q^2)-\phi_{[1]}^2\}}{48\Mpl^4r^5}\notag \\
&+\frac{ \phi_{[1]}^{2}\bigg[ 12\{16M^{4}+(P^2+Q^2) {f_{3,\phi}} \} {\Mpl}^{4}-12 \{6(P^{2}+Q^{2})+\phi_{[1]}^{2}\} M^{2} {\Mpl}^{2}+3{(P^{2}+Q^{2}) \{(3(P^{2}+Q^{2})+\phi_{[1]}^{2}}\}\bigg]}{120 {\Mpl}^{6} r^{6}}\notag \\
&+\mathcal{O}\bigg(\frac{1}{r^{7}}\bigg)\,,\\
V=
&V_{\infty}+\frac{Q}{r}-\frac{Q {\phi_{[1]}}^{2}}{12 {\Mpl}^{2} r^{3}}-\frac{Q {\phi_{[1]}} ({\phi_{[1]}} M-6 {\Mpl}^{2} f_3)}{6 {\Mpl}^{2} r^{4}}+\frac{{\phi_{[1]}}^{2} Q \{128 f_{3,\phi} {\Mpl}^{4}-48 M^{2} {\Mpl}^{2}+4 (P^{2}+ Q^{2})+3 {\phi_{[1]}}^{2}\}}{160 {\Mpl}^{4} r^{5}}\notag \\
&+\frac{{\phi_{[1]}}^{2} Q \{60 {\Mpl}^{4} M f_{3,\phi}+30 {\Mpl}^{4} f_{3,\phi\phi} {\phi_{[1]}}-48 M^{3} {\Mpl}^{2}+9 M (P^{2}+ Q^{2})+8 M {\phi_{[1]}}^{2}\}}{90 {\Mpl}^{4} r^{6}}+\mathcal{O}\bigg(\frac{1}{r^{7}}\bigg)\,,\\
\phi=
&\phi_{\infty}+\frac{{\phi_{[1]}}}{r}+\frac{{\phi_{[1]}} M}{r^{2}}+\frac{{\phi_{[1]}} \{16 M^{2} {\Mpl}^{2}-2( P^{2}+ Q^{2})-{\phi_{[1]}}^{2}\}}{12 {\Mpl}^{2} r^{3}}\notag \\
&+\frac{12 M^{3} {\Mpl}^{2} {\phi_{[1]}}+3 {\Mpl}^{2}( P^{2}+Q^2) f_3-3 M (P^{2}+Q^2) {\phi_{[1]}}-2 M {\phi_{[1]}}^{3}}{6 {\Mpl}^{2} r^{4}}+\mathcal{O}\bigg(\frac{1}{r^{5}}\bigg)\,, \label{f3-sol-inf-phi-dependence}
\end{align}
where we have set $f_{[1]}= h_{[1]}=-2M$ and $V_{[1]}=Q$. 
Compared with the shift-symmetric case in Secs.~\ref{sec3}--\ref{shift_g3_sec}, both $\phi_{[1]}$ and $\phi_\infty$ appear as independent integration constants of the asymptotic expansion and are not fixed by $M$, $Q$, or $P$ at this stage.
Crucially, unlike in the shift-symmetric case, the constant $\phi_\infty$ cannot be removed by a constant shift of the scalar field, and thus constitutes a physical parameter affecting the solution through the $\phi$-dependence of $f_3$.
The explicit dependence of the expansion coefficients on both $\phi_{[1]}$ and $\phi_\infty$ in Eq.~\eqref{f3-sol-inf-phi-dependence} indicates that the scalar field is not merely a secondary response but carries independent degrees of freedom, classifying this solution as primary hair~\cite{Herdeiro:2015waa}. 

It is worth noting that by imposing the shift symmetry on these results, i.e., taking the limits $f_{3,\phi}\to 0$ and $f_{3,\phi\phi}\to 0$, and setting $\phi_{[1]}=0$ required by the current conservation $J_\phi=0$ in the shift-symmetric case, we recover the shift-symmetric asymptotic solutions given in Eqs.~\eqref{solf3-shift-inf-f}-\eqref{solf3-shift-inf-phi}.
Comparing the orders of magnitude, we observe that the interaction function enters the metric functions at $\mathcal{O}(r^{-6})$ via $f_{3,\phi}$ in the $\phi$-dependent case, whereas it appears at $\mathcal{O}(r^{-8})$ in the shift-symmetric case.
Similarly, for the vector field $V$, the leading interaction term is $\mathcal{O}(r^{-4})$ here, compared to $\mathcal{O}(r^{-7})$ in the shift-symmetric case.
For the scalar field $\phi$, the interaction terms start at $\mathcal{O}(r^{-4})$ in both cases, but the leading behavior is governed by $\phi_{[1]}/r$ in the $\phi$-dependent case versus $1/r^4$ in the shift-symmetric case. 
Thus, the effect of the interaction function here manifest itself at the lower order compared to the shift-symmetric case.

\subsection{Near-horizon solution for the \texorpdfstring{$f_3$}{f3}-\rm{IIa} case}
\label{f3IIasec}

We now derive the near-horizon solution for the specific choice of a purely $\phi$-dependent coupling,
\be
f_3=f_3(\phi)\,.
\label{f3-phi-depend-sol1-function}
\ee
We refer to the solution specified by this choice as the $f_3$-IIa solution.
Substituting the near-horizon ansatz~\eqref{expand_horizon_fh_V_phi} into the field equations~\eqref{eq_f_all_function}-\eqref{eq_J_V_all_function} and solving them iteratively, we obtain the linear-order coefficients as

\be
f^{(1)}= h^{(1)}=\frac{1-\mu}{r_h}, \qquad 
V^{(1)}=\frac{\sqrt{2 {\Mpl}^{2} {r_h}^{2} \mu -P^{2}}}{{r_h}^{2}}, \qquad 
\phi^{(1)}=-\frac{4 ({\Mpl}^{2} {r_h}^{2} \mu -P^{2}) f_3}{-{r_h}^{5}+2 P^{2} f_{3,\phi} {r_h}}\,,
\label{f3-sol1-linear}
\ee
where we have chosen the branch $V^{(1)} >0$. 
At second order,
\begin{align}
f^{(2)}=
&\frac{12 (1-\mu ) ({\Mpl}^{2} {r_h}^{2} \mu -P^{2})^{2} f_3^{2}}{{\Mpl}^{2}{r_h}^{6} (2 P^{2} f_{3,\phi}-{r_h}^{4}) }+\frac{2 \mu -1}{{r_h}^{2}}\,,\\
h^{(2)}=
&\frac{4 (\mu-1 ) ({\Mpl}^{2} {r_h}^{2} \mu -P^{2})^{2} f_3^{2}}{{\Mpl}^{2} {r_h}^{6} (2 P^{2} f_{3,\phi}-{r_h}^{4})}+\frac{2 \mu -1}{{r_h}^{2}}\,,\\
V^{(2)}=
&-\frac{4 \{{\Mpl}^{2} {r_h}^{2} (\mu -2) +P^{2}\} ({\Mpl}^{2} {r_h}^{2} \mu -P^{2}) \sqrt{2 {\Mpl}^{2} {r_h}^{2} \mu -P^{2}}\, f_3^{2}}{{r_h}^{7} (2 P^{2} f_{3,\phi}-{r_h}^{4}) {\Mpl}^{2}}-\frac{\sqrt{2 {\Mpl}^{2} {r_h}^{2} \mu -P^{2}}}{{r_h}^{3}}\,,\\
\phi^{(2)}=
&\frac{32 (1-\mu ) (2 {\Mpl}^{2} {r_h}^{2} \mu -P^{2})( 2P^{2}f_{3,\phi}-{r_h}^{4}) ( P^{2}-{\Mpl}^{2} {r_h}^{2} \mu ) f_3^{3}}{{r_h}^{4} (2 P^{2} f_{3,\phi}-{r_h}^{4})^{3}} 
-\frac{12 P^{2} ({\Mpl}^{2} {r_h}^{2} \mu -P^{2})^2   f_3^{2}f_{3,\phi\phi}}{{r_h}^{2} (2 P^{2} f_{3,\phi}-{r_h}^{4})^{3}} \notag \\
&+\frac{ \{8P^{2} {r_h}^{4} \mu  {\Mpl}^{2} f_{3,\phi}^{2}-4{r_h}^{6} ({\Mpl}^{2} {r_h}^{2} \mu +5 P^{2}) f_{3,\phi}+10 {r_h}^{10}\} ({\Mpl}^{2} {r_h}^{2} \mu -P^{2}) f_3}{{r_h}^{4} (2 P^{2} f_{3,\phi}-{r_h}^{4})^{3}}\,.
\label{f3-sol1-phi2}
\end{align}
This solution reduces to the shift-symmetric solution $f_3$-Ia in the limit $f_3=\beta_3$.
It can therefore be regarded as a dyonic and $\phi$-dependent extension of the solutions in the pure electric configuration with a constant interaction $f_3=\beta_3$ obtained in Ref.~\cite{Heisenberg:2018vti}.
In the $P\to 0$ limit, this solution $f_3$-IIa continuously approaches a purely electric hairy BH solution with a nonvanishing scalar profile.
We observe that the interaction function $f_3$ affects the scalar field $\phi$ already at linear order, whereas its contribution to $f$, $h$, and $V$ appears starting from the second order.
Furthermore, using the current conservation law of the vector field~\eqref{eq_J_V_all_function}, we confirm that the relation between the horizon and asymptotic charges takes the same form as in Eq.~\eqref{J_Vcurrent_relation_f3-sol1-shift}.

\begin{figure}[!htbp]
\includegraphics[width=0.55\linewidth]{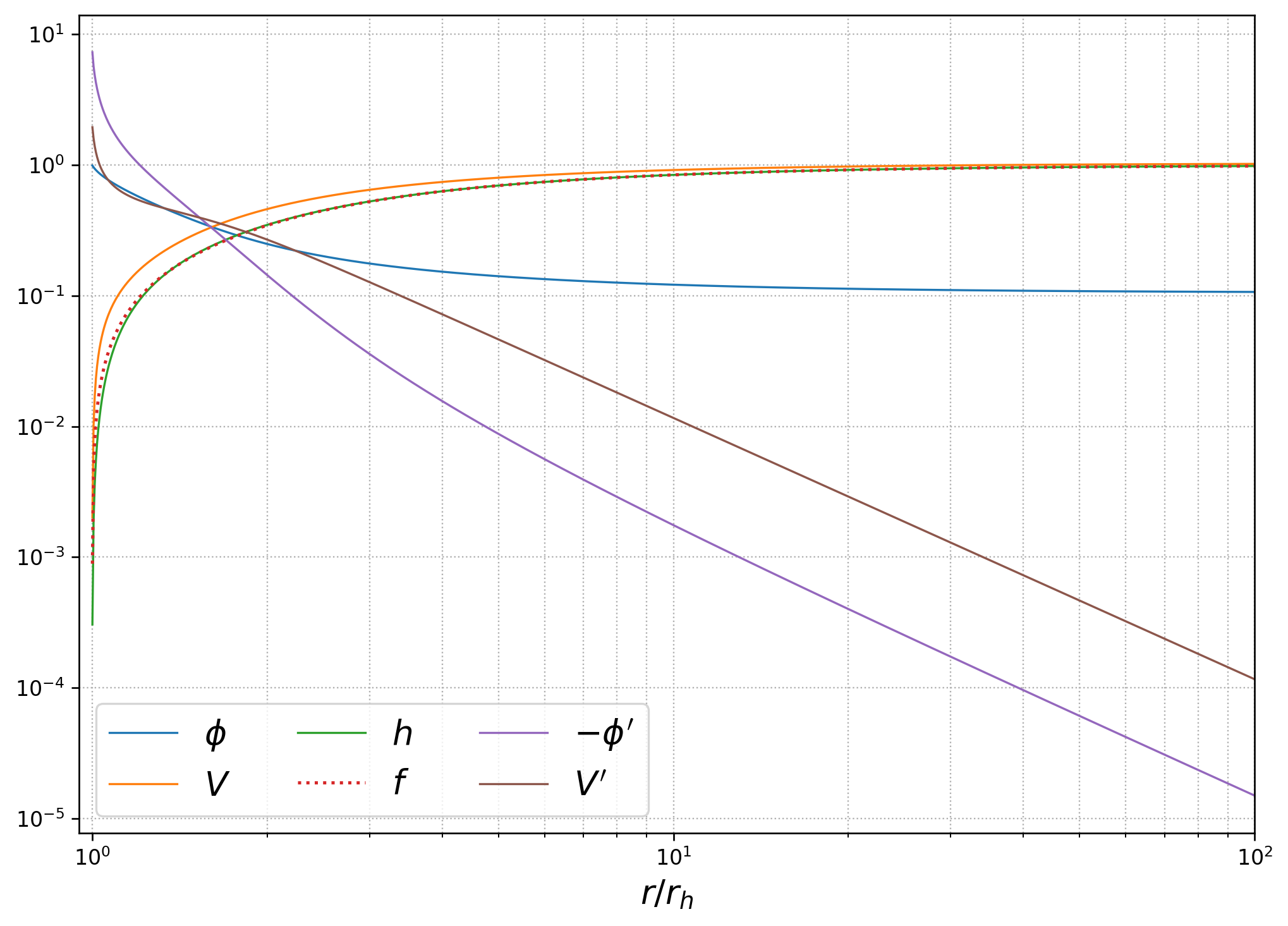}
\caption{\RaggedRight
Radial profiles for the $f_3$-IIa solution. 
We numerically solve for the radial dependence of the metric functions $f$ and $h$, the vector field $V$, and the scalar field $\phi$, along with their derivatives $V'$ and $\phi'$, in the exterior region of the event horizon. 
The fields $V$ and $\phi$ are normalized by $\Mpl$, while their radial derivatives $V'$ and $\phi'$ are normalized by $\Mpl/r_h$. 
Boundary conditions are imposed at $r=1.001\,r_h$ consistent with the near-horizon expansions in Eqs.~\eqref{f3-sol1-linear}-\eqref{f3-sol1-phi2}, with the parameters set to $\mu=0.7$, $\phi^{(0)}=\Mpl$, $P=0.1\,\Mpl\,r_h$, and $C_3=-3.0$.
}
\label{Fig-f3-2a}
\end{figure}
We examine the connection between the near-horizon and asymptotic solutions by numerical integration, specifying the interaction function as
\be
f_3=\frac{r_h^2}{M_{\mathrm{pl}}}\,\bigg(C_3\,\frac{\phi}{\Mpl} \bigg)\,,
\ee
where $C_3$ is a dimensionless parameter.
Following the procedure in Fig.~\ref{Fig-f3-1a-1}, we employ the two-step integration method to enforce $f\to1$ as $r\to\infty$.
We integrate the field equations~\eqref{eq_f_all_function}-\eqref{eq_J_V_all_function} outwards from the horizon using boundary conditions consistent with the near-horizon expansions~\eqref{f3-sol1-linear}-\eqref{f3-sol1-phi2}. 
Figure~\ref{Fig-f3-2a} shows the results for the parameters $\mu=0.7$, $\phi^{(0)}=M_{\mathrm{pl}}$, $P=0.1\,M_{\mathrm{pl}}\,r_h$, and $C_3=-3.0$. 
For coupling constants in the range $|C_3|\lesssim\mathcal{O}(1)$, the numerical profiles remain finite and smooth over the integration domain and approach the asymptotic expansions in Eqs.~\eqref{f3-sol1-inf-f}-\eqref{f3-sol-inf-phi-dependence} at large radii.
Notably, the scalar field $\phi$ converges to a finite asymptotic value $\phi_{\infty}\simeq 0.1$.
Unlike the shift-symmetric case, this asymptotic value $\phi_{\infty}$ has physical significance and cannot be shifted away.

\subsection{Near-horizon solution for the \texorpdfstring{$f_3$}{f3}-\rm{IIb} case}
\label{f3IIbsec}

We now turn to the second branch of near-horizon solutions, denoted as the $f_3$-IIb solution.
This solution exists only if the coupling function $f_3$ depends explicitly on the kinetic term. 
Hence, employing the general choice~\eqref{f3-phi-function}, we substitute the ansatz~\eqref{expand_horizon_fh_V_phi} into the field equations and solve them iteratively. 
At linear order, the metric and vector field coefficients are given by
\be
f^{(1)}= h^{(1)}=\frac{1-\mu}{r_h}\,, \qquad 
V^{(1)}=\frac{\sqrt{2 {\Mpl}^{2} {r_h}^{2} \mu -P^{2}}}{{r_h}^{2}}\,, 
\label{f3-sol2-f1h1V1}
\ee
where we have chosen the branch $V^{(1)}>0$. 
The scalar field coefficient is derived as
\begin{align}
    \phi^{(1)}
=\frac{\sqrt{\Delta_h^2+8P^2f_3f_{3,X}(1-\mu)\left(P^2-\Mpl^2\mu r_h^2\right)}-\Delta_h}{(\mu-1)f_{3,X}P^2} \,,\label{f3-sol2-phi1}
\end{align}
where we define
\be
\Delta_h = r_h^5-2P^2r_h f_{3,\phi}\,.
\ee
The corresponding second-order coefficients are lengthy and thus provided in the Supplemental Material. 
We have chosen the plus sign in front of the square root in Eq.~\eqref{f3-sol2-phi1}. 
We consider the functional limit in which the kinetic dependence is removed, $f_3(\phi,X)\to f_3(\phi)$.
All relevant derivatives with respect to $X$, including $f_{3,X}$, $f_{3,XX}$, and $f_{3,\phi X}$, are taken to vanish consistently in this limit.
For
\begin{equation}
\Delta_h>0 ,
\label{f3-IIb-to-IIa-condition}
\end{equation}
the square root in Eq.~\eqref{f3-sol2-phi1} approaches $\Delta_h$. 
The two terms in the numerator then cancel, and the expression seems to take an indeterminate form.
Nevertheless, expanding along this functional limit gives a finite result that agrees with the $f_3$-IIa solution in Eq.~\eqref{f3-sol1-linear}. 
We have also verified that the near-horizon coefficients through second order reduce to those of the $f_3$-IIa solution in this functional limit.
For $\Delta_h<0$, the analogous cancellation instead occurs for the algebraic branch defined by the minus sign in front of the square root. 
For the subsequent analysis, we retain the plus-sign branch because its near-horizon coefficients through second order reduce to those of the $f_3$-Ib solution when the explicit $\phi$ dependence is removed, $f_3(\phi,X)\to f_3(X)$. 
In this limit, $f_{3,\phi}=0$ and hence $\Delta_h=r_h^5>0$, consistently selecting the plus-sign branch.
Finally, using the current conservation law~\eqref{eq_J_V_all_function} of the vector field, we recover the same relation between the horizon and asymptotic charges as in Eq.~\eqref{J_Vcurrent_relation_f3-sol1-shift}.

To examine the connection between the near-horizon and asymptotic solutions, we perform numerical integration following the same two-step procedure as in Fig.~\ref{Fig-f3-1a-1}.
We specify the interaction function as $f_3=r_h^2 \phi /\Mpl(\bar{C}_3+\bar{D}_3 r_h^2/\Mpl^2 X)$ where $\bar{C}_3$ and $\bar{D}_3$ are dimensionless parameters.
Using the near-horizon expansions Eqs.~\eqref{f3-sol2-f1h1V1}-\eqref{f3-sol2-phi1} and the second-order coefficients provided in the Supplemental Material as boundary conditions, we integrate the equations outwards from the horizon. 
For the parameters $\mu=0.7$, $\phi^{(0)}=M_{\mathrm{pl}}$, $P=0.1\,M_{\mathrm{pl}}\,r_h$, $\bar{C}_3=-1.0$, and $\bar{D}_3=1.0$, the numerical profiles remain finite and smooth over the integration domain and approach the asymptotic expansions in Eqs.~\eqref{f3-sol1-inf-f}--\eqref{f3-sol-inf-phi-dependence} at large radii. 
The resulting profiles are qualitatively similar to those shown in Fig.~\ref{Fig-f3-2a}.

\section{\texorpdfstring{$\phi$}{phi}-dependent solutions in the \texorpdfstring{$g_3$}{g3} sector}
\label{phi_g3_sec}

In this section, we investigate the solutions driven by the interaction function $g_3$ in the presence of $\phi$-dependence.
We specify the interaction functions as
\be
f_3=0\,,\qquad g_3=g_3(\phi,X)\,,\qquad f_4=0\,.
\label{g3-phi-depend-inf-function}
\ee
We first derive the asymptotic solutions at spatial infinity, and proceed to the near-horizon solutions.

\subsection{Asymptotic solution}
\label{phi_g3_inf}

Substituting the choice~\eqref{g3-phi-depend-inf-function} into the field equations~\eqref{eq_f_all_function}-\eqref{eq_J_V_all_function} and using the asymptotic expansions~\eqref{expand_f_h_V_phi}, we obtain the iterative solutions at spatial infinity as 
\begin{align}
f=
&1-\frac{2 M}{r}+\frac{P^{2}+Q^{2}}{2 {\Mpl}^{2} r^{2}}+\frac{M {\phi_{[1]}}^{2}}{6 {\Mpl}^{2} r^{3}}+\frac{{\phi_{[1]}}^{2} \{4 M^{2} {\Mpl}^{2}-(P^{2}+Q^{2})\}}{12 {\Mpl}^{4} r^{4}} \label{g3-inf-f}\notag \\
&+\frac{{\phi_{[1]}} \{144 {\phi_{[1]}} M^{3} {\Mpl}^{2}+96 P^{2} g_3 {\Mpl}^{2}-52 {\phi_{[1]}} M (P^{2}+Q^2)-9 M {\phi_{[1]}}^{3}\}}{240 {\Mpl}^{4} r^{5}}+\mathcal{O}\bigg(\frac{1}{r^{6}}\bigg)\,,\\
h=
&1-\frac{2 M}{r}+\frac{P^{2}+Q^{2}+{\phi_{[1]}}^{2}}{2 {\Mpl}^{2} r^{2}}+\frac{M {\phi_{[1]}}^{2}}{2 {\Mpl}^{2} r^{3}}+\frac{{\phi_{[1]}}^{2} \{8 M^{2} {\Mpl}^{2}-(P^{2}+Q^{2})\}}{12 {\Mpl}^{4} r^{4}} \notag \\
&-\frac{{\phi_{[1]}} \{M {\phi_{[1]}}^{3}+12({P^{2}}+{Q^{2}}-4M^{2} {\Mpl}^{2}) M {\phi_{[1]}}-96 P^{2} g_3 {\Mpl}^{2}\}}{48{\Mpl}^{4} r^{5}}+\mathcal{O}\bigg(\frac{1}{r^{6}}\bigg)\,,\\
V=
&V_{\infty}+\frac{Q}{r}-\frac{Q {\phi_{[1]}}^{2}}{12 {\Mpl}^{2} r^{3}}-\frac{M Q {\phi_{[1]}}^{2}}{6 {\Mpl}^{2} r^{4}}-\frac{Q {\phi_{[1]}}^{2} \{48 M^{2} {\Mpl}^{2}-4( P^{2}+ Q^{2})-3 {\phi_{[1]}}^{2}\}}{160 {\Mpl}^{4} r^{5}} \notag \\
&-\frac{Q {\phi_{[1]}} \{48 {\phi_{[1]}} M^{3} {\Mpl}^{2}+12 P^{2} g_3 {\Mpl}^{2}-9 {\phi_{[1]}} M (P^{2}+Q^2)-8 M {\phi_{[1]}}^{3}\}}{90 {\Mpl}^{4} r^{6}}+\mathcal{O}\bigg(\frac{1}{r^{7}}\bigg)\,,\\
\phi=
&\phi_{\infty}+\frac{{\phi_{[1]}}}{r}+\frac{{\phi_{[1]}} M}{r^{2}}+\frac{{\phi_{[1]}} \{16 M^{2} {\Mpl}^{2}-2 (P^{2}+ Q^{2})-{\phi_{[1]}}^{2}\}}{12 {\Mpl}^{2} r^{3}}+\frac{M {\phi_{[1]}} \{12 M^{2} {\Mpl}^{2}-3 (P^{2}+Q^2)-2 {\phi_{[1]}}^{2}\}}{6 {\Mpl}^{2} r^{4}} \notag \\
&+\frac{1}{480 {\Mpl}^{4} r^{5}}\bigg[9{\phi_{[1]}}^{5}-16 \{29 M^{2} {\Mpl}^{2}-2 (P^{2}+ Q^{2})\} {\phi_{[1]}}^{3}+24 \{64 M^{4} {\Mpl}^{4}-24 M^{2} (P^{2}+Q^{2}) {\Mpl}^{2} \notag \\
&+(P^{2}+Q^{2})^{2}\} {\phi_{[1]}}-192 P^{2} g_3 M {\Mpl}^{4} \bigg]+\mathcal{O}\bigg(\frac{1}{r^{6}}\bigg)\,,
\label{g3-sol-inf-phi-dependence}
\end{align}
where we have set $ f_{[1]}= h_{[1]}=-2M$ and $ V_{[1]}=Q$. 
As in the $\phi$-dependent $f_3$ case, both $\phi_{[1]}$ and $\phi_\infty$ appear as independent integration constants of the asymptotic expansion and are not fixed by $M$, $Q$, or $P$ at this stage.
Since $\phi_\infty$ cannot be removed by a constant shift due to the existence of the explicit $\phi$-dependence in Eq.~\eqref{g3-phi-depend-inf-function}, it constitutes a physical parameter, and the solution is classified as having primary hair. 
If we impose the shift symmetry on these results, i.e., taking $g_3(\phi, X) \to g_3(X)$ and assuming that $\phi_{[1]}$ satisfies the secondary-hair relation given in Eq.~\eqref{relation_g3_shift_charge}, we correctly recover the shift-symmetric solutions Eqs.~\eqref{g3-shift-inf-f}-\eqref{g3-shift-inf-phi}. 
Regarding the asymptotic behavior, the interaction term $g_3$ evaluated at spatial infinity explicitly appears in $f$, $h$, and $\phi$ starting at $\mathcal{O}(r^{-5})$, and in $V$ at $\mathcal{O}(r^{-6})$. 
While this explicit appearance order is consistent with the shift-symmetric case, it is important to note that in the shift-symmetric limit, the leading scalar coefficient $\phi_{[1]}$ itself is sourced by the interaction as Eq.~\eqref{relation_g3_shift_charge} via the conservation of the scalar current, whereas here $\phi_{[1]}$ is a free parameter.

\subsection{Near-horizon solution for the \texorpdfstring{$g_3$}{g3}-\rm{IIa} case}
\label{g3IIasec}

We derive the near-horizon solution for the purely $\phi$-dependent coupling in the $g_3$ sector, specified by
\be
g_3=g_3(\phi)\,.
\label{g3-phi-depend-sol1-function}
\ee
We refer to this as the $g_3$-IIa solution.
Solving the field equations~\eqref{eq_f_all_function}-\eqref{eq_J_V_all_function} iteratively with the ansatz~\eqref{expand_horizon_fh_V_phi}, we find the linear-order coefficients as
\be
f^{(1)}=
h^{(1)}=\frac{1-\mu}{r_h}\,, \quad 
V^{(1)}=\frac{1}{r_h^2}\sqrt{\frac{2 \Mpl^{2} r_h^2[\Mpl^2 r_h^8\mu+2P^4g_3^2(2-\mu)]}{\Mpl^2 r_h^8+4P^4g_3^2}-P^2}\,,\quad
\phi^{(1)}=\frac{2 (3\mu -2) {\Mpl}^{2} {g_3} P^{2} {r_h}^{3}}{({\Mpl}^{2} {r_h}^{8}+4 P^{4} {g_3}^{2}) (1-\mu)}\,,
\label{g3-sol1-horizon}
\ee
where $g_3$ denotes $g_3(\phi^{(0)})$. 
The corresponding second-order coefficients are provided in the Supplemental Material.
We observe that $\phi$ and $V$ receive contributions from the interaction function already at the linear order, whereas $f$ and $h$ are affected starting from the second order.
As expected, the above coefficients reproduce those in the shift symmetric case given in Eq.~\eqref{g3-sol1-shift-first-order}, by taking the limit $g_3(\phi)\to \tilde{\beta}_3=$ constant. 
In the $P\to 0$ limit, the solution remains finite and smoothly reduces to the purely electric RN solution. 
Furthermore, the current conservation law~\eqref{eq_J_V_all_function} yields the same relation as in Eq.~\eqref{J_Vcurrent_relation_g3-sol1-shift}.

To examine the connection between the near-horizon and asymptotic solutions, we perform numerical integration.
We specify the interaction function as $g_3=(r_h^2/\Mpl)(\tilde{C}_3\phi/\Mpl)$, where $\tilde{C}_3$ is a dimensionless parameter. 
The parameters are set to be $\mu=0.5$, $\phi^{(0)}=M_{\mathrm{pl}}$, $P=0.5\,M_{\mathrm{pl}}\,r_h$, $\tilde{C}_3=1.0$. 
Following the procedure in Fig.~\ref{Fig-f3-1a-1}, we employ the two-step integration.
Using boundary conditions consistent with Eq.~\eqref{g3-sol1-horizon} and the second-order coefficients given in the Supplemental Material, we obtain numerical profiles that remain finite and smooth over the integration domain for the parameter values specified above.
At large radii, these profiles approach the asymptotic expansions in Eqs.~\eqref{g3-inf-f}-\eqref{g3-sol-inf-phi-dependence} and are qualitatively similar to those shown in Fig.~\ref{Fig-g3-1a}.
We have checked numerically that this behavior persists for the parameter values examined within $|\tilde{C}_3|\lesssim 1.0\,$.
%

\subsection{Near-horizon solution for the \texorpdfstring{$g_3$}{g3}-\rm{IIb} case}
\label{g3IIbsec}

Finally, we consider the general case~\eqref{g3-phi-depend-inf-function} where $g_3$ depends on not only $\phi$ but also $X$ explicitly, referred to as the $g_3$-IIb solution. 
Assuming $g_{3} \neq 0$ and $g_{3,X} \neq 0$ at the horizon, the iterative solution to the field equations~\eqref{eq_f_all_function}-\eqref{eq_J_V_all_function} with the ansatz~\eqref{expand_horizon_fh_V_phi} yields the same linear-order coefficients 
as Eqs.~\eqref{g3-sol2-shift-f1h1V1} and~\eqref{g3-sol2-shift-phi1} for the $g_3$-Ib case. 
Hence, $\phi$ and $V$ are modified at the linear order, while the metric functions are affected from the second order, analogous to the $g_3$-IIa case.
We note that, although the expressions are the same, the quantity $g_3$ appearing in the linear-order coefficients here contain the explicit $\phi$-dependence evaluated at the horizon. 
Moreover, the $\phi$-dependence of $g_3$ manifests itself in the second-order coefficients given in the Supplemental Material, through the derivatives of the interaction function, $g_{3,\phi}$ and $g_{3,\phi X}$, which are absent in the shift symmetric case. 
In the absence of explicit $\phi$ dependence, this solution reduces to the $g_3$-Ib solution. 
As in Sec.~\ref{g3Ibsec}, the algebraic solutions can be classified into two analogous branches. 
Since $g_3$ again enters the background equations~\eqref{eq_f_all_function}--\eqref{defJ_V} only through the combination $P^{2}g_3$, the same reasoning as in Sec.~\ref{g3Ibsec} applies, i.e., we restrict the analysis to the branch continuously connected to the purely electric configuration. 
We retain the branch characterized by $\phi^{(1)}_{-}$, for which the near-horizon coefficients calculated here approach their RN values as $P\to0$, with $R_{h,-}\to0$. 
For the branch characterized by $\phi^{(1)}_{+}$, the horizon Ricci scalar again remains finite in this limit so that the divergence of $\phi^{(1)}_{+}$ does not indicate a curvature singularity. 
This branch nevertheless does not admit a coefficient-by-coefficient connection to the RN expansion, and we do not use it in the subsequent solution analysis. 

The conserved current relation of the vector field is again consistent with Eq.~\eqref{J_Vcurrent_relation_g3-sol1-shift}. 
\begin{figure}[!htbp]
\includegraphics[width=0.55\linewidth]{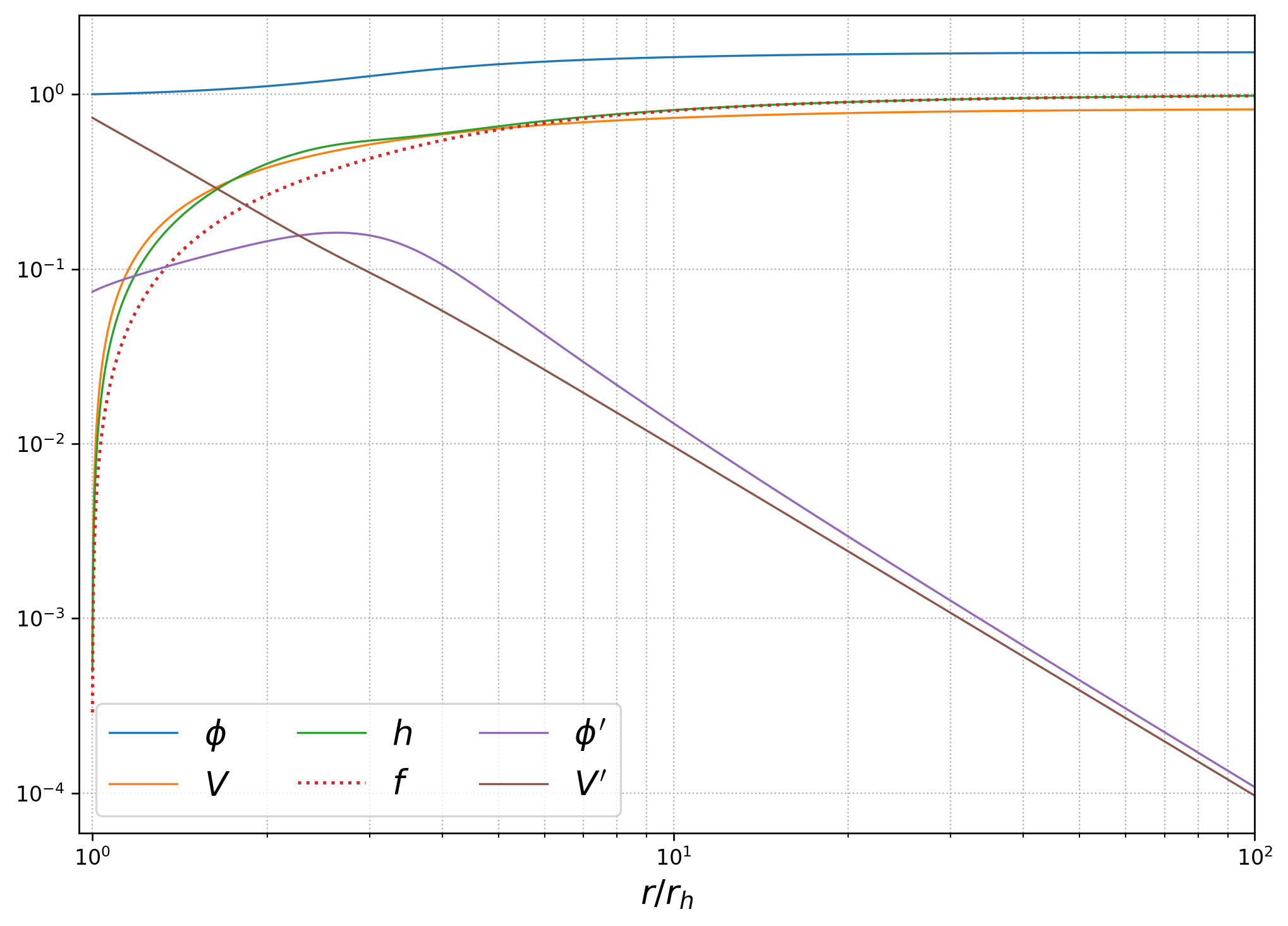}
\caption{\RaggedRight
Radial profiles for the positive $V^{(1)}$-branch of the $g_3$-IIb solution.
We numerically solve for the radial dependence of the fields from the horizon to the exterior region.
Normalization is the same as in Fig.~\ref{Fig-f3-2a}.
Boundary conditions are imposed at $r=1.001\,r_h$ consistent with Eqs.~\eqref{g3-sol2-shift-f1h1V1} and~\eqref{g3-sol2-shift-phi1} and the Supplemental Material, with parameters $\mu=0.5$, $\phi^{(0)}=M_{\mathrm{pl}}$, $P=0.5\,M_{\mathrm{pl}}\,r_h$, $\hat{C}_3=-10$, and $\hat{D}_3=3000$.
} \label{Fig-g3-2b_up_branch_Ver2}
\end{figure}

We present numerical results for the $g_3$-IIb solution by considering the interaction function
\be
g_3=\frac{r_h^2\phi}{\Mpl^2}
\left(
\hat{C}_3+\hat{D}_3\frac{r_h^2}{\Mpl^2}X
\right)\,,
\ee
where $\hat{C}_3$ and $\hat{D}_3$ are dimensionless parameters. 
For the retained scalar branch characterized by $\phi^{(1)}_{-}$, the near-horizon expansion admits two choices distinguished by the sign of $V^{(1)}$.
For each choice, we employ the same two-step integration procedure and impose boundary conditions consistent with the near-horizon expansion.

We first consider the positive-$V^{(1)}$ branch.
For $\hat{C}_3\hat{D}_3<0$, the numerical profiles remain finite and smooth over the integration domain and approach the asymptotic series at large radii for $|\hat{C}_3|\lesssim\mathcal{O}(10)$ and $|\hat{D}_3|\lesssim\mathcal{O}(10^3)$.
For $\hat{C}_3\hat{D}_3>0$, the same behavior is found when both couplings satisfy $|\hat{C}_3|\lesssim\mathcal{O}(1)$ and $|\hat{D}_3|\lesssim\mathcal{O}(1)$.
In the latter parameter region, the radial profiles are qualitatively similar to those of the $g_3$-Ib solution.
For $\hat{C}_3\hat{D}_3<0$ and $|\hat{C}_3|\lesssim\mathcal{O}(1)$, the behavior of $\phi'$ can be classified into two cases according to the magnitude of $|\hat{D}_3|$. When $|\hat{D}_3|$ is comparable to $|\hat{C}_3|$, namely, $|\hat{D}_3|=\mathcal{O}(1)$, the numerical profiles are similar to those of the $g_3$-Ia solution, and the effects specific to the explicit $\phi$ and $X$ dependences are difficult to distinguish from the radial profiles. By contrast, for $|\hat{D}_3|=\mathcal{O}(10)$--$\mathcal{O}(10^3)$, $\phi'$ remains approximately constant in the vicinity of the horizon, as in the $g_3$-Ib solution. This behavior can be understood analytically from the scaling $\left|\phi^{(2)}\right| \propto |\hat{C}_3|^8|\hat{D}_3|^{-1/2}\,,$ which shows that a large kinetic coupling suppresses the variation of $\phi'$ near the horizon unless $|\hat{C}_3|$ is also large.

For a relatively large coupling, $|\hat{C}_3|=\mathcal{O}(10)$, a correspondingly large kinetic coupling, $|\hat{D}_3|=\mathcal{O}(10^3)$, is required to obtain a smooth connection between the near-horizon and asymptotic solutions. In this regime, the metric profiles exhibit significant deviations from those obtained for $|\hat{C}_3|=\mathcal{O}(1)$. Figure~\ref{Fig-g3-2b_up_branch_Ver2} displays the positive-$V^{(1)}$ branch for $\hat{C}_3=-10$ and $\hat{D}_3=3000$, with $\mu=0.5$, $\phi^{(0)}=\Mpl$, and $P=0.5\Mpl r_h$. In this case, the second-order coefficient of the near-horizon expansion satisfies $\phi^{(2)}>0$, leading to a pronounced increase in $\phi'$. The resulting variation in $\phi'$ enhances $h'$ more strongly than $f'$, since the relevant terms scale as $h'\propto(\phi')^2$ and $f'\propto(\phi')^{-1}$. Consequently, $h$ exhibits a significant deviation from $f$. The explicit expressions for $f'$ and $h'$ are given in the Supplemental Material. We have numerically confirmed the same qualitative trend for other parameter choices yielding $\phi^{(2)}>0$.
The negative-$V^{(1)}$ branch exhibits qualitatively the same parameter dependence and radial behavior as the positive-$V^{(1)}$ branch, apart from the opposite sign in the vector-field sector. 
%

\section{\texorpdfstring{$\phi$}{phi}-dependent solutions in the \texorpdfstring{$f_4$}{f4} sector}
\label{phi_f4_sec}

In this section, we investigate the $f_4$ interaction sector. 
Unlike the previous sectors, the functional form of $f_4(\phi, X)$ is first constrained by the condition~\eqref{degeneracy_condition}, $\tilde{f}_4(\phi)=3f_{4,X}(\phi,X)/2$, to avoid the appearance of higher-order derivatives. 
Since the left-hand side is independent of $X$, the interaction function $f_4$ must be linear in $X$. 
Integrating this condition yields the general form
\begin{equation}
f_4(\phi,X)=g_4(\phi)+\frac{2}{3}\,\tilde{f}_4(\phi)\,X\,,
\label{eq:f4_linearX}
\end{equation}
where $g_4(\phi)$ appears as an integration constant with respect to $X$. 
For later convenience, we define a new arbitrary function $\tilde{g}_4(\phi) = 2\tilde{f}_4(\phi)/3$, allowing us to express the $f_4$ sector in terms of two independent $\phi$-dependent functions, $g_4(\phi)$ and $\tilde{g}_4(\phi)$, as
\begin{equation}
f_4(\phi,X)=g_4(\phi)+\tilde{g}_4(\phi)\,X\,.
\label{eq:f4_linearX_g4tilde}
\end{equation}
In the following, we examine the physical consequences of these two couplings separately.

\subsection{Asymptotic solution for the \texorpdfstring{$g_4$}{g4} sector}
\label{g4infsec}

We first focus on the purely $\phi$-dependent interaction $g_4(\phi)$ by setting:
\be
f_3=0\,,\qquad g_3=0\,,\qquad f_4=g_4(\phi)\,.
\label{g4-phi-depend-inf-function}
\ee
Substituting this setup into the field equations~\eqref{eq_f_all_function}-\eqref{eq_J_V_all_function} and employing the asymptotic expansions~\eqref{expand_f_h_V_phi}, we obtain the iterative solutions at spatial infinity as
\begin{align}
f=
&1-\frac{2 M}{r}+\frac{P^{2}+Q^{2}}{2 {\Mpl}^{2} r^{2}}+\frac{{\phi_{[1]}}^{2} M}{6 {\Mpl}^{2} r^{3}}+\frac{4 \{{\phi_{[1]}}^{2} M^{2}-6 {g_4} (P^{2}+Q^{2})\} {\Mpl}^{2}-{\phi_{[1]}}^{2} (P^{2}+Q^{2})}{12 {\Mpl}^{4} r^{4}}+\mathcal{O}\bigg(\frac{1}{r^{5}}\bigg)\,,
\label{g4-sol1-inf-f}\\
h=
&1-\frac{2 M}{r}+\frac{P^{2}+Q^{2}+{\phi_{[1]}}^{2}}{2 {\Mpl}^{2} r^{2}}+\frac{{\phi_{[1]}}^{2} M}{2 {\Mpl}^{2} r^{3}}+\frac{\{8 M^{2} {\Mpl}^{2}-(P^{2}+Q^{2})\} {\phi_{[1]}}^{2}-96 {\Mpl}^{2} P^{2} {g_4}}{12 {\Mpl}^{4} r^{4}}+\mathcal{O}\bigg(\frac{1}{r^{5}}\bigg)\,,\\
V=
&V_{\infty}+\frac{Q}{r}-\frac{Q {\phi_{[1]}}^{2}}{12 {\Mpl}^{2} r^{3}}-\frac{Q M (24 {\Mpl}^{2} {g_4}+{\phi_{[1]}}^{2})}{6 {\Mpl}^{2} r^{4}}+\mathcal{O}\bigg(\frac{1}{r^{5}}\bigg)\,,\\
\phi=
&\phi_{\infty}+\frac{{\phi_{[1]}}}{r}+\frac{{\phi_{[1]}} M}{r^{2}}+\frac{{\phi_{[1]}} \{16 M^{2} {\Mpl}^{2}-2 (P^{2}+ Q^{2})-{\phi_{[1]}}^{2}\}}{12 {\Mpl}^{2} r^{3}}+\frac{M {\phi_{[1]}} \{12 M^{2} {\Mpl}^{2}-3 (P^{2}+ Q^{2})-2 {\phi_{[1]}}^{2}\}}{6 {\Mpl}^{2} r^{4}}\notag \\
&+\frac{1}{480 {\Mpl}^{4} r^{5}}\bigg[9 {\phi_{[1]}}^{5}+16 \{-29 M^{2} {\Mpl}^{2}+2 (P^{2}+ Q^{2})\} {\phi_{[1]}}^{3}+24 \bigg(64 M^{4} {\Mpl}^{4}-4 \{6 (P^{2}+Q^{2}) M^{2}\notag \\
&-(5 P^{2} +Q^{2}) {g_4}\} {\Mpl}^{2}+(P^{2}+Q^{2})^{2}\bigg){\phi_{[1]}}+192 M {\Mpl}^{4} {g_{4,\phi}}( P^2-Q^2)\bigg]+\mathcal{O}\bigg(\frac{1}{r^{6}}\bigg)\,,
\label{g4-sol-inf-phi-dependence}
\end{align}
where we have set the integration constants as $f_{[1]}= h_{[1]}=-2M$ and $V_{[1]}=Q$. 
As for the general $\phi$-dependent case of the interaction functions $f_3$ and $g_3$, both $\phi_{[1]}$ and $\phi_\infty$ remain as independent integration constants, characterizing the solution with primary scalar hair. 
The interaction-dependent terms enter the metric and vector fields at $\mathcal{O}(r^{-4})$, while their leading-order contribution to the scalar field appears at $\mathcal{O}(r^{-5})$. 
Crucially, as evident from the $1/r^5$ coefficient in Eq.~\eqref{g4-sol-inf-phi-dependence}, the presence of the term proportional to $g_{4,\phi}(P^2-Q^2)$ explicitly breaks the $P$--$Q$ duality.

\subsection{Near-horizon solution for the \texorpdfstring{$g_4$}{g4}-IIa case}
\label{g4IIasec}

We now consider the near-horizon solutions for the purely $\phi$-dependent coupling~\eqref{g4-phi-depend-inf-function}, referred to as the $g_4$-IIa solution. 
Iteratively solving the field equations~\eqref{eq_f_all_function}-\eqref{eq_J_V_all_function} with the ansatz~\eqref{expand_horizon_fh_V_phi} around the horizon $r=r_h$ yields the following linear-order coefficients,
\begin{align}
f^{(1)}=
& h^{(1)}=\frac{1-\mu}{r_h}\,, 
\label{g4-sol1-horizon-f1h1}\\
V^{(1)}=
&\frac{\sqrt{({r_h}^{2}+8 {g_4}) \{2 {\Mpl}^{2} \mu  {r_h}^{4}+4 P^{2} ( \mu-1)(2g_4- {g_{4,\phi}} r_h{\phi^{(1)}}) -P^{2} {r_h}^{2}\}}}{({r_h}^{2}+8 {g_4}) {r_h}^{2}}\,,
\label{g4-sol1-horizon-V1}\\
\phi^{(1)}=
&\frac{8 {g_{4,\phi}} ({\Mpl}^{2} {r_h}^{4}+4P^{2} {g_4}) \bigg[\{4{g_4} (\mu -2)-{r_h}^{2}\} (\mu-1 ) P^{2}-{{\Mpl}^{2} \mu  {r_h}^{4}}\bigg]}{{r_h} (\mu-1 ) \bigg[8{g_{4,\phi}}^{2} P^4\{8{g_4} (\mu -2)-{r_h}^{2}\} -4{r_h}^{4}P^2{ \{4 {\Mpl}^{2} {g_{4,\phi}}^{2}+({r_h}^{2} +8 {g_4})g_4\} }-{\Mpl}^{2} {r_h}^{8}{ ({r_h}^{2}+8 {g_4})}\bigg]}\,,
\label{g4-sol1-horizon-phi1}
\end{align}
where we have chosen the $V^{(1)}>0$ branch. 
The corresponding second-order coefficients are provided in the Supplemental Material. 
A notable property of this solution is its behavior in the $P \to 0$ limit. 
Unlike the shift-symmetric configurations studied in Ref.~\cite{Heisenberg:2018vti}, the scalar hair in our setup remains non-vanishing even for purely electric BHs ($P=0$). 
This originates from the explicit $\phi$-dependence in $g_4(\phi)$, which provides a non-trivial source for the scalar field through $g_{4,\phi}$ even in the absence of the magnetic charge. 
From the current conservation law~\eqref{eq_J_V_all_function}, the asymptotic charge $Q$ is related to the horizon quantities by
\be
\left(r_h^{2}+8g_4\right)V^{(1)}=-Q\,.
\label{relation_g4}
\ee

To test whether the $g_4$-IIa near-horizon solution can be extended to spatial infinity, we perform numerical integration for representative choices of the coupling function and model parameters. 
We specify the interaction function as $g_4=(r_h^2c_4/\Mpl)\phi$, where $c_4$ is a dimensionless parameter.
Starting from the horizon with $\mu=0.7$, $\phi^{(0)}=\Mpl$, $P=0.5\,\Mpl r_h$, and $c_4=0.01$, we integrate the field equations outward using the two-step procedure.
The resulting profiles remain finite and smooth over the integration domain and approach the asymptotic expansions in Eqs.~\eqref{g4-sol1-inf-f}--\eqref{g4-sol-inf-phi-dependence} at large radii.
We have checked numerically that this behavior persists for the parameter values considered in the range $|c_4|\lesssim \mathcal{O}(10^{-2})$, with profiles qualitatively similar to those of the $g_3$-Ia solution.

\subsection{Asymptotic solution for the \texorpdfstring{$\tilde{g}_4$}{tildeg4} sector}
\label{tg4infsec}

Next, we investigate the quartic SVT sector involving the kinetic coupling by setting
\be
f_3=0\,,\qquad g_3=0\,,\qquad f_4=\tilde{g}_4(\phi)X\,,\label{g4tld-inf-function}
\ee
Substituting this choice into the field equations~\eqref{eq_f_all_function}-\eqref{eq_J_V_all_function} and employing the asymptotic expansions~\eqref{expand_f_h_V_phi}, we derive the iterative solutions at spatial infinity. 
Due to the complexity of the resulting expressions, they are provided in the Supplemental Material. 
The interaction contributes to the asymptotic series starting at $\mathcal{O}(r^{-8})$ in the metric functions $f$ and $h$, and at $\mathcal{O}(r^{-7})$ in the vector field $V$ and the scalar field $\phi$. 
Similar to the $g_4$ sector, these higher-order terms explicitly break the $P$--$Q$ duality.

\subsection{Near-horizon solution for the \texorpdfstring{$\tilde{g}_4$}{tildeg4}-IIb case}
\label{tg4IIbsec}

We derive the near-horizon solutions for the $X$-dependent coupling~\eqref{g4tld-inf-function}, referred to as the $\tilde{g}_4$-IIb solution. 
Solving the field equations~\eqref{eq_f_all_function}-\eqref{eq_J_V_all_function} iteratively around the horizon $r=r_h$ with the ansatz~\eqref{expand_horizon_fh_V_phi}, we obtain the linear-order coefficients as
\be
f^{(1)}=
h^{(1)}=\frac{1-\mu}{r_h}\,, \qquad 
V^{(1)}=\frac{\sqrt{2 {\Mpl}^{2} {r_h}^{2} \mu -P^{2}}}{{r_h}^{2}}\,, \qquad 
\phi^{(1)}=\frac{4 {\tilde{g}_{4}} (P^2-2 {\Mpl}^{2} {r_h}^{2}) \mu -{r_h}^{6}}{3 P^{2} {\tilde{g}_{4,\phi}} {r_h} (\mu-1 )}\,,
\label{g4tld-horizon-linear}
\ee
where we have chosen the $V^{(1)}>0$ branch. 
The corresponding second- and third-order coefficients are provided in the Supplemental Material. 
For generic values of $\tilde g_4$ at the horizon, the coefficient $\phi^{(1)}$ diverges in the purely electric limit $P\to0$.
To determine whether this divergent behavior is accompanied by a geometric singularity, we examine the Ricci scalar. For this nontrivial local branch, substituting the near-horizon expansion coefficients into the Ricci scalar and expanding the result around $P=0$ yields
\begin{equation}
R(r_h)
=-\frac{2\tilde g_{4}\left(r_h^4+8\mu \Mpl^{2}\tilde g_{4}\right)^2}{9r_h^4\tilde g_{4,\phi}^{2}\Mpl^2}\frac{1}{P^2}+\mathcal{O}(P^0)\,.
\label{eq:g4tilde_Ricci}
\end{equation}
For generic parameter values, the coefficient of the $P^{-2}$ term is nonzero, and hence $R(r_h)$ becomes unbounded as $P\to0$. 
Therefore, this nontrivial local branch does not admit a purely electric limit with a finite horizon Ricci scalar. 
We emphasize that this conclusion follows from the local near-horizon analysis of the limiting family $P\to0$ and does not imply the presence of a curvature singularity at the horizon for any fixed $P\neq0$. 
We also stress that the criterion applied here differs from the one used for the $g_3$ sectors in Secs.~\ref{g3Ibsec} and \ref{g3IIbsec}. 
There, the coupling $g_3$ enters the background equations only through the combination $P^{2}g_3$, so that the sector decouples at $P=0$ and the purely electric configuration is necessarily the RN solution. 
The branch selection is therefore a restriction to solutions continuously connected to that configuration. 
In the present sector, by contrast, the $f_4$ contributions to Eqs.~\eqref{eq_f_all_function}--\eqref{defJ_V} do not carry an overall factor of $P^{2}$, and the purely electric limit need not be the RN solution. 
Indeed, the $g_4$-IIa branch of Sec.~\ref{g4IIasec} retains scalar hair even at $P=0$. 
The present statement about the $\tilde g_4$-IIb branch is consequently not a selection criterion of the same kind, but the
stronger observation that a curvature invariant itself becomes unbounded along the limiting family.
There are, however, two exceptional parameter choices,
\begin{equation}
\tilde g_{4}=0\,,
\qquad
\tilde g_{4}
=-\frac{r_h^4}{8\mu \Mpl^{2}}\,,
\end{equation}
for which the coefficient of the leading $P^{-2}$ divergence in the Ricci scalar vanishes. For $\tilde g_{4}=0$, the coefficient of the term linear in $(r-r_h)$ in the near-horizon expansion of the Ricci scalar becomes unbounded as $P\to0$. For $\tilde g_{4}=-r_h^4/(8\mu \Mpl^{2})$, the cancellation persists at the next order in the Ricci-scalar expansion, but higher-order field coefficients either become unbounded or fail to approach their RN values. Thus, neither exceptional choice yields a smooth purely electric limit of the full near-horizon expansion.

Apart from these nontrivial local branches, the near-horizon equations also admit the no-hair branch $\phi'=0$.
Among the $\phi$-dependent sectors considered in this work, such a no-hair branch appears only in the $\tilde{g}_4$ sector. 
This property can be understood from the scalar-field equation, which can be written schematically as
\be
\phi'\,\mathcal{A}=\phi''\,\mathcal{B}\,,
\label{g4tld-scalar-eom-schematic}
\ee
where $\mathcal{A}$ and $\mathcal{B}$ are functions of the background quantities. 
The constant-scalar configuration $\phi'=\phi''=0$ therefore satisfies the scalar-field equation identically. 
This factorized structure of the scalar-field equation is absent in the other $\phi$-dependent interaction sectors considered here.
From the vector current conservation law~\eqref{eq_J_V_all_function}, we obtain the relation between the asymptotic charge $Q$ and horizon quantities as
\be
r_h^{2} V^{(1)}=-Q\,,
\ee
which is analogous to the relation~\eqref{relation_g4} in the $g_4$-IIa case.
This follows by the replacement $g_4\to \tilde{g}_4 X$ together with the evaluation $X=0$ at $r=r_h$.

To examine whether the local near-horizon $\tilde{g}_4$-IIb solution can be extended to spatial infinity, we specify the interaction function as
$
\tilde{g}_4=({r_h}^4/\Mpl^2)( \tilde{c}_4\phi/{\Mpl})\,,
$
where $\tilde{c}_4$ is a dimensionless parameter.
We solve Eqs.~\eqref{eq_f_all_function}-\eqref{eq_J_V_all_function}, together with the radial derivative of Eq.~\eqref{eq_h_all_function}, algebraically for $f''$, $f'$, $h'$, $\phi''$, and $V''$.
The resulting expression for $\phi''$ has the near-horizon structure
\be
\phi''=\frac{K(r_h)}{r-r_h}+{\cal O}\left((r-r_h)^0\right)\,,
\ee
where $K(r_h)$ is composed of the background quantities evaluated at the horizon. 
For the exact $\tilde{g}_4$-IIb near-horizon solution, substitution of Eq.~\eqref{g4tld-horizon-linear} gives $K(r_h)=0$. 
Thus, this expression does not imply an analytical divergence of $\phi''$ on the local branch.

The corresponding initial-value formulation is nevertheless numerically ill-conditioned near the horizon. 
Truncation of the near-horizon expansion and finite numerical precision can leave a small residual in the numerator, which is strongly amplified by the factor $1/(r-r_h)$ when the integration is initialized close to the horizon. 
In our calculations, this amplification prevented a robust outward integration using the present formulation. 
The unsuccessful integration does not establish either the existence or the absence of a global solution. 
Accordingly, with the present formulation, our result for this branch is restricted to the local near-horizon expansion, and we do not include it among the globally constructed solutions. 

\section{Conclusions}
\label{conclusion_sec}

%
In this paper, we have systematically constructed and classified static, spherically symmetric BH solutions carrying both electric and magnetic charges in the $U(1)$ gauge-invariant scalar-vector-tensor theories described by the action~\eqref{eq:specific_action}. 
Previous studies investigated hairy BH solutions either in purely electric configurations with general interaction functions~\cite{Heisenberg:2018vti} or in dyonic configurations restricted to the quadratic sector~\cite{Taniguchi:2024ear}. 
Extending these studies to the cubic and quartic SVT sectors, we have clarified how the magnetic charge modifies scalar-hair formation and the resulting spacetime structure across the different interaction sectors.

In Sec.~\ref{sec2}, we used the reduced action to identify interaction functions whose contributions remain regular at spatial infinity. 
From the full covariant equations, we derived the condition~\eqref{eq:covariant_HD_condition} that eliminates higher-order derivatives and independently recovered the same condition from the symmetry-reduced action. 
For the quartic sector considered in Sec.~\ref{phi_f4_sec}, this condition leads to the linear form of $f_4$ in $X$ given in Eq.~\eqref{eq:f4_linearX} and ensures that the resulting background equations are second order.

In Sec.~\ref{sec3}, we used the conservation of the scalar Noether current $J_\phi$ to determine the conditions for scalar-hair formation in the shift-symmetric sectors. 
In the $f_4$ sector, $J_\phi$ is proportional to $\phi'$, so the regular horizon condition $J_\phi=0$ enforces the trivial solution $\phi'=0$. 
In the $f_3$ sector, $\phi'$ cannot be factored out of $J_\phi$, allowing a nontrivial scalar profile to arise as an algebraic root. 
In the $g_3$ sector, the magnetic charge supports a nonzero conserved current and scalar hair with the nonvanishing asymptotic coefficient $\phi_{[1]}$ given in Eq.~\eqref{relation_g3_shift_charge}.

The solution families connected to spatial infinity also differ according to whether the shift symmetry is preserved. 
In the shift-symmetric $f_3$ and $g_3$ sectors discussed in Secs.~\ref{shift_f3_sec} and \ref{shift_g3_sec}, respectively, the scalar hair is secondary, with its asymptotic coefficient fixed by the horizon data through the conservation of the Noether current. 
For the asymptotically flat solutions with explicit $\phi$ dependence discussed in Secs.~\ref{phi_f3_sec}--\ref{phi_f4_sec}, the asymptotic expansion admits independent scalar quantities such as $\phi_\infty$ and $\phi_{[1]}$, allowing primary hair.

The magnetic charge produces interaction-dependent corrections that generically break the electric-magnetic duality in the asymptotic scalar and metric fields. 
Most notably, the cubic coupling $\tilde{f}_3$, or equivalently $g_3$, does not contribute in the purely electric configuration studied in Ref.~\cite{Heisenberg:2018vti}, but is activated by the magnetic charge and generates new hairy solutions through Eqs.~\eqref{eq_f_all_function}-\eqref{eq_J_V_all_function}. 
The behavior of the near-horizon expansions in the vanishing magnetic-charge limit depends on the interaction sector and the solution branch. 
The retained $f_3$-Ia (Sec.~\ref{f3Iasec}), $f_3$-Ib (Sec.~\ref{f3Ibsec}), $f_3$-IIa (Sec.~\ref{f3IIasec}), $f_3$-IIb (Sec.~\ref{f3IIbsec}), and $g_4$-IIa (Sec.~\ref{g4IIasec}) branches admit smooth purely electric limits through the orders evaluated and retain nontrivial scalar hair. 
The near-horizon expansions of the $g_3$-Ia (Sec.~\ref{g3Iasec}) and $g_3$-IIa (Sec.~\ref{g3IIasec}) solutions and the retained $g_3$-Ib (Sec.~\ref{g3Ibsec}) and $g_3$-IIb (Sec.~\ref{g3IIbsec}) branches instead approach the RN expansion through the orders evaluated, with the nontrivial scalar profile vanishing in this limit. 
For the $g_3$ branches, this classification reflects the decoupling of the $g_3$ sector at $P=0$ and amounts to selecting the branch continuously connected to the purely electric configuration.
The alternative local branches have a finite horizon Ricci scalar and are therefore outside the scope of the present analysis rather than physically excluded.
The nontrivial local $\tilde g_4$-IIb branch (Sec.~\ref{tg4IIbsec}) does not approach the full RN near-horizon expansion smoothly and is not included among the globally constructed hairy BH solutions.
We also identified distinct asymptotic behaviors for the scalar field depending on the interaction type and symmetry.
In the shift-symmetric $f_3$ sector, the scalar hair decays rapidly as $\mathcal{O}(r^{-4})$ [see Eq.~\eqref{solf3-shift-inf-phi}] as a consequence of the vanishing scalar current discussed in Sec.~\ref{shift_hair_con_f3}. 
By contrast, in the $\phi$-dependent $f_3$ sector, the scalar hair decays more slowly as $\mathcal{O}(r^{-1})$ [see Eq.~\eqref{f3-sol-inf-phi-dependence}].
The scalar hair induced by the $g_3$ interaction also decays as $\mathcal{O}(r^{-1})$ in both the shift-symmetric case~\eqref{g3-shift-inf-phi} and the $\phi$-dependent case~\eqref{g3-sol-inf-phi-dependence}.

For the branches for which the numerical integration was successful, we obtained smooth profiles connecting the near-horizon and asymptotic regimes for the representative parameter values examined.
Specifically, for the $g_3$-IIb case (Sec.~\ref{g3IIbsec}), we observed that the scalar kinetic coupling induces a strong backreaction on the metric, leading to a significant deviation between the gravitational potentials $f$ and $h$ as shown in Fig.~\ref{Fig-g3-2b_up_branch_Ver2}. 

Several interesting avenues for future research remain.
Since we have established the existence of hairy BH solutions, a natural next step is to analyze their linear stability against perturbations following the line of Ref.~\cite{Taniguchi:2025bmc,DeFelice:2024bdq}.
Based on the perturbation analysis, investigating the quasi-normal modes will be crucial for characterizing the ringdown phase of gravitational waves and identifying specific signatures of the magnetic charge and scalar hair. 
Furthermore, studying the optical appearance of these dyonic BHs, particularly the corrections to the shadow radius by analyzing null geodesics, would provide another powerful tool for testing these theories against observations. 
We leave these issues for future work.

\section*{Acknowledgments}
We would like to thank K.~Taniguchi for valuable discussions and insightful comments.
R K. is supported by the Grant-in-Aid for Scientific Research (C) of the JSPS No.~23K03421. 

\bibliographystyle{mybibstyle}
\bibliography{dyonicSVT}

@article{Bekenstein:1995un,
    author = "Bekenstein, J. D.",
    title = "{Novel \textquoteleft{}\textquoteleft{}no-scalar-hair\textquoteright{}\textquoteright{} theorem for black holes}",
    doi = "10.1103/PhysRevD.51.R6608",
    journal = "Phys. Rev. D",
    volume = "51",
    number = "12",
    pages = "R6608",
    year = "1995"
}

@article{Taniguchi:2024ear,
    author = "Taniguchi, Kitaro and Takagishi, Shinta and Kase, Ryotaro",
    title = "{Hairy black holes in extended Einstein-Maxwell-scalar theories with magnetic charge and kinetic couplings}",
    eprint = "2403.17484",
    archivePrefix = "arXiv",
    primaryClass = "gr-qc",
    doi = "10.1103/PhysRevD.110.044006",
    journal = "Phys. Rev. D",
    volume = "110",
    number = "4",
    pages = "044006",
    year = "2024"
}

@article{Gibbons:1987ps,
    author = "Gibbons, G. W. and Maeda, Kei-ichi",
    title = "{Black Holes and Membranes in Higher Dimensional Theories with Dilaton Fields}",
    reportNumber = "UTAP-48-87, LPTENS-87-10",
    doi = "10.1016/0550-3213(88)90006-5",
    journal = "Nucl. Phys. B",
    volume = "298",
    pages = "741--775",
    year = "1988"
}

@article{Shapere:1991ta,
    author = "Shapere, Alfred D. and Trivedi, Sandip and Wilczek, Frank",
    title = "{Dual dilaton dyons}",
    reportNumber = "IASSNS-HEP-91-33",
    doi = "10.1142/S0217732391003122",
    journal = "Mod. Phys. Lett. A",
    volume = "6",
    pages = "2677--2686",
    year = "1991"
}

@article{Heisenberg:2018vti,
    author = "Heisenberg, Lavinia and Tsujikawa, Shinji",
    title = "{Hairy black hole solutions in $U(1)$ gauge-invariant scalar-vector-tensor theories}",
    eprint = "1802.07035",
    archivePrefix = "arXiv",
    primaryClass = "gr-qc",
    doi = "10.1016/j.physletb.2018.03.059",
    journal = "Phys. Lett. B",
    volume = "780",
    pages = "638--646",
    year = "2018"
}

@article{Hui:2012qt,
    author = "Hui, Lam and Nicolis, Alberto",
    title = "{No-Hair Theorem for the Galileon}",
    eprint = "1202.1296",
    archivePrefix = "arXiv",
    primaryClass = "hep-th",
    doi = "10.1103/PhysRevLett.110.241104",
    journal = "Phys. Rev. Lett.",
    volume = "110",
    pages = "241104",
    year = "2013"
}

@article{DeFelice:2024bdq,
    author = "De Felice, Antonio and Kase, Ryotaro and Tsujikawa, Shinji",
    title = "{Scrutinizing black hole stability in cubic vector Galileon theories}",
    eprint = "2409.15606",
    archivePrefix = "arXiv",
    primaryClass = "gr-qc",
    reportNumber = "YITP-24-109, WUCG-24-08",
    doi = "10.1088/1475-7516/2024/10/072",
    journal = "JCAP",
    volume = "10",
    pages = "072",
    year = "2024"
}

@article{Herdeiro:2015waa,
    author = "Herdeiro, Carlos A. R. and Radu, Eugen",
    editor = "Herdeiro, Carlos A. R. and Cardoso, Vitor and Lemos, Jose P. S. and Mena, Filipe C.",
    title = "{Asymptotically flat black holes with scalar hair: a review}",
    eprint = "1504.08209",
    archivePrefix = "arXiv",
    primaryClass = "gr-qc",
    doi = "10.1142/S0218271815420146",
    journal = "Int. J. Mod. Phys. D",
    volume = "24",
    number = "09",
    pages = "1542014",
    year = "2015"
}

@article{Woodard:2015zca,
    author = "Woodard, Richard P.",
    title = "{Ostrogradsky's theorem on Hamiltonian instability}",
    eprint = "1506.02210",
    archivePrefix = "arXiv",
    primaryClass = "hep-th",
    reportNumber = "UFIFT-QG-15-03",
    doi = "10.4249/scholarpedia.32243",
    journal = "Scholarpedia",
    volume = "10",
    number = "8",
    pages = "32243",
    year = "2015"
}

@article{Heisenberg:2018acv,
    author = "Heisenberg, Lavinia",
    title = "{Scalar-Vector-Tensor Gravity Theories}",
    eprint = "1801.01523",
    archivePrefix = "arXiv",
    primaryClass = "gr-qc",
    doi = "10.1088/1475-7516/2018/10/054",
    journal = "JCAP",
    volume = "10",
    pages = "054",
    year = "2018"
}

@article{Maldacena:2020skw,
    author = "Maldacena, Juan",
    title = "{Comments on magnetic black holes}",
    eprint = "2004.06084",
    archivePrefix = "arXiv",
    primaryClass = "hep-th",
    doi = "10.1007/JHEP04(2021)079",
    journal = "JHEP",
    volume = "04",
    pages = "079",
    year = "2021"
}

@article{Taniguchi:2025bmc,
    author = "Taniguchi, Kitaro and Nishimura, Shunta and Tsukamoto, Naoki and Kase, Ryotaro",
    title = "{Linear perturbations of dyonic black holes in the lowest-order U(1) gauge-invariant scalar-vector-tensor theories}",
    eprint = "2504.21279",
    archivePrefix = "arXiv",
    primaryClass = "gr-qc",
    doi = "10.1103/718g-ncgn",
    journal = "Phys. Rev. D",
    volume = "112",
    number = "12",
    pages = "124043",
    year = "2025"
}

@article{Nomura:2020tpc,
    author = "Nomura, Kimihiro and Yoshida, Daisuke and Soda, Jiro",
    title = "{Stability of magnetic black holes in general nonlinear electrodynamics}",
    eprint = "2004.07560",
    archivePrefix = "arXiv",
    primaryClass = "gr-qc",
    reportNumber = "KOBE-COSMO-20-08",
    doi = "10.1103/PhysRevD.101.124026",
    journal = "Phys. Rev. D",
    volume = "101",
    number = "12",
    pages = "124026",
    year = "2020"
}

@article{Chen:2025aom,
    author = "Chen, Che-Yu and De Felice, Antonio and Tsujikawa, Shinji and Sano, Taishi",
    title = "{Vector Horndeski black holes in nonlinear electrodynamics}",
    eprint = "2509.23134",
    archivePrefix = "arXiv",
    primaryClass = "gr-qc",
    reportNumber = "RIKEN-iTHEMS-Report-25, YITP-25-152, WUCG-25-11",
    doi = "10.1103/fjqh-7gb2",
    journal = "Phys. Rev. D",
    volume = "113",
    number = "2",
    pages = "024027",
    year = "2026"
}

@article{Fernandes:2019kmh,
    author = "Fernandes, Pedro G. S. and Herdeiro, Carlos A. R. and Pombo, Alexandre M. and Radu, Eugen and Sanchis-Gual, Nicolas",
    title = "{Charged black holes with axionic-type couplings: Classes of solutions and dynamical scalarization}",
    eprint = "1908.00037",
    archivePrefix = "arXiv",
    primaryClass = "gr-qc",
    doi = "10.1103/PhysRevD.100.084045",
    journal = "Phys. Rev. D",
    volume = "100",
    number = "8",
    pages = "084045",
    year = "2019"
}

@article{Horndeski:1974wa,
    author = "Horndeski, Gregory Walter",
    title = "{Second-order scalar-tensor field equations in a four-dimensional space}",
    doi = "10.1007/BF01807638",
    journal = "Int. J. Theor. Phys.",
    volume = "10",
    pages = "363--384",
    year = "1974"
}

@article{Kolevatov:2016ppi,
    author = "Kolevatov, R. and Mironov, S.",
    title = "{Cosmological bounces and Lorentzian wormholes in Galileon theories with an extra scalar field}",
    eprint = "1607.04099",
    archivePrefix = "arXiv",
    primaryClass = "hep-th",
    reportNumber = "INR-TH-2016-024",
    doi = "10.1103/PhysRevD.94.123516",
    journal = "Phys. Rev. D",
    volume = "94",
    number = "12",
    pages = "123516",
    year = "2016"
}

@article{Kobayashi:2012wm,
    author = "Kobayashi, Tsutomu and Yamaguchi, Masahide and Yokoyama, Jun'ichi",
    title = "{Generalized G-inflation: Inflation with the most general second-order field equations}",
    eprint = "1105.5723",
    archivePrefix = "arXiv",
    primaryClass = "hep-th",
    reportNumber = "KUNS-2339, RESCEU-9-11",
    doi = "10.1143/PTP.126.511",
    journal = "Prog. Theor. Phys.",
    volume = "126",
    pages = "511--529",
    year = "2011"
}

@article{Heisenberg:2014rta,
    author = "Heisenberg, Lavinia",
    title = "{Generalization of the Proca Action}",
    eprint = "1402.7026",
    archivePrefix = "arXiv",
    primaryClass = "hep-th",
    doi = "10.1088/1475-7516/2014/05/015",
    journal = "JCAP",
    volume = "05",
    pages = "015",
    year = "2014"
}

@article{Goulart:2016cuv,
    author = "Goulart, Prieslei",
    title = "{Dyonic black holes and dilaton charge in string theory}",
    eprint = "1611.03093",
    archivePrefix = "arXiv",
    primaryClass = "hep-th",
    month = "11",
    year = "2016"
}

@article{Shepherd:2015dse,
    author = "Shepherd, Ben L. and Winstanley, Elizabeth",
    title = "{Dyons and dyonic black holes in ${\mathfrak {su}}(N)$ Einstein-Yang-Mills theory in anti{\textendash}de Sitter spacetime}",
    eprint = "1512.03010",
    archivePrefix = "arXiv",
    primaryClass = "gr-qc",
    doi = "10.1103/PhysRevD.93.064064",
    journal = "Phys. Rev. D",
    volume = "93",
    number = "6",
    pages = "064064",
    year = "2016"
}

@article{Meng:2018wza,
    author = "Meng, Kun",
    title = "{Hairy black holes of Lovelock{\textendash}Born{\textendash}Infeld-scalar gravity}",
    eprint = "1804.10951",
    archivePrefix = "arXiv",
    primaryClass = "gr-qc",
    doi = "10.1016/j.physletb.2018.07.029",
    journal = "Phys. Lett. B",
    volume = "784",
    pages = "56--61",
    year = "2018"
}

@article{Israel:1967wq,
    author = "Israel, Werner",
    title = "{Event horizons in static vacuum space-times}",
    doi = "10.1103/PhysRev.164.1776",
    journal = "Phys. Rev.",
    volume = "164",
    pages = "1776--1779",
    year = "1967"
}

@article{Carter:1971zc,
    author = "Carter, B.",
    title = "{Axisymmetric Black Hole Has Only Two Degrees of Freedom}",
    doi = "10.1103/PhysRevLett.26.331",
    journal = "Phys. Rev. Lett.",
    volume = "26",
    pages = "331--333",
    year = "1971"
}

@article{Ruffini:1971bza,
    author = "Ruffini, Remo and Wheeler, John A.",
    title = "{Introducing the black hole}",
    doi = "10.1063/1.3022513",
    journal = "Phys. Today",
    volume = "24",
    number = "1",
    pages = "30",
    year = "1971"
}

@article{Hawking:1971vc,
    author = "Hawking, S. W.",
    title = "{Black holes in general relativity}",
    doi = "10.1007/BF01877517",
    journal = "Commun. Math. Phys.",
    volume = "25",
    pages = "152--166",
    year = "1972"
}

@article{Chew:2022enh,
    author = "Chew, Xiao Yan and Yeom, Dong-han and Bl{\'a}zquez-Salcedo, Jose Luis",
    title = "{Properties of scalar hairy black holes and scalarons with asymmetric potential}",
    eprint = "2210.01313",
    archivePrefix = "arXiv",
    primaryClass = "gr-qc",
    doi = "10.1103/PhysRevD.108.044020",
    journal = "Phys. Rev. D",
    volume = "108",
    number = "4",
    pages = "044020",
    year = "2023"
}

@article{Chew:2023olq,
    author = "Chew, Xiao Yan and Lim, Kok-Geng",
    title = "{Scalar hairy black holes with an inverted Mexican-hat potential}",
    eprint = "2307.13972",
    archivePrefix = "arXiv",
    primaryClass = "gr-qc",
    doi = "10.1103/PhysRevD.109.064039",
    journal = "Phys. Rev. D",
    volume = "109",
    number = "6",
    pages = "064039",
    year = "2024"
}

@article{Ghosh:2023kge,
    author = "Ghosh, Rajes and Sk, Selim and Sarkar, Sudipta",
    title = "{Hairy black holes: Nonexistence of short hairs and a bound on the light ring size}",
    eprint = "2306.14193",
    archivePrefix = "arXiv",
    primaryClass = "gr-qc",
    doi = "10.1103/PhysRevD.108.L041501",
    journal = "Phys. Rev. D",
    volume = "108",
    number = "4",
    pages = "L041501",
    year = "2023"
}

@article{Chew:2024rin,
    author = "Chew, Xiao Yan and Yeom, Dong-han",
    title = {{Hairy Reissner-Nordstr{\"o}m black holes with asymmetric vacua}},
    eprint = "2401.09039",
    archivePrefix = "arXiv",
    primaryClass = "gr-qc",
    doi = "10.1103/PhysRevD.110.044036",
    journal = "Phys. Rev. D",
    volume = "110",
    number = "4",
    pages = "044036",
    year = "2024"
}

@article{Gibbons:1982ih,
    author = "Gibbons, G. W.",
    title = "{Antigravitating Black Hole Solitons with Scalar Hair in N=4 Supergravity}",
    doi = "10.1016/0550-3213(82)90170-5",
    journal = "Nucl. Phys. B",
    volume = "207",
    pages = "337--349",
    year = "1982"
}

@article{Garfinkle:1990qj,
    author = "Garfinkle, David and Horowitz, Gary T. and Strominger, Andrew",
    title = "{Charged black holes in string theory}",
    reportNumber = "UCSB-TH-90-66",
    doi = "10.1103/PhysRevD.43.3140",
    journal = "Phys. Rev. D",
    volume = "43",
    pages = "3140",
    year = "1991",
    note = "[Erratum: Phys.Rev.D 45, 3888 (1992)]"
}

@article{Campbell:1991rz,
    author = "Campbell, Bruce A. and Kaloper, Nemanja and Olive, Keith A.",
    title = "{Axion hair for dyon black holes}",
    reportNumber = "UMN-TH-934-91, CERN-TH-6086-91",
    doi = "10.1016/0370-2693(91)90474-5",
    journal = "Phys. Lett. B",
    volume = "263",
    pages = "364--370",
    year = "1991"
}

@article{Lee:1991jw,
    author = "Lee, Ki-Myeong and Weinberg, Erick J.",
    title = "{Charge black holes with scalar hair}",
    reportNumber = "CU-TP-515",
    doi = "10.1103/PhysRevD.44.3159",
    journal = "Phys. Rev. D",
    volume = "44",
    pages = "3159--3163",
    year = "1991"
}

@article{Reuter:1991cb,
    author = "Reuter, M.",
    title = "{A Mechanism generating axion hair for Kerr black holes}",
    reportNumber = "DESY-91-049, ITP-UH-3-91",
    doi = "10.1088/0264-9381/9/3/014",
    journal = "Class. Quant. Grav.",
    volume = "9",
    pages = "751--756",
    year = "1992"
}

@article{Monni:1995vu,
    author = "Monni, S. and Cadoni, M.",
    title = "{Dilatonic black holes in a S duality model}",
    eprint = "hep-th/9511067",
    archivePrefix = "arXiv",
    reportNumber = "INFN-CA-TH-9523",
    doi = "10.1016/0550-3213(96)00090-9",
    journal = "Nucl. Phys. B",
    volume = "466",
    pages = "101--111",
    year = "1996"
}

@article{Stefanov:2007qw,
    author = "Stefanov, Ivan Zh. and Yazadjiev, Stoytcho S. and Todorov, Michail D.",
    title = "{Scalar-tensor black holes coupled to Born-Infeld nonlinear electrodynamics}",
    eprint = "0704.3784",
    archivePrefix = "arXiv",
    primaryClass = "gr-qc",
    doi = "10.1103/PhysRevD.75.084036",
    journal = "Phys. Rev. D",
    volume = "75",
    pages = "084036",
    year = "2007"
}

@article{Stefanov:2007bn,
    author = "Stefanov, Ivan Zh. and Yazadjiev, Stoytcho S. and Todorov, Michail D.",
    title = "{Scalar-tensor black holes coupled to Euler-Heisenberg nonlinear electrodynamics}",
    eprint = "0708.3203",
    archivePrefix = "arXiv",
    primaryClass = "gr-qc",
    doi = "10.1142/S0217732307023560",
    journal = "Mod. Phys. Lett. A",
    volume = "22",
    pages = "1217--1231",
    year = "2007"
}

@article{Stefanov:2007eq,
    author = "Stefanov, Ivan Zh. and Yazadjiev, Stoytcho S. and Todorov, Michail D.",
    title = "{Phases of 4D scalar-tensor black holes coupled to Born-Infeld nonlinear electrodynamics}",
    eprint = "0708.4141",
    archivePrefix = "arXiv",
    primaryClass = "gr-qc",
    doi = "10.1142/S0217732308028351",
    journal = "Mod. Phys. Lett. A",
    volume = "23",
    pages = "2915--2931",
    year = "2008"
}

@article{Sheykhi:2014gia,
    author = "Sheykhi, A. and Hajkhalili, S.",
    title = "{Dilaton black holes coupled to nonlinear electrodynamic field}",
    eprint = "1504.04009",
    archivePrefix = "arXiv",
    primaryClass = "gr-qc",
    doi = "10.1103/PhysRevD.89.104019",
    journal = "Phys. Rev. D",
    volume = "89",
    number = "10",
    pages = "104019",
    year = "2014"
}

@article{Sheykhi:2014ipa,
    author = "Sheykhi, A. and Kazemi, A.",
    title = "{Higher dimensional dilaton black holes in the presence of exponential nonlinear electrodynamics}",
    eprint = "1506.01786",
    archivePrefix = "arXiv",
    primaryClass = "gr-qc",
    doi = "10.1103/PhysRevD.90.044028",
    journal = "Phys. Rev. D",
    volume = "90",
    number = "4",
    pages = "044028",
    year = "2014"
}

@article{Sheykhi:2015ira,
    author = "Sheykhi, A. and Naeimipour, F. and Zebarjad, S. M.",
    title = "{Phase transition and thermodynamic geometry of topological dilaton black holes in gravitating logarithmic nonlinear electrodynamics}",
    doi = "10.1103/PhysRevD.91.124057",
    journal = "Phys. Rev. D",
    volume = "91",
    number = "12",
    pages = "124057",
    year = "2015"
}

@article{Herdeiro:2018wub,
    author = "Herdeiro, Carlos A. R. and Radu, Eugen and Sanchis-Gual, Nicolas and Font, Jos{\'e} A.",
    title = "{Spontaneous Scalarization of Charged Black Holes}",
    eprint = "1806.05190",
    archivePrefix = "arXiv",
    primaryClass = "gr-qc",
    doi = "10.1103/PhysRevLett.121.101102",
    journal = "Phys. Rev. Lett.",
    volume = "121",
    number = "10",
    pages = "101102",
    year = "2018"
}

@article{Myung:2018vug,
    author = "Myung, Yun Soo and Zou, De-Cheng",
    title = {{Instability of Reissner{\textendash}Nordstr{\"o}m black hole in Einstein-Maxwell-scalar theory}},
    eprint = "1808.02609",
    archivePrefix = "arXiv",
    primaryClass = "gr-qc",
    doi = "10.1140/epjc/s10052-019-6792-6",
    journal = "Eur. Phys. J. C",
    volume = "79",
    number = "3",
    pages = "273",
    year = "2019"
}

@article{Myung:2019oua,
    author = "Myung, Yun Soo and Zou, De-Cheng",
    title = "{Stability of scalarized charged black holes in the Einstein{\textendash}Maxwell{\textendash}Scalar theory}",
    eprint = "1904.09864",
    archivePrefix = "arXiv",
    primaryClass = "gr-qc",
    doi = "10.1140/epjc/s10052-019-7176-7",
    journal = "Eur. Phys. J. C",
    volume = "79",
    number = "8",
    pages = "641",
    year = "2019"
}

@article{Hod:2019ulh,
    author = "Hod, Shahar",
    title = {{Spontaneous scalarization of charged Reissner-Nordstr{\"o}m black holes: Analytic treatment along the existence line}},
    eprint = "2002.01948",
    archivePrefix = "arXiv",
    primaryClass = "gr-qc",
    doi = "10.1016/j.physletb.2019.135025",
    journal = "Phys. Lett. B",
    volume = "798",
    pages = "135025",
    year = "2019"
}

@article{Dehghani:2019cuf,
    author = "Dehghani, M. and Setare, M. R.",
    title = "{Dilaton black holes with power law electrodynamics}",
    eprint = "1906.11063",
    archivePrefix = "arXiv",
    primaryClass = "gr-qc",
    doi = "10.1103/PhysRevD.100.044022",
    journal = "Phys. Rev. D",
    volume = "100",
    number = "4",
    pages = "044022",
    year = "2019"
}

@article{Filippini:2019cqk,
    author = "Filippini, Francesco and Tasinato, Gianmassimo",
    title = "{On long range axion hairs for black holes}",
    eprint = "1903.02950",
    archivePrefix = "arXiv",
    primaryClass = "gr-qc",
    doi = "10.1088/1361-6382/ab4371",
    journal = "Class. Quant. Grav.",
    volume = "36",
    number = "21",
    pages = "215015",
    year = "2019"
}

@article{Boskovic:2018lkj,
    author = "Boskovic, Mateja and Brito, Richard and Cardoso, Vitor and Ikeda, Taishi and Witek, Helvi",
    title = "{Axionic instabilities and new black hole solutions}",
    eprint = "1811.04945",
    archivePrefix = "arXiv",
    primaryClass = "gr-qc",
    doi = "10.1103/PhysRevD.99.035006",
    journal = "Phys. Rev. D",
    volume = "99",
    number = "3",
    pages = "035006",
    year = "2019"
}

@article{Astefanesei:2019pfq,
    author = "Astefanesei, D. and Herdeiro, C. and Pombo, A. and Radu, E.",
    title = "{Einstein-Maxwell-scalar black holes: classes of solutions, dyons and extremality}",
    eprint = "1905.08304",
    archivePrefix = "arXiv",
    primaryClass = "hep-th",
    doi = "10.1007/JHEP10(2019)078",
    journal = "JHEP",
    volume = "10",
    pages = "078",
    year = "2019"
}

@article{Fernandes:2020gay,
    author = "Fernandes, Pedro G. S.",
    title = "{Einstein{\textendash}Maxwell-scalar black holes with massive and self-interacting scalar hair}",
    eprint = "2003.01045",
    archivePrefix = "arXiv",
    primaryClass = "gr-qc",
    doi = "10.1016/j.dark.2020.100716",
    journal = "Phys. Dark Univ.",
    volume = "30",
    pages = "100716",
    year = "2020"
}

@article{Priyadarshinee:2021rch,
    author = "Priyadarshinee, Supragyan and Mahapatra, Subhash and Banerjee, Indrani",
    title = "{Analytic topological hairy dyonic black holes and thermodynamics}",
    eprint = "2108.02514",
    archivePrefix = "arXiv",
    primaryClass = "hep-th",
    doi = "10.1103/PhysRevD.104.084023",
    journal = "Phys. Rev. D",
    volume = "104",
    number = "8",
    pages = "084023",
    year = "2021"
}

@article{Priyadarshinee:2023cmi,
    author = "Priyadarshinee, Supragyan and Mahapatra, Subhash",
    title = "{Analytic three-dimensional primary hair charged black holes and thermodynamics}",
    eprint = "2305.09172",
    archivePrefix = "arXiv",
    primaryClass = "gr-qc",
    doi = "10.1103/PhysRevD.108.044017",
    journal = "Phys. Rev. D",
    volume = "108",
    number = "4",
    pages = "044017",
    year = "2023"
}

@article{Promsiri:2023yda,
    author = "Promsiri, Chatchai and Tangphati, Takol and Hirunsirisawat, Ekapong and Ponglertsakul, Supakchai",
    title = "{Scalarization of planar anti{\textendash}de Sitter charged black holes in Einstein-Maxwell-scalar theory}",
    eprint = "2302.04654",
    archivePrefix = "arXiv",
    primaryClass = "gr-qc",
    doi = "10.1103/PhysRevD.108.024015",
    journal = "Phys. Rev. D",
    volume = "108",
    number = "2",
    pages = "024015",
    year = "2023"
}

@article{Belkhadria:2023ooc,
    author = "Belkhadria, Zakaria and Pombo, Alexandre M.",
    title = "{Mixed scalarization of charged black holes: From spontaneous to nonlinear scalarization}",
    eprint = "2311.15850",
    archivePrefix = "arXiv",
    primaryClass = "gr-qc",
    doi = "10.1103/PhysRevD.110.044014",
    journal = "Phys. Rev. D",
    volume = "110",
    number = "4",
    pages = "044014",
    year = "2024"
}

@article{Al-Badawi:2025urb,
    author = "Al-Badawi, Ahmad and Ahmed, Faizuddin and Sakall{\i}, Izzet",
    title = "{Geometric, thermodynamic and perturbative properties of Frolov black holes surrounded by a cloud of strings and a global monopole}",
    doi = "10.1088/1572-9494/adfd3f",
    journal = "Commun. Theor. Phys.",
    volume = "78",
    number = "2",
    pages = "025401",
    year = "2026"
}

@article{Liang:2025hzr,
    author = "Liang, Yizhi and Tao, Jun and Yang, Rui",
    title = "{Black holes in f(R, T) gravity coupled with Euler{\textendash}Heisenberg electrodynamics}",
    eprint = "2506.14860",
    archivePrefix = "arXiv",
    primaryClass = "gr-qc",
    reportNumber = "CTP-SCU/2025014",
    doi = "10.1016/j.physletb.2025.140092",
    journal = "Phys. Lett. B",
    volume = "872",
    pages = "140092",
    year = "2026"
}

@article{Ahmed:2025did,
    author = "Ahmed, Faizuddin and Al-Badawi, Ahmad and Sakall{\i}, {\.I}zzet",
    title = "{Photon spheres, gravitational lensing/mirroring, and greybody radiation in deformed AdS-Schwarzschild black holes with phantom global monopole}",
    eprint = "2503.12092",
    archivePrefix = "arXiv",
    primaryClass = "gr-qc",
    doi = "10.1016/j.dark.2025.101988",
    journal = "Phys. Dark Univ.",
    volume = "49",
    pages = "101988",
    year = "2025"
}

@article{LIGOScientific:2016aoc,
    author = "Abbott, B. P. and others",
    collaboration = "LIGO Scientific, Virgo",
    title = "{Observation of Gravitational Waves from a Binary Black Hole Merger}",
    eprint = "1602.03837",
    archivePrefix = "arXiv",
    primaryClass = "gr-qc",
    reportNumber = "LIGO-P150914",
    doi = "10.1103/PhysRevLett.116.061102",
    journal = "Phys. Rev. Lett.",
    volume = "116",
    number = "6",
    pages = "061102",
    year = "2016"
}

@article{EventHorizonTelescope:2019dse,
    author = "Akiyama, Kazunori and others",
    collaboration = "Event Horizon Telescope",
    title = "{First M87 Event Horizon Telescope Results. I. The Shadow of the Supermassive Black Hole}",
    eprint = "1906.11238",
    archivePrefix = "arXiv",
    primaryClass = "astro-ph.GA",
    doi = "10.3847/2041-8213/ab0ec7",
    journal = "Astrophys. J. Lett.",
    volume = "875",
    pages = "L1",
    year = "2019"
}

@article{EventHorizonTelescope:2019ths,
    author = "Akiyama, Kazunori and others",
    collaboration = "Event Horizon Telescope",
    title = "{First M87 Event Horizon Telescope Results. IV. Imaging the Central Supermassive Black Hole}",
    eprint = "1906.11241",
    archivePrefix = "arXiv",
    primaryClass = "astro-ph.GA",
    doi = "10.3847/2041-8213/ab0e85",
    journal = "Astrophys. J. Lett.",
    volume = "875",
    number = "1",
    pages = "L4",
    year = "2019"
}

@article{EventHorizonTelescope:2022wkp,
    author = "Akiyama, Kazunori and others",
    collaboration = "Event Horizon Telescope",
    title = "{First Sagittarius A* Event Horizon Telescope Results. I. The Shadow of the Supermassive Black Hole in the Center of the Milky Way}",
    eprint = "2311.08680",
    archivePrefix = "arXiv",
    primaryClass = "astro-ph.HE",
    doi = "10.3847/2041-8213/ac6674",
    journal = "Astrophys. J. Lett.",
    volume = "930",
    number = "2",
    pages = "L12",
    year = "2022"
}

@article{Will:2014kxa,
    author = "Will, Clifford M.",
    title = "{The Confrontation between General Relativity and Experiment}",
    eprint = "1403.7377",
    archivePrefix = "arXiv",
    primaryClass = "gr-qc",
    doi = "10.12942/lrr-2014-4",
    journal = "Living Rev. Rel.",
    volume = "17",
    pages = "4",
    year = "2014"
}

@article{Dirac:1931kp,
    author = "Dirac, Paul Adrien Maurice",
    title = "{Quantised singularities in the electromagnetic field,}",
    reportNumber = "RX-722",
    doi = "10.1098/rspa.1931.0130",
    journal = "Proc. Roy. Soc. Lond. A",
    volume = "133",
    number = "821",
    pages = "60--72",
    year = "1931"
}

@article{Gibbons:1990um,
    author = "Gibbons, G. W.",
    editor = "Barrow, John D. and Henriques, A. B. and Lago, M. T. V. T. and Longair, M. S.",
    title = "{Selfgravitating magnetic monopoles, global monopoles and black holes}",
    eprint = "1109.3538",
    archivePrefix = "arXiv",
    primaryClass = "gr-qc",
    reportNumber = "DAMTP-R-90-31",
    doi = "10.1007/3-540-54293-0_24",
    journal = "Lect. Notes Phys.",
    volume = "383",
    pages = "110--138",
    year = "1991"
}

@article{Ortiz:1991eu,
    author = "Ortiz, Miguel E.",
    title = "{Curved space magnetic monopoles}",
    reportNumber = "MIT-CTP-2038",
    doi = "10.1103/PhysRevD.45.R2586",
    journal = "Phys. Rev. D",
    volume = "45",
    pages = "R2586--R2589",
    year = "1992"
}

@article{Lee:1991vy,
    author = "Lee, Ki-Myeong and Nair, V. P. and Weinberg, Erick J.",
    title = "{Black Holes in Magnetic Monopoles}",
    eprint = "hep-th/9112008",
    archivePrefix = "arXiv",
    reportNumber = "CU-TP-539, FERMILAB-PUB-91-312-A",
    doi = "10.1103/PhysRevD.45.2751",
    journal = "Phys. Rev. D",
    volume = "45",
    pages = "2751--2761",
    year = "1992"
}

@article{Lee:1991qs,
    author = "Lee, Ki-Myeong and Nair, V. P. and Weinberg, Erick J.",
    title = "{A Classical Instability of Reissner-Nordstrom Solutions and the Fate of Magnetically Charged Black Holes}",
    eprint = "hep-th/9111045",
    archivePrefix = "arXiv",
    reportNumber = "CU-TP-540, FERMILAB-PUB-91-326-A-T",
    doi = "10.1103/PhysRevLett.68.1100",
    journal = "Phys. Rev. Lett.",
    volume = "68",
    pages = "1100--1103",
    year = "1992"
}

@article{Breitenlohner:1991aa,
    author = "Breitenlohner, Peter and Forgacs, Peter and Maison, Dieter",
    title = "{Gravitating monopole solutions}",
    reportNumber = "MPI-PH-91-91",
    doi = "10.1016/0550-3213(92)90682-2",
    journal = "Nucl. Phys. B",
    volume = "383",
    pages = "357--376",
    year = "1992"
}

@article{Bronnikov:2000vy,
    author = "Bronnikov, Kirill A.",
    title = "{Regular magnetic black holes and monopoles from nonlinear electrodynamics}",
    eprint = "gr-qc/0006014",
    archivePrefix = "arXiv",
    doi = "10.1103/PhysRevD.63.044005",
    journal = "Phys. Rev. D",
    volume = "63",
    pages = "044005",
    year = "2001"
}

@article{Ayon-Beato:2000mjt,
    author = "Ayon-Beato, Eloy and Garcia, Alberto",
    title = "{The Bardeen model as a nonlinear magnetic monopole}",
    eprint = "gr-qc/0009077",
    archivePrefix = "arXiv",
    doi = "10.1016/S0370-2693(00)01125-4",
    journal = "Phys. Lett. B",
    volume = "493",
    pages = "149--152",
    year = "2000"
}

@article{Doneva:2017bvd,
    author = "Doneva, Daniela D. and Yazadjiev, Stoytcho S.",
    title = "{New Gauss-Bonnet Black Holes with Curvature-Induced Scalarization in Extended Scalar-Tensor Theories}",
    eprint = "1711.01187",
    archivePrefix = "arXiv",
    primaryClass = "gr-qc",
    doi = "10.1103/PhysRevLett.120.131103",
    journal = "Phys. Rev. Lett.",
    volume = "120",
    number = "13",
    pages = "131103",
    year = "2018"
}

@article{Silva:2017uqg,
    author = "Silva, Hector O. and Sakstein, Jeremy and Gualtieri, Leonardo and Sotiriou, Thomas P. and Berti, Emanuele",
    title = "{Spontaneous scalarization of black holes and compact stars from a Gauss-Bonnet coupling}",
    eprint = "1711.02080",
    archivePrefix = "arXiv",
    primaryClass = "gr-qc",
    doi = "10.1103/PhysRevLett.120.131104",
    journal = "Phys. Rev. Lett.",
    volume = "120",
    number = "13",
    pages = "131104",
    year = "2018"
}

@article{Tsukamoto:2024asy,
    author = "Tsukamoto, Naoki and Kase, Ryotaro",
    title = "{Neutral particle collisions near Gibbons-Maeda-Garfinkle-Horowitz-Strominger black holes after shadow observations}",
    eprint = "2409.04990",
    archivePrefix = "arXiv",
    primaryClass = "gr-qc",
    doi = "10.1103/PhysRevD.110.104063",
    journal = "Phys. Rev. D",
    volume = "110",
    number = "10",
    pages = "104063",
    year = "2024"
}

@article{Zhang:2024cbw,
    author = "Zhang, Chao and Kase, Ryotaro",
    title = "{Even-parity stability of hairy black holes in U(1) gauge-invariant scalar-vector-tensor theories}",
    eprint = "2404.11910",
    archivePrefix = "arXiv",
    primaryClass = "gr-qc",
    doi = "10.1103/PhysRevD.110.044047",
    journal = "Phys. Rev. D",
    volume = "110",
    number = "4",
    pages = "044047",
    year = "2024"
}

@inproceedings{Horowitz:1992jp,
    author = "Horowitz, Gary T.",
    title = "{The dark side of string theory: Black holes and black strings.}",
    eprint = "hep-th/9210119",
    archivePrefix = "arXiv",
    reportNumber = "UCSBTH-92-32",
    month = "10",
    year = "1992"
}

@article{Heisenberg:2017xda,
    author = "Heisenberg, Lavinia and Kase, Ryotaro and Minamitsuji, Masato and Tsujikawa, Shinji",
    title = "{Hairy black-hole solutions in generalized Proca theories}",
    eprint = "1705.09662",
    archivePrefix = "arXiv",
    primaryClass = "gr-qc",
    doi = "10.1103/PhysRevD.96.084049",
    journal = "Phys. Rev. D",
    volume = "96",
    number = "8",
    pages = "084049",
    year = "2017"
}

@article{Heisenberg:2017hwb,
    author = "Heisenberg, Lavinia and Kase, Ryotaro and Minamitsuji, Masato and Tsujikawa, Shinji",
    title = "{Black holes in vector-tensor theories}",
    eprint = "1706.05115",
    archivePrefix = "arXiv",
    primaryClass = "gr-qc",
    doi = "10.1088/1475-7516/2017/08/024",
    journal = "JCAP",
    volume = "08",
    pages = "024",
    year = "2017"
}

@article{Kase:2018owh,
    author = "Kase, Ryotaro and Minamitsuji, Masato and Tsujikawa, Shinji",
    title = "{Black holes in quartic-order beyond-generalized Proca theories}",
    eprint = "1803.06335",
    archivePrefix = "arXiv",
    primaryClass = "gr-qc",
    doi = "10.1016/j.physletb.2018.05.078",
    journal = "Phys. Lett. B",
    volume = "782",
    pages = "541--550",
    year = "2018"
}

@article{Kase:2018voo,
    author = "Kase, Ryotaro and Minamitsuji, Masato and Tsujikawa, Shinji and Zhang, Ying-Li",
    title = "{Black hole perturbations in vector-tensor theories: The odd-mode analysis}",
    eprint = "1801.01787",
    archivePrefix = "arXiv",
    primaryClass = "gr-qc",
    doi = "10.1088/1475-7516/2018/02/048",
    journal = "JCAP",
    volume = "02",
    pages = "048",
    year = "2018"
}

@article{Heisenberg:2018mgr,
    author = "Heisenberg, Lavinia and Kase, Ryotaro and Tsujikawa, Shinji",
    title = "{Odd-parity stability of hairy black holes in $U(1)$ gauge-invariant scalar-vector-tensor theories}",
    eprint = "1804.00535",
    archivePrefix = "arXiv",
    primaryClass = "gr-qc",
    doi = "10.1103/PhysRevD.97.124043",
    journal = "Phys. Rev. D",
    volume = "97",
    number = "12",
    pages = "124043",
    year = "2018"
}

@article{Kase:2023kvq,
    author = "Kase, Ryotaro and Tsujikawa, Shinji",
    title = "{Black hole perturbations in Maxwell-Horndeski theories}",
    eprint = "2301.10362",
    archivePrefix = "arXiv",
    primaryClass = "gr-qc",
    reportNumber = "WUCG-23-02, WUCG-23-01",
    doi = "10.1103/PhysRevD.107.104045",
    journal = "Phys. Rev. D",
    volume = "107",
    number = "10",
    pages = "104045",
    year = "2023"
}

@article{Gorji:2025lqr,
    author = "Gorji, Mohammad Ali and Mukohyama, Shinji and Petrov, Pavel and Yamaguchi, Masahide",
    title = "{Linear higher-order Maxwell-Einstein-Scalar theories}",
    eprint = "2509.16526",
    archivePrefix = "arXiv",
    primaryClass = "hep-th",
    doi = "10.1007/JHEP04(2026)090",
    journal = "JHEP",
    volume = "04",
    pages = "090",
    year = "2026"
}

@article{Mironov:2025dzz,
    author = "Mironov, S. and Shtennikova, A. and Valencia-Villegas, M.",
    title = "{Ghost-free, gauge invariant SVT generalizations of Horndeski theory}",
    eprint = "2509.16850",
    archivePrefix = "arXiv",
    primaryClass = "hep-th",
    doi = "10.1140/epjc/s10052-025-15125-6",
    journal = "Eur. Phys. J. C",
    volume = "85",
    number = "12",
    pages = "1378",
    year = "2025"
}

\end{document}